\documentclass{article}

\usepackage{PRIMEarxiv}

\usepackage[utf8]{inputenc} 
\usepackage[T1]{fontenc}    
\usepackage{hyperref}       
\usepackage{url}            
\usepackage{booktabs}       
\usepackage{multirow}
\usepackage{amsmath,amssymb,amsfonts}%
\usepackage{amsthm}%
\usepackage{mathrsfs}%
\usepackage{nicefrac}       
\usepackage{microtype}      
\usepackage{lipsum}
\usepackage{fancyhdr}       
\usepackage{graphicx}       
\graphicspath{{media/}}     
\usepackage[finalnew]{trackchanges} 

\title{An Effective Self-supervised Learning Method for Various Seismic Noise Attenuation
}

\author{
  Shijun Cheng \\
  Formerly at Innovation Academy for Precision Measurement Science and Technology, \\ Chinese Academy of Sciences, Wuhan 430077, China. \\
  Presently with King Abdullah University of Science and Technology, Thuwal 23955-6900, Saudi Arabia. \\
  \texttt{sjcheng.academic@gmail.com} \\
  \And
  Zhiyao Cheng and Chao Jiang,  \\
  Changchun University of Science and Technology, Changchun 130022, China. \\
 \texttt{zhiyaocheng@mails.cust.edu.cn, jiangchao@mails.cust.edu.cn} \\
 \And
 Weijian Mao and Qingchen Zhang \\
  Innovation Academy for Precision Measurement Science and Technology, \\ Chinese Academy of Sciences, Wuhan 430077, China. \\
  \texttt{wjmao@whigg.ac.cn, qczh@apm.ac.cn} \\
}

\begin{document}
\maketitle

\begin{abstract}
Faced with the scarcity of clean label data in real scenarios, seismic denoising methods based on supervised learning (SL) often encounter performance limitations. Specifically, when a model trained on synthetic data is directly applied to field data, its performance would drastically decline due to significant differences in feature distributions between the two. To address this challenge, we develop an effective self-supervised strategy. This strategy, while relying on a single denoising network model, adeptly attenuates various types of seismic noise. The strategy comprises two main phases: 1. The warm-up phase. By using prior knowledge or extracting information from real data, we introduce additional noise to the original noisy data, constructing a noisier data with intensified noise. This data serves as the input, with the original noisy data acting as pseudo-labels. This facilitates rapid pre-training of the network to capture a certain noise characteristics and boosts network stability, setting the stage for the subsequent phase. 2. Iterative data refinement (IDR) phase. During this phase, we use the predictions of the original noisy data from the network trained in the previous epoch as the pseudo-labels. We continue to add noise to the predictions, creating a new noisier-noisy dataset for the current epoch of network training. Through this iterative process, we progressively reduce the discrepancy between the original noisy data and the desired clean data. Ultimately, the network's predictions on the original noisy data become our denoised results. Validations under scenarios with random noise, backscattered noise, and blending noise reveal that our method not only matches the traditional SL techniques on synthetic data but significantly outperforms them on field data.
\end{abstract}

\keywords{Self-supervised learning  \and Seismic noise attenuation \and Iterative data refinement}
\section{Introduction}
Seismic data is invariably susceptible to contamination by noise due to the complex dynamic nature of the Earth's subsurface and the intricate nature of its acquisition \cite{krohn2008introduction}. Various factors contribute to this persistent contamination, ranging from natural sources like wind, ocean waves, and geological heterogeneity to human activities such as construction and industrial processes. As a result, the noise pollution manifests itself in various forms, such as random noise, trace-wise noise, backscattered noise, and ground roll. Denoising, in this context, becomes a critical step in all seismic processing workflows \cite{yilmaz2001seismic}, as it offer an effective means to enhance the integrity of seismic signals, improve data quality, and elevate the signal-to-noise ratio (SNR).

In the realm of traditional seismic data denoising, there's a comprehensive array of methods which can be grouped into three main categories: filter-based, sparse transformation-based, and modal decomposition-based techniques. Each category offers its own set of mathematical approaches and tools. Filter-based denoising, characterized by its judicious application of noise and signal distribution differences in time or frequency domains, seeks to exploit spectral characteristics and temporal smoothing for noise reduction, where the representative examples include median filtering \cite{liu20062d, liu20091d} and f-x deconvolution \cite{chase1992random, abma1995lateral}. Sparse transformation-based denoising, on the other hand, utilizes either analytic or learning-based methods to unveil the sparse representation of seismic data in transformed domains, enabling the selective removal of noise through coefficient thresholding. Commonly, the analytic method leverages a fixed set of basis to transform the seismic data from the t-x domain, such as Fourier transform \cite{yilmaz2001seismic}, wavelet transform \cite{zhang2003physical, jian2006denoising}, curvelet transform \cite{neelamani2008coherent, naghizadeh2018ground}, shearlet transform \cite{hosseini2015adaptive, zhang2018multicomponent, zhang2019strong}, and seislet transform \cite{liu2009high, fomel2010seislet}. In contrast, the learning-based method for sparse representation involves the process of adaptively learning or training a dictionary of basis functions directly from the seismic data. Examples of such approach is K-singular value decomposition \cite{cheng2018seismic, chen2020fast, chen2022retrieving}, data-driven tight frame \cite{yu2015interpolation, wang2019adaptive}, double-sparsity dictionary \cite{chen2016double, zhang2019seismic}, and etc. Alternatively, modal decomposition-based denoising strategies delve into the inherent modes or components within the seismic data, endeavoring to discriminate and retain signal-carrying modes while discarding perturbing noise, such as empirical mode decomposition \cite{bekara2009random, chen2014random, gomez2016simple} and singular value decomposition \cite{kendall2005svd, kreimer2012tensor}. It is commendable that these methods have significantly propelled the advancement of seismic denoising techniques. However, these methods often necessitate substantial reliance on expert experience for manual parameter tuning. Take, for example, sparse representation-based seismic denoising, which typically entails the establishment of a reliable threshold coefficient. Even minor deviations in these coefficients can exert a profound influence on the final denoising results. Consequently, there exists a growing consensus within the broader seismological community for the pursuit of automated and intelligent approaches, aimed at mitigating the burden of manual intervention.

Recently, the rapid advancement of deep learning (DL) technology has ushered in a new dawn for intelligent seismic denoising. A novel seismic denoising paradigm based on deep neural networks (DNNs) has been emerged. This approach typically involves the construction of extensive labeled datasets containing both noisy and clean seismic signals, and then leverages supervised learning (SL) algorithms to train DNN. The trained DNNs exhibit adeptness in discerning intricate patterns within the data, thereby effectively applying the learned knowledge to effectively denoise seismic recordings. Currently, numerous studies have utilized the synthetic data to train convolutional neural networks (CNNs) due to the challenges associated with acquiring labels for field data \cite{yu2019deep, liu2018random, liu2019poststack, zhu2019seismic, dong2019desert, dong2020denoising, saad2020deep, tang2023simultaneous, cheng2023meta, cheng2023elastic}. Although these approaches have demonstrated commendable denoising efficacy on synthetic data, their performance often encounters substantial degradation when applied to field data. Consequently, to enhance the denoising capabilities of DNNs on field data, some researches have sought to extract noise from observed field data and incorporate it into synthetic data for network training \cite{zhao2018low, yao2020dnresnext, brusova2021innovative, harsuko2022storseismic}. Although these approaches has the potential to improve denoising performance to a certain extent, there persists a significant data bias between synthetic and field data, rendering the signal characteristics learned directly from synthetic data less amenable to field data. Alternatively, some researchers leverage the conventional seismic denoising methods to provide labeled data for noisy field data \cite{liu2018random, mandelli2019interpolation, yuqing2019random}. Nevertheless, this approach merely enhances denoising efficiency and typically remains commensurate with the performance of the underlying conventional seismic denoising methods it relies upon \cite{birnie2021potential, dong2023potential}. Consequently, the imperative question of how to mitigate the reliance on labeled data during DNNs training remains an exigent concern.

To address this challenge, some preliminary studies have been presented to train DNNs using unsupervised learning methods, or often referred to as self-supervised learning (SSL). This approach involves training models exclusively on noisy data, thereby eliminating the need for clean labeled data. One pioneering avenue of research explores the application of autoencoders. In this paradigm, raw seismic data undergoes encoding into a lower-dimensional latent space via an encoder network, followed by the reconstruction of denoised results from the compressed latent representation using a decoder network. Autoencoders have demonstrated promise in attenuating both random \cite{yang2021unsupervised, saad2021fully} and coherent noise \cite{markovic2022diffraction} within seismic data. A second strategy is rooted in the Noise2Noise framework \cite{lehtinen2018noise2noise}, wherein the denoising process is learned by capitalizing on pairs of noisy data \cite{shao2022noisy2noisy, wang2023self, zhao2023sample2sample}. Specifically, two noisy seismic data instances are constructed, each sharing identical seismic signals but exhibiting disparate noise levels. The network is optimized by minimizing errors between paired noisy data instances. A third stratagem embraces a mask-based SSL framework, which encompasses both input data masking and network masking. The first modality presupposes strong inter-signal correlations and noise independence, thereby permitting the masked signals to be accurately recovered through contextual information, effectively eliminating the noise. This method has been applied to the attenuation of random \cite{birnie2021potential, cao2022self, sun2022seismic, birnie2023explainable} and coherent noise \cite{abedi2022multidirectional, liu2023trace}. Meanwhile, network masking, achieved through the masking of the neural network's receptive fields, has demonstrated performance in removing the blending-noise \cite{wang2022self, luiken2023integrating}. Most recently, another novel SSL method has emerged, utilizing Gabor filters as a substitute for the conventional convolution kernels within CNNs \cite{liu2023gabor}. The adoption of Gabor filters is driven by their capacity to encompass essential parameters, such as wavelength, frequency, and phase. Consequently, this allows for the embedding of a priori wave physics information pertaining to the desired signals for extraction. As a result, noise can be removed through an SSL paradigm. This method has been empirically validated as efficacious in the attenuation of both pseudo-random noise and ground roll. While some of the aforementioned SSL methods have demonstrated notable efficacy in seismic denoising, the expanse of this domain beckons us to delve further into the exploration of alternative strategies to address the formidable challenges posed by the diverse types and intensities of noise prevalent in real-world scenarios.

In this study, we adopt a Noisier2Noise strategy \cite{moran2020noisier2noise} to develop an SSL seismic denoising workflow. This procedure entails the direct introduction of noise onto the already noisy seismic data, where the noise can be extract from field data or determined by prior knowledge, to create noisier-noisy seismic data pairs. In this way, we can effectively obviate the prerequisite for paired noise data, as conventionally encountered in the Noise2Noise method. We leverage the classic U-Net \cite{ronneberger2015u}, which has found ubiquitous application across various seismic processing workflows \cite{yang2019deep, zhang2021least, wu2019faultseg3d, cheng2023meta}, as our denoising DNN model. The U-Net model will be first trained on the noisier-noisy dataset as the warm-up phase. This phase serves the dual purpose of enhancing network stability while partially capturing specific noise characteristics. After the warm-up training stage, to mitigate the data bias between the noisier-noisy dataset and the noisy-clean dataset, which has been reported as potentially fundamental to the constraints on SSL performance \cite{zhang2022idr}, we iteratively employ the denoising network trained in the preceding iteration to refine the raw noisy seismic data, effectively generating new pseudo-labeled data. The corresponding input is deduced by adding noise to the pseudo-labeled data, and thus we can create a new noisier-noisy dataset. Through this iterative refinement process, we progressively refine the quality of the raw noisy seismic data. Once reaching the predefined maximum number of iterations, the network's predictions on the raw noisy data constitute our denoised results. To comprehensively evaluate the efficacy of our method, we subject it to various seismic noise types, encompassing random noise, backscattered noise, and blending noise. Our evaluation include both synthetic and field data to validate our method's performance through comparisons with traditional denoising approaches and SL denoising methods. 
\section{Method}
\subsection{Supervised learning vs Noisier2Noise}
The recorded seismic data can be represented as \cite{yu2019deep}
\begin{equation}\label{eq1}
y=x+n,
\end{equation}
where $x$ is the clean seismic data, $y$ is the noisy seismic data, contaminated by noise $n$. The ultimate objective in developing denoising algorithms is to eliminate the influence of noise $n$ from the noisy data $y$, thereby approximating the clean data $x$ as closely as possible. Presently, the concept of neural network (NN)-based denoising algorithms is to leverage machine learning techniques to learn the underlying nonlinear relationships between the noisy data and the clean data as follows 
\begin{equation}\label{eq2}
x=\mathrm{NN}(y;\boldsymbol{\theta}),
\end{equation}
where $\boldsymbol{\theta}$, denoting the trainable network parameters, is optimized through extensive training on large datasets. 

\begin{figure}[htp]
\centering
\includegraphics[width=0.3\textwidth]{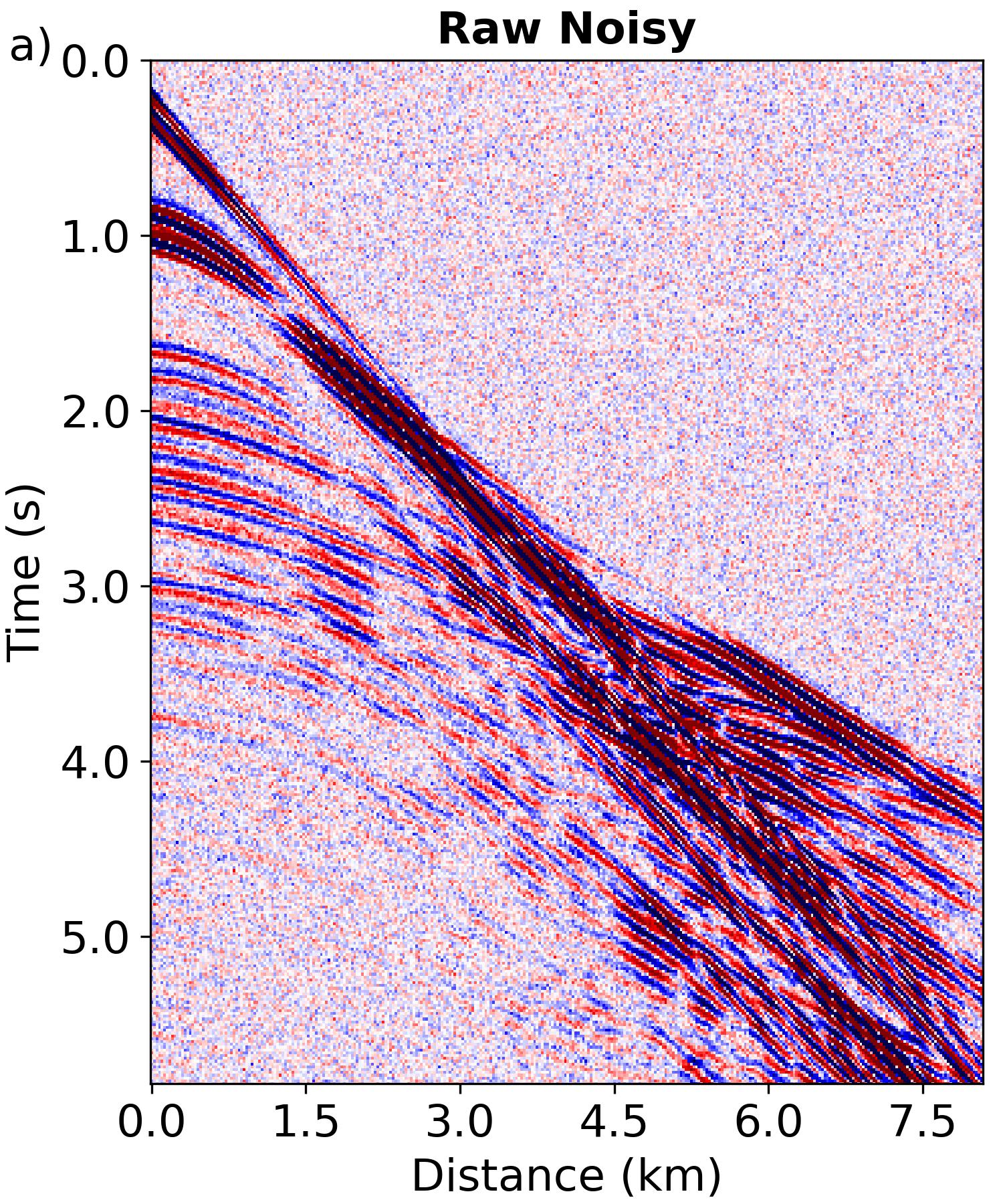}
\hspace{1cm}
\includegraphics[width=0.3\textwidth]{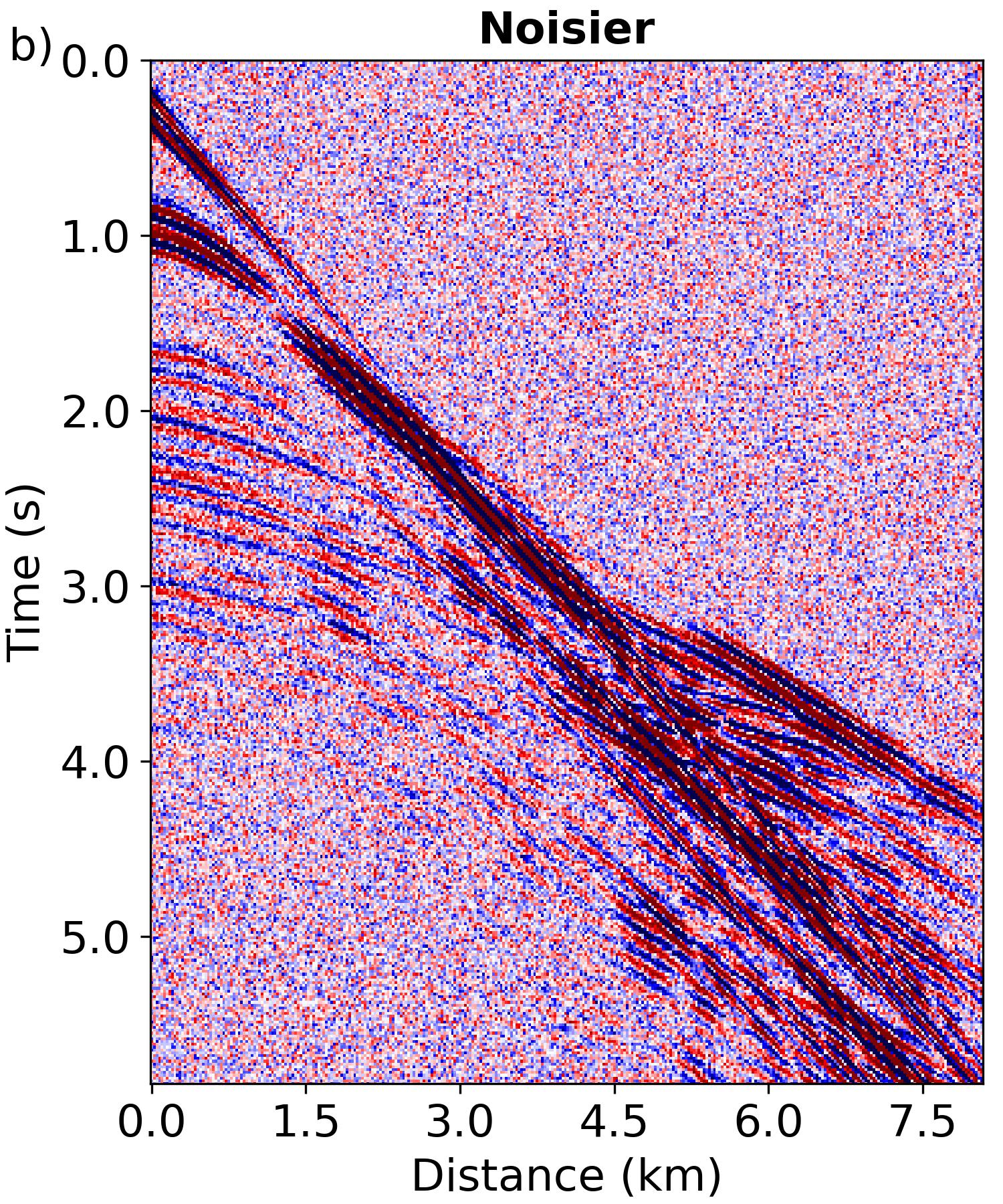}
\caption{An illustration of the Noisier2Noise method. (a) denotes the raw noisy data. (b) is a noisier data compared to raw noisy data, which comes from directly injecting noise into the raw noisy data. (a) and (b) can be regarded as the label and input of the denoising network, respectively.}
\label{fig1}
\end{figure}

Typically, NNs are trained in an SL fashion. However, due to the unavailability of labels corresponding to noisy field data, synthetic noisy data is often generated by adding noise to simulated clean data. This synthetic training data, denoted as noise-clean data $\{(x_i+n_i,x_i)\}_{i=1}^N$, is employed to train the NN, which is subsequently applied to denoise test data, which can be either synthetic or field data.  In stark contrast, an SSL Noisier2Noise approach, which is illustrated in Fig. \ref{fig1}, directly injects noise into the raw noisy data (Fig. \ref{fig1}a) to create data with stronger noise. In this paradigm, the data with stronger noise (see Fig. \ref{fig1}b) is considered as the input to the network, while the raw noisy data serves as pseudo-label data. Training data generated in this manner is referred to as noisier-noisy data $\{(y_i+n_i,y_i)\}_{i=1}^N$. It's important to note that the added noise should match the type of noise to be removed, which can be determined through prior knowledge, such as random noise, or extracted directly from field data.

While utilizing the Noisier2Noise approach effectively eliminates the need for labeled data, a notable data bias exists between the noisier-noisy data and noise-clean data. This bias can significantly impact the denoising capability of the neural network trained on noisy training data when applied to the raw noisy data. In the following, we will compare the performance of NNs trained on synthetic noisy-clean seismic data and noisier-noisy seismic data to elucidate this critical observation. We create three distinct datasets, namely the noisy-clean seismic dataset and two noisier-noisy datasets. The noisy-clean seismic dataset is generated by introducing random noise with levels ranging from 5 to 40 into synthetic data, employing the following equation:
\begin{equation}\label{eq3}
y_i=x_i+0.01\epsilon \cdot std(x_i) \cdot rand(0,1), ~ i=1, \cdot \cdot \cdot, N,
\end{equation}
where $\epsilon$ denotes the noise level, $std(x_i)$ represents the standard deviation of clean data $x_i$, and $rand(0,1)$ is the standard normal distribution. The raw noisy data in the two noisier-noisy datasets are obtained by adding random noise with the levels of 10 (namely noisier-noisy 1) and 20 (namely noisier-noisy 2), respectively. Correspondingly, the pseudo-label data, similar to the noisy-clean data, are generated by introducing random noise with the levels ranging from 5 to 40 to the raw noisy data.

We train U-Nets from scratch on all three datasets, and the architecture and training settings will be presented in the subsequent third section. The trained U-Nets are tested on seven noisy data, each sharing a common clean data, with noise levels of 5, 15, 25, 35, 45, 55, and 65 respectively. To quantitatively assess the denoising performance, we employ the SNR as our evaluation metric as follows:
\begin{equation}\label{eq4}
\text{SNR} = 10\text{log}\frac{\lvert\lvert x \lvert\lvert^2}{\lvert\lvert x - \widehat{x} \lvert\lvert^2},
\end{equation}
where $\widehat{x}$ represents the denoising results.

\begin{figure}[htp]
\centering
\includegraphics[width=0.3\textwidth]{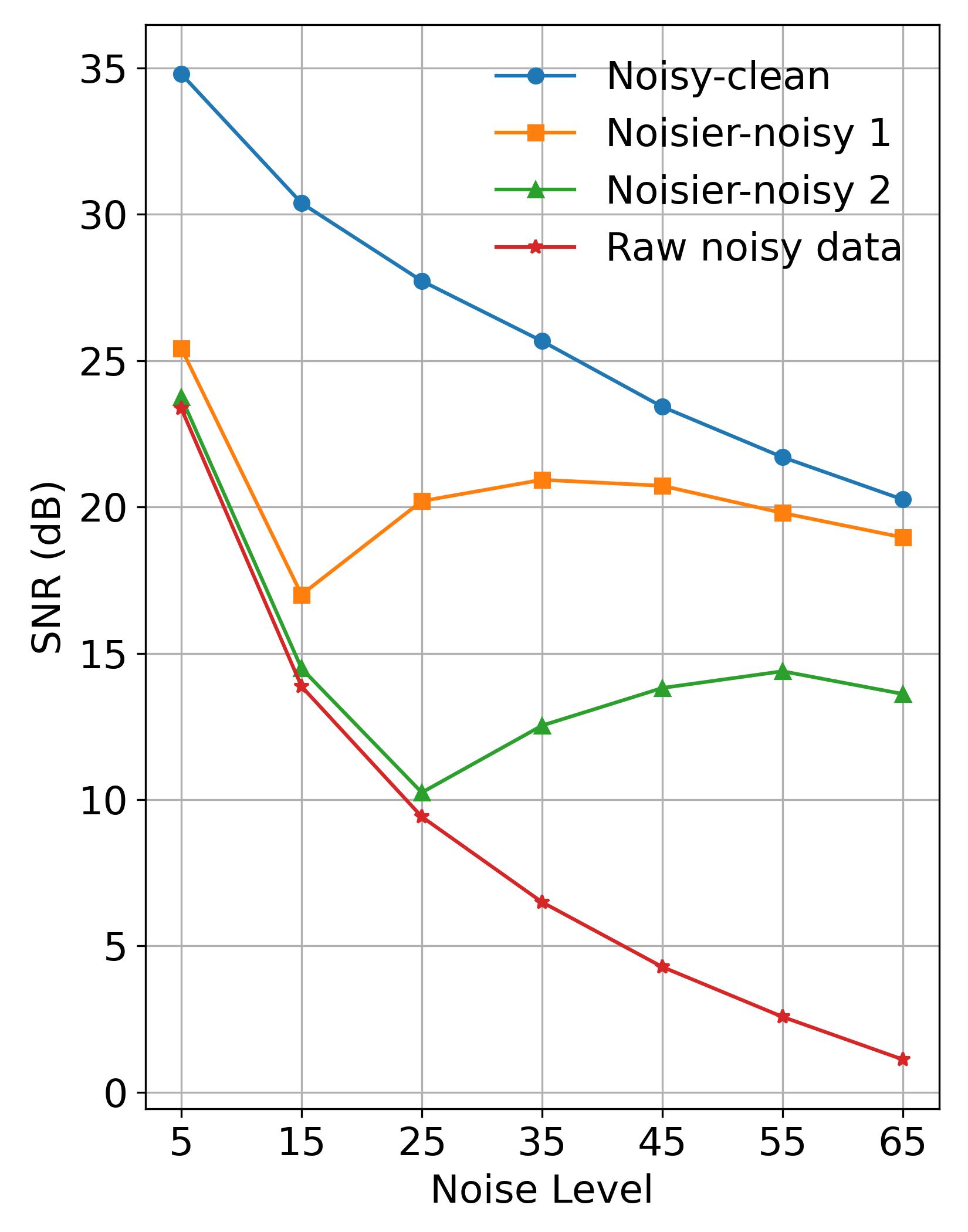}
\caption{The denoising performance comparison of denoising U-Nets trained on noisy-clean dataset and two noisier-noisy dataset across seven different noise levels, with the red line representing the SNR of noisy test data. }
\label{fig2}
\end{figure} 

Fig. \ref{fig2} presents a comparative analysis of the denoising performance of NNs trained on three distinct datasets when evaluated on test data. The red line in the figure corresponds to the SNR of the noisy test data across seven different noise levels. A salient observation is that the performance of denoising networks trained on datasets characterized by a reduced data bias, such as “Noisier-noisy 1”, incrementally approximates that of a NN educated on a noisy-clean dataset. This sheds light on the intrinsic superiority of SL over SSL, particularly when performing within the realms of synthetic data. Furthermore, an additional discovery is that the network trained on noisier-noisy dataset can have certain denoising capabilities. This prowess becomes pronounced, especially when the noise level aligns coherently with the one encountered during their training phase. For example, the raw noisy data for “Noisier-noisy 1” is engineered by superimposing random noise, quantified at a level of 10, onto clean data, supplemented by an extended noise level ranging between 5 and 40. Hence, the bandwidth of noise levels that a denoising network, trained on the “Noisier-noisy 1” dataset, can effectively learn is conceivable from 15 to 50. Consequently, there's a perceptible augmentation in the denoising efficacy of networks commencing from a noise level of 15, attributable to their prowess in capturing the noise characteristics proffered by this specific dataset.

These observations suggest the potential of leveraging the Noisier2Noise strategy to mitigate certain seismic noise interferences. The scope of such noise attenuation is determined by the overlapping noise characteristics between the noisier-noisy and noisy-clean datasets. Hence, if we can reduce the data bias between the noisier-noisy and noisy-clean datasets, aligning the noise distribution features of both datasets becomes feasible, thereby substantially augmenting the efficacy of SSL denoising. Inspired by this idea, the subsequent section will utilize a iterative data refinement strategy to progressively create a dataset that, while characterized by reduced bias, closely approximates the attributes of an exemplary noisy-clean dataset.

\subsection{Self-supervised seismic denoising using iterative data refinement}
\begin{figure*}[!t]
\centering
\includegraphics[width=0.9\textwidth]{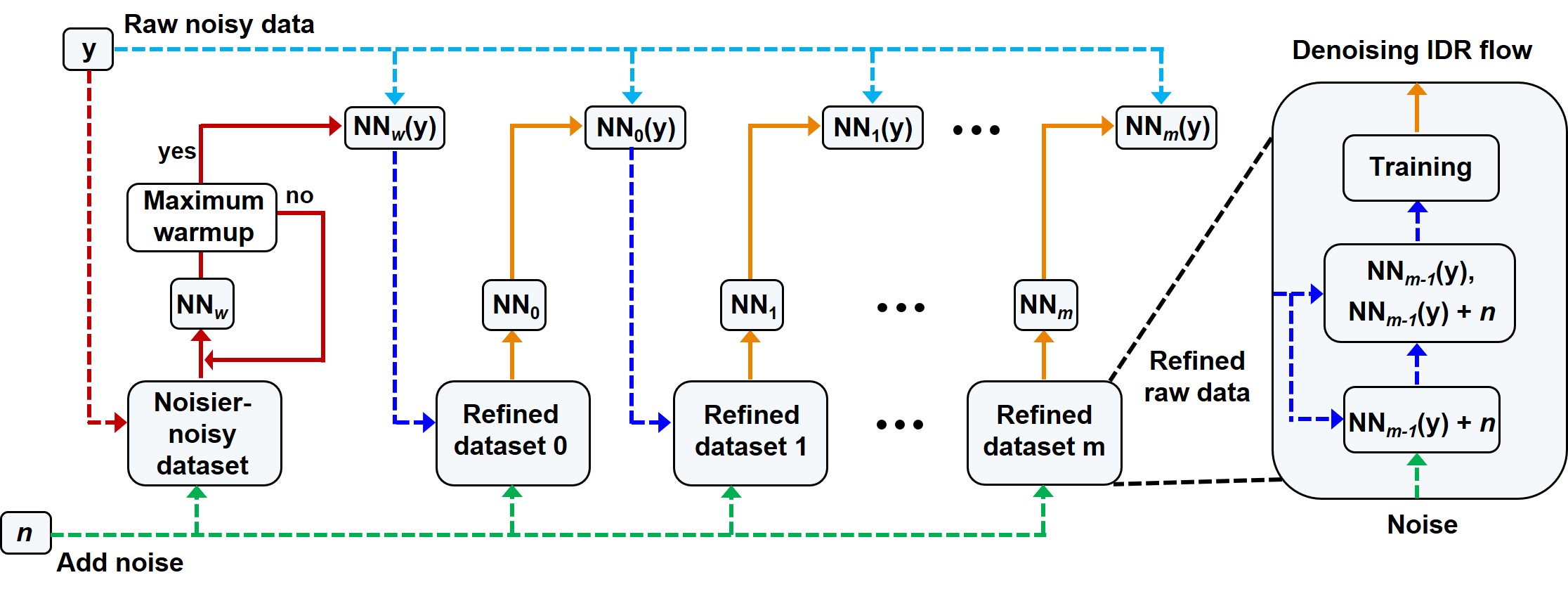}
\caption{An illustration of the self-supervised seismic denoising workflow. }
\label{fig3}
\end{figure*}

In this section, we will detail the paradigm of employing iterative data refinement (IDR) \cite{zhang2022idr} for the SSL seismic noise attenuation. The comprehensive workflow of this methodology is delineated in Fig. \ref{fig3}.

Firstly, we inject the additional noise $n_i$ into the raw noisy seismic data $y_i$, aiming to produce a noisier-noisy dataset $\{(y_i+n_i,y_i)\}_{i=1}^N$. An insight from the previous section indicates that a U-Net trained on the noisier-noisy dataset is endowed with a modicum of denoising capability. Therefore, we can instantiate a U-Net with randomized initialization and subject it to a preliminary warm-up training phase on the noisier-noisy dataset. This UNet, having undergone this warm-up training, is equipped to discern particular noise characteristics and can eliminate a degree of noise on the raw data. Consequently, we employ the pre-trained U-Net, denoted as $\text{NN}_w$, to denoise the raw noisy data. The denoised output serves as the initial pseudo-label data for the subsequent IDR phase, while the initial input is derived by superimposing noise upon these denoised results. In doing so, we generate a new, refined version of the noisier-noisy dataset:
\begin{equation}\label{eq5}
\{(\text{NN}_w(y_i)+n_i, ~ \text{NN}_w(y_i))\}_{i=1}^N.
\end{equation}

Then, we utilize the pre-trained U-Net to perform a fresh round of training on the new refined noisier-noisy dataset above:
\begin{equation}\label{eq6}
\text{NN}_0 \leftarrow \{(\text{NN}_w(y_i)+n_i, ~ \text{NN}_w(y_i))\}_{i=1}^N.
\end{equation}
Building upon the first discovery from the previous section, it becomes evident that NNs trained on datasets with reduced noise bias exhibit enhanced denoising capabilities. Owing to the denoising proficiency of the pre-trained UNet, this refined noisier-noisy dataset manifests a diminished data bias in relation to the noisy-clean data. As a result, the network model $\text{NN}_0$, trained on new dataset $\{(\text{NN}_w(y_i)+n_i, ~ \text{NN}_w(y_i))\}_{i=1}^N$, is imbued with superior denoising performance compared to the network model $\text{NN}_w$ pre-trained on the former dataset $\{(y_i+n_i,y_i)\}_{i=1}^N$.

Subsequently, we can iteratively execute the aforementioned process to enhance the denoising model and create successive versions of the noisier-noisy dataset. As an exemple, during the refinement round $m^{th}$, we can leverage the denoising network model from the preceding round, denoted as $\text{NN}_{m-1}$, to construct a new noisier-noisy dataset $\{(\text{NN}_{m-1},(y_i)+n_i, ~ \text{NN}_{m-1}(y_i))\}_{i=1}^N$. This generated dataset is further used to train the network model $\text{NN}_{m}$: 
\begin{equation}\label{eq7}
\text{NN}_m \leftarrow \{(\text{NN}_{m-1}(y_i)+n_i, ~ \text{NN}_{m-1}(y_i))\}_{i=1}^N.
\end{equation}
In every such training round serves dual purposes: Firstly, we bolster the denoising proficiency of our model, and secondly, as a direct benefit of this enhancement, we methodically narrow the feature gap that exists between the noisier-noisy dataset and the noisy-clean dataset. This meticulous progression ultimately facilitates our SSL seismic denoising architecture to reach a performance threshold that aligns closely with its SL denoising counterparts.

We emphasize that during the IDR stage, each round of refinement employs just a single epoch of training. If we are to ensure network convergence in every refinement iteration, it would result in a significant time cost. As a result, we abandon exhaustive optimization in each refinement and, instead, increase the number of iterations for data refinement. In our implementation, we observe that this approach does not compromise denoising performance.

\subsection{Network architecture and training}
\begin{figure*}[!t]
\centering
\includegraphics[width=0.9\textwidth]{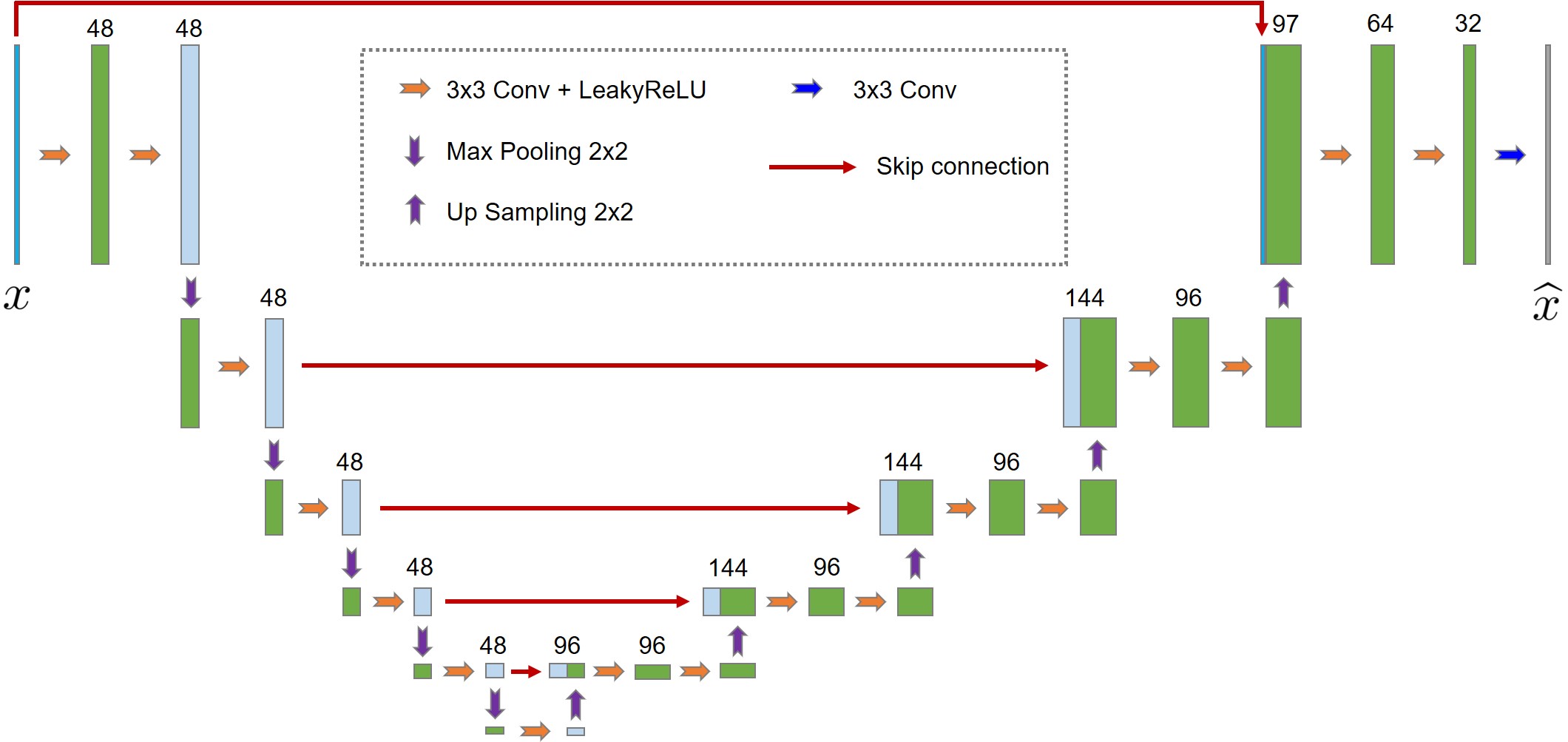}
\caption{The neural network architecture used in our study, where the input is the noisy seismic data $x$, and the output is the denoising results $\widehat{x}$. }
\label{fig4}
\end{figure*}

Fig. \ref{fig4} present the NN architecture utilized for our SSL seismic denoising framework, which is a modified rendition of the traditional U-Net and is designed specifically to optimize the denoising of seismic data. 

Beginning with the input layer, the input noisy data, denoted as $x$, flows through a series of convolutional blocks. Each block is characterized by a $3\text{x}3$ convolution (Conv) layer followed by a Leaky Rectified Linear Unit (LeakyReLU) activation. The depth of the feature maps in these convolutional blocks is symbolized by the numeric annotations, starting from a depth of 48 and expanding as we progress deeper into the architecture. Incorporated within the network are max-pooling layers of $2\text{x}2$ dimensions. These layers function to reduce the spatial dimensions of the feature maps, thus capturing the hierarchical patterns within the data. It should be noted that unlike the classic U-Net \cite{ronneberger2015u}, we adopt 5 downsampling attempts to capture features at more scales. Correspondingly, to upscale the feature representations back to the original spatial dimensions, the architecture integrates up-sampling layers of 2x2 dimensions. 

Also, we inherit a pivotal component from the classic U-Net architecture—skip connections, as illustrated by the red arrows. These connections allow the transfer of high-resolution feature maps from the encoder section of the network to the decoder section. This mechanism ensures the retention of fine-grained details, which are pivotal for high-fidelity seismic denoising tasks. However, a subtle distinction lies in the skip connection at the maximum scale. Instead of adopting feature maps processed through convolutional layers, we directly assimilate the input noisy data. This approach is motivated by the understanding that, within deep neural networks, certain signals inherent in the raw data might get diminished or lost after numerous convolution and pooling operations. By directly harnessing the raw input, we can sidestep this potential erosion of information. The final segment of our architecture sees the feature maps undergoing a 3x3 Conv operation before producing the denoised output, labeled as $\widehat{x}$.

Our denoising network is optimized by utilizing the most common loss function, namely mean absolute error (MAE) as follows:
\begin{equation}\label{eq8}
\begin{gathered}
\mathcal{L}\left (L,O \right)=\frac{1}{N}\displaystyle \sum^{N}_{i=1}{\left|L_{i}-O_{i} \right|},
\end{gathered}
\end{equation}
where $O$ represents the output of network, and $L$ is the corresponding pseudo label. During our training phase, we employed the AdamW optimization technique \cite{loshchilov2017decoupled}. Our implementation leverages the PyTorch framework, and the training is executed on a GeForce RTX 8000 graphics processing unit.

\section{Numerical Examples}
Next, we will evaluate the performance of our method in attenuating the following three common types of seismic noise: random noise, backscattered noise, and blending noise. We present the results both on synthetic and field data to demonstrate the effectiveness of our approach. Our denoising results will be compared with those from SL and curvelet-based denoising techniques. Within the SL approach, we utilize the same network architecture and training configurations as in SSL to preclude any unfair comparisons stemming from setup disparities. We emphasize that, due to the unavailability of clean label data for the field noisy data, our denoising network, once trained on synthetic data, is directly applied to the field data to produce denoised results. To quantitatively assess the denoising efficacy, we employ the SNR and MAE as our evaluation metrics. 

It's important to note that the method for acquiring the noise added when constructing the noisier-noisy dataset varies depending on the noise attenuation task. If, based on prior knowledge, seismic data is determined to be contaminated by random noise, we utilize Equation \ref{eq3} to generate the noise data. The backscattered noise is extracted outside of first-arrival in field data and is then respectively injected into both synthetic and field data. The blending noise is first extracted outside of first-arrival from noisy synthetic and field data, and subsequently reintroduced into their respective raw noisy datasets. 

Meanwhile, in our denoising network training, the noise added when constructing the noisier-noisy dataset is multiplied by a random scale factor to account for varying degrees of natural noise contamination. In the case of random noise attenuation, we generate random noise using term $0.01\epsilon \cdot std(x_i) \cdot rand(0,1)$ from Equation \ref{eq3}; hence, this scale factor is equivalent to $\epsilon$. For the backscattered noise and the blending noise, since they are directly extracted from the noisy data, we simply multiply the extracted noise by the scale factor. For clarity in subsequent descriptions, we will denote the range of this scale factor as $s=[s_1,s_2]$, representing the intensity range of the added noise. For example, during the warm-up phase, $s=[5,10]$ indicates that we amplify the noise by a random factor between 5 to 10 times before adding it to the original noisy data to form the noisier-noisy dataset. In the IDR phase, $s=[5,10]$ means that we incorporate noise into the pseudo-label data, which is predicted by the network trained in the previous epoch, to create the noisier-noisy dataset for the current training epoch.

\subsection{Random noise attenuation}
\subsubsection{\textbf{Synthetic data}}

We first evaluate our method on the attenuation of random noise. From the Marmousi model, we extract several layer models and simulate 2400 shots of synthetic seismic recordings. From these synthetic seismic data, we carve out 12000 data blocks, each of size 256x256, to constitute a clean dataset. The raw noisy data are generated by leveraging Equation \ref{eq3} to introduce random noise, with the level ranging from 5 to 40, into the clean dataset. During our SSL training, in the warm-up and IDR phases, we utilize the same noise level, i.e., $s=[5,40]$. For the SL training, we remove portions containing test data from both the clean and raw noisy datasets. Subsequently, the clean data serves as the label, while the raw noisy data functions as input to train the denoising network. Within our SSL implementation, the denoising network undergoes a total of 150 training epochs, of which the warm-up phase encompasses 30 epochs. The SL procedure maintains an identical training duration of 150 epochs. In both SSL and SL frameworks, the initial learning rate is 3e-4, which is decreased by a factor of 0.8 at the 25, 50, and 75 epochs. 

\begin{figure}[htp]
\centering
\includegraphics[width=0.3\textwidth]{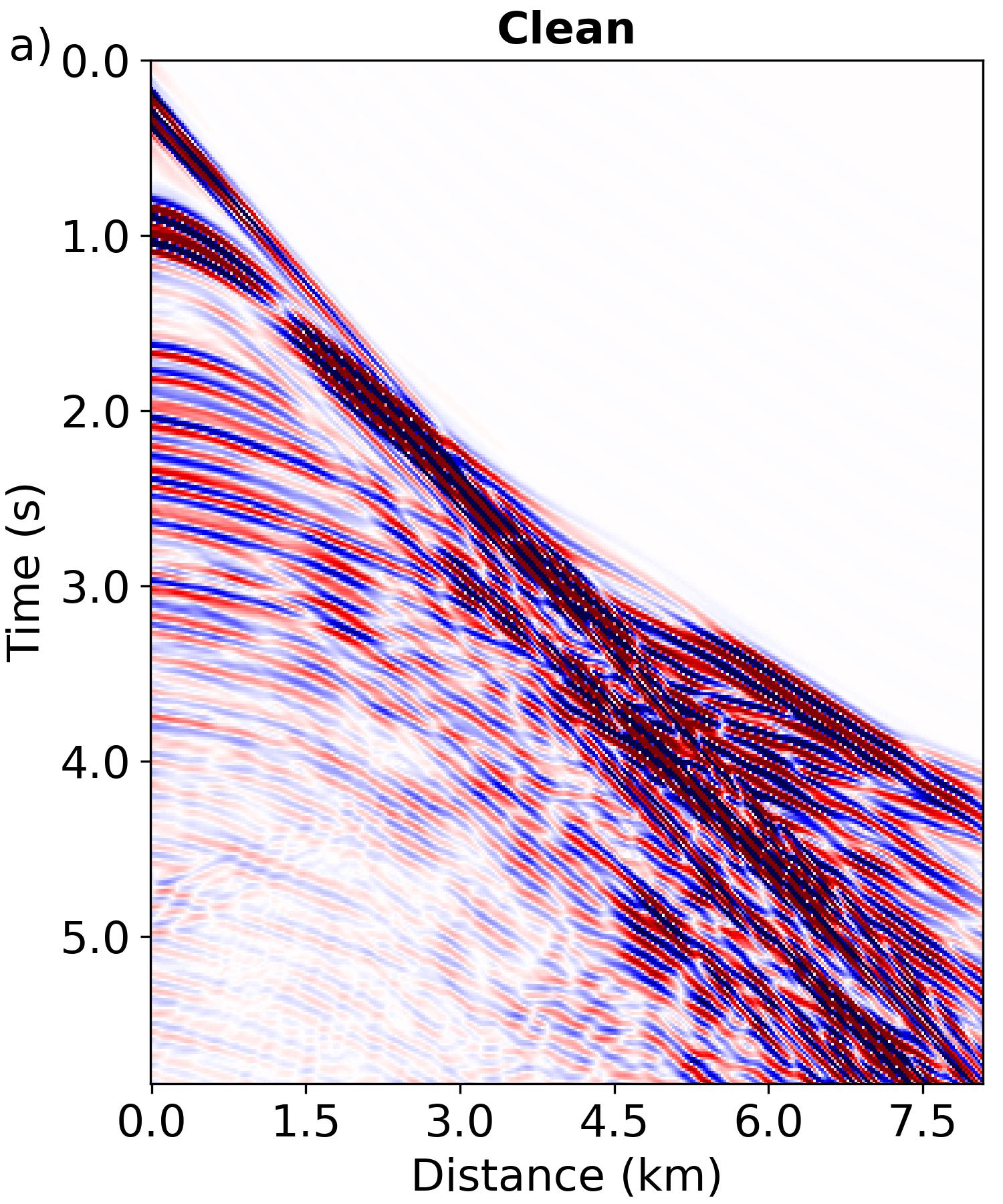}
\hspace{1cm}
\includegraphics[width=0.3\textwidth]{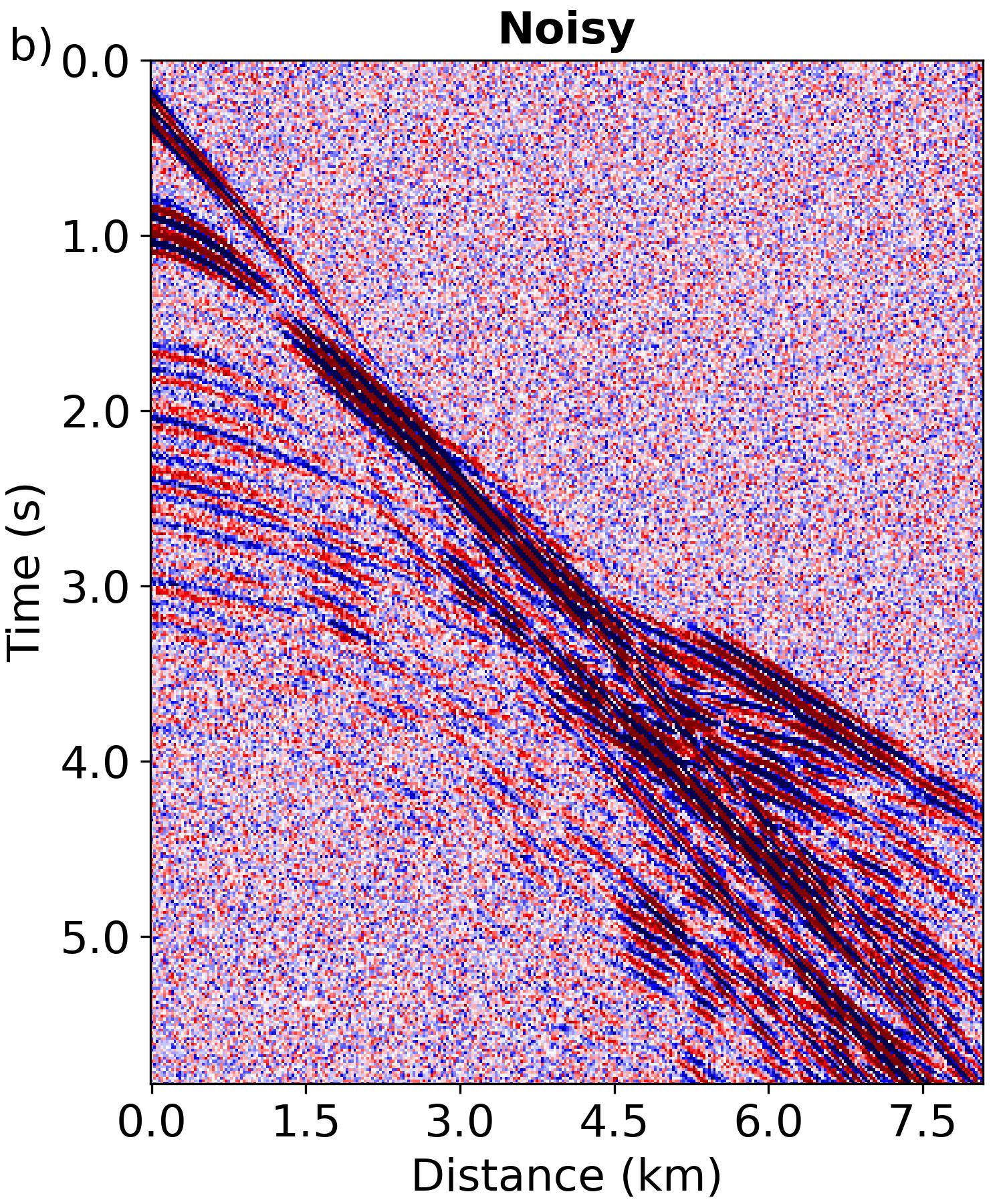}
\caption{The clean (a) and noisy (b) data of the synthetic test data, where the noisy data is generated adding the random noise with the level of 30 into the clean data. }
\label{fig5}
\end{figure}

\begin{figure*}[!t]
\centering
\includegraphics[width=0.3\textwidth]{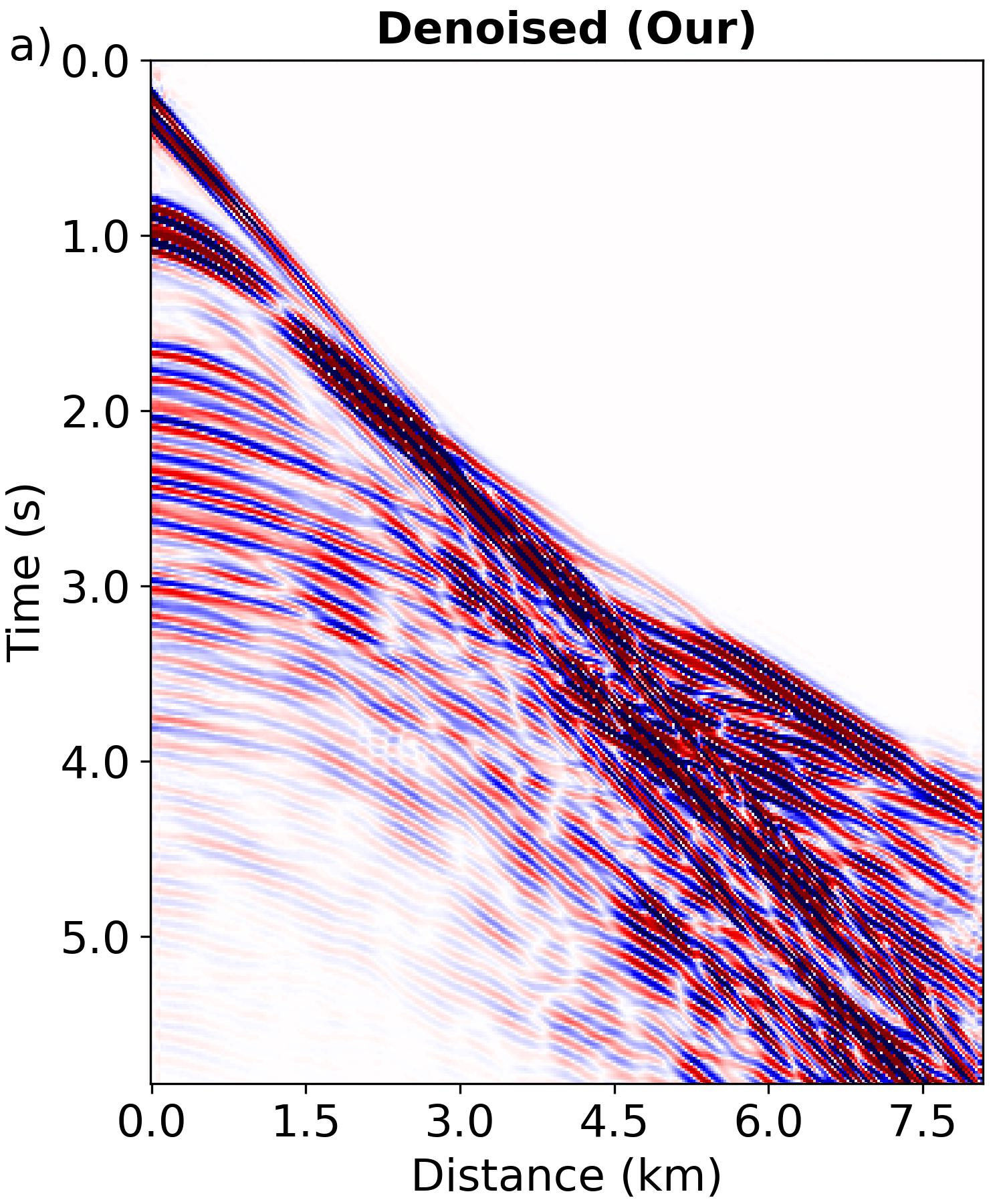} 
\includegraphics[width=0.3\textwidth]{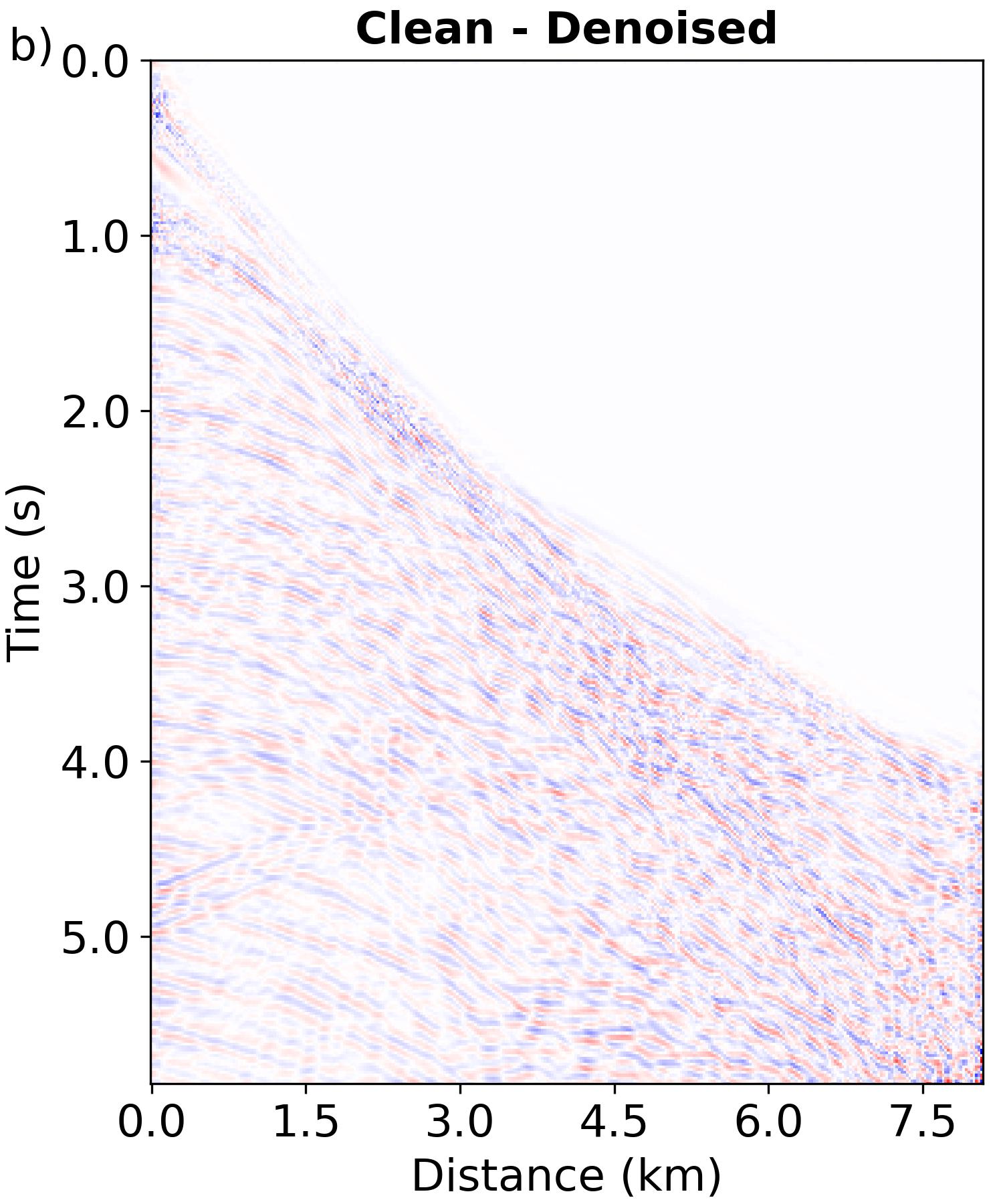}
\includegraphics[width=0.3\textwidth]{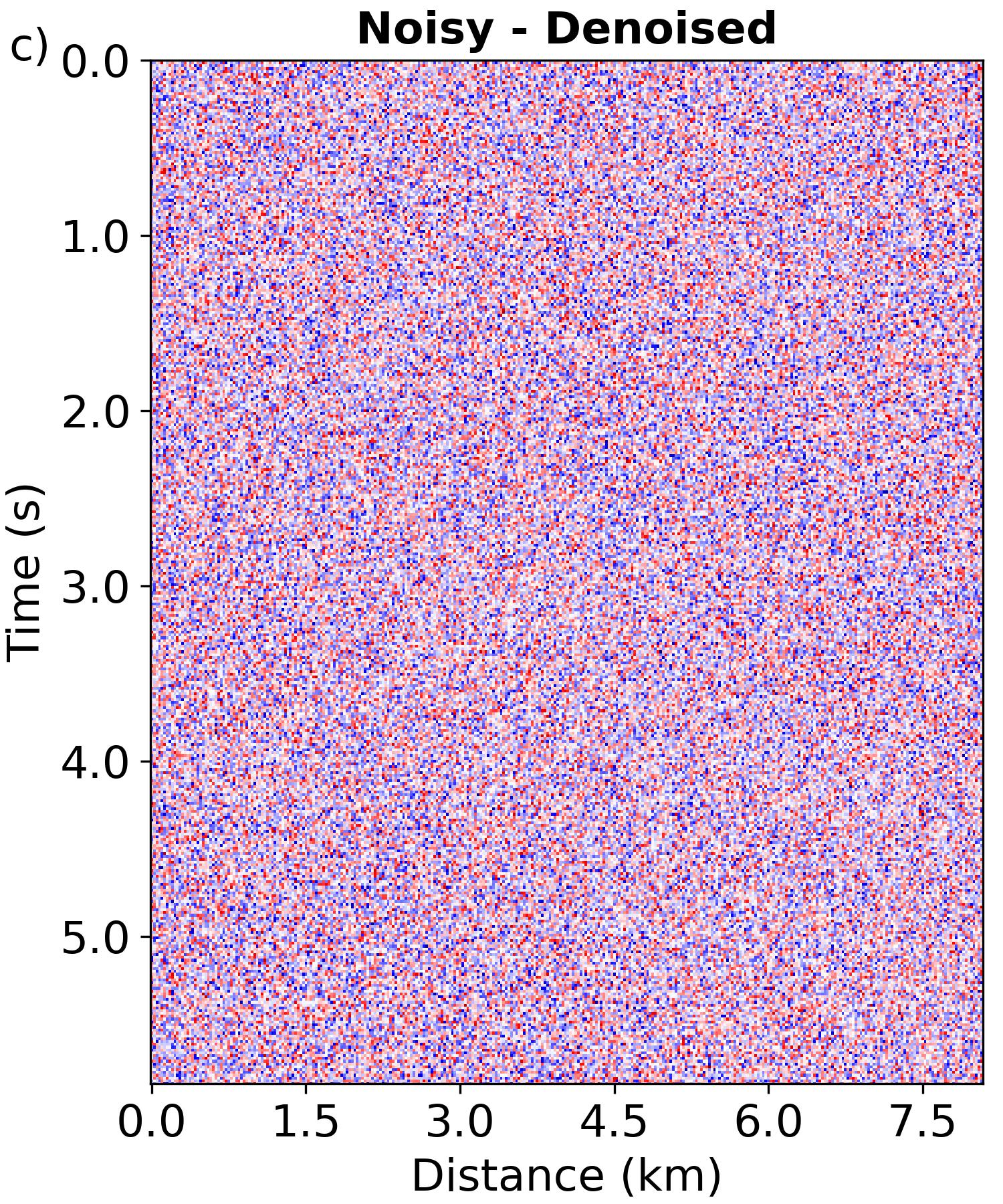} \\
\includegraphics[width=0.3\textwidth]{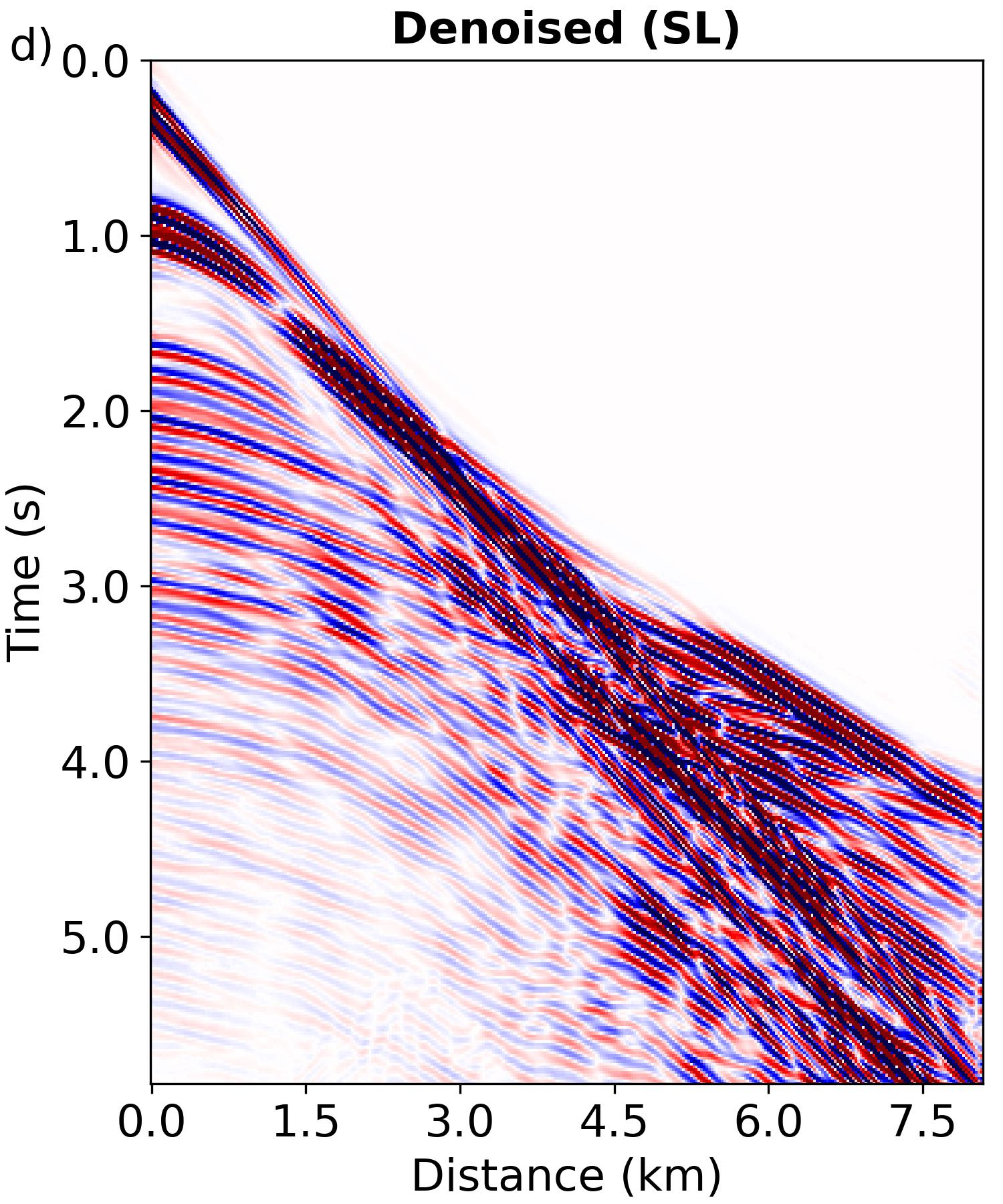} 
\includegraphics[width=0.3\textwidth]{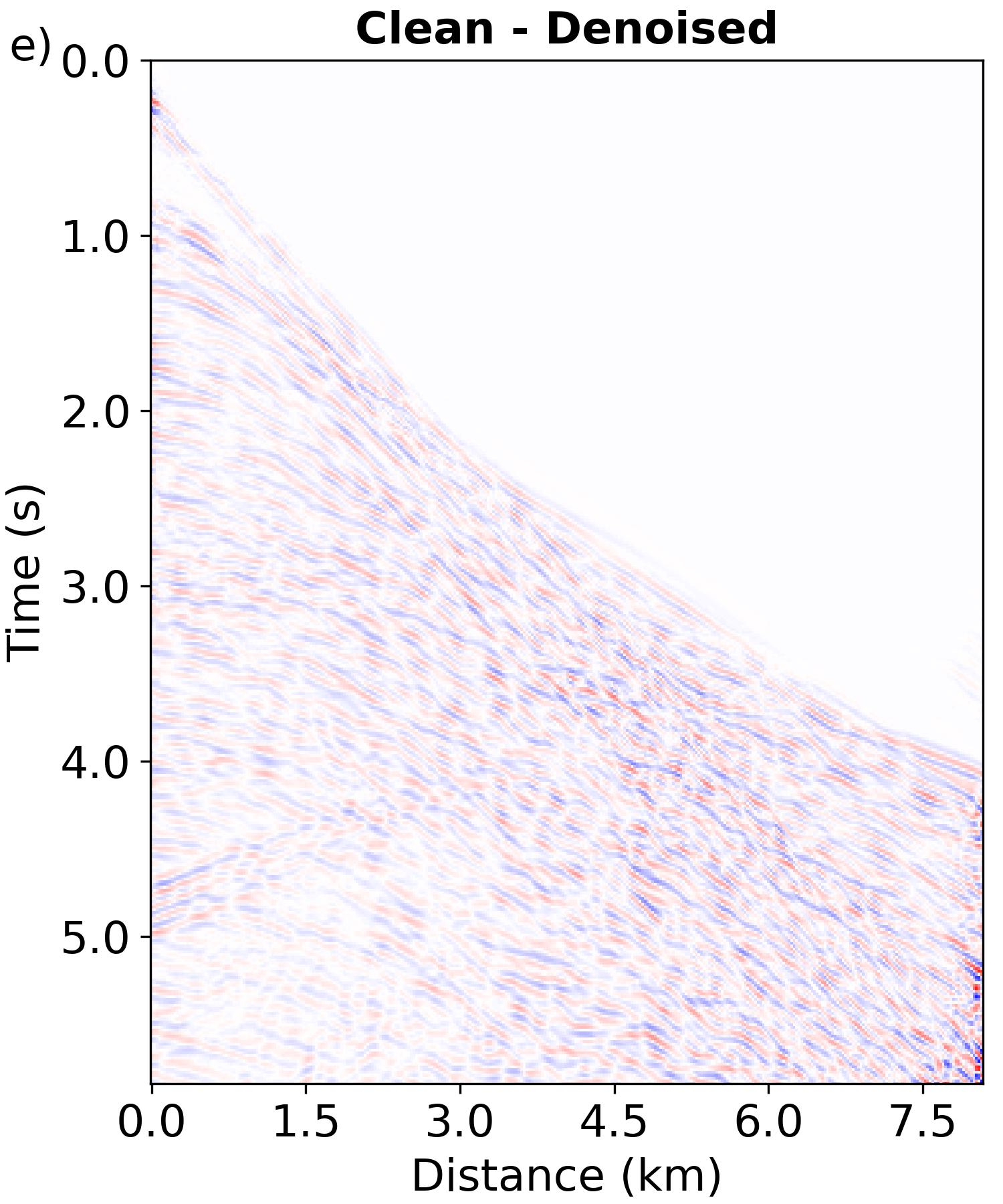} 
\includegraphics[width=0.3\textwidth]{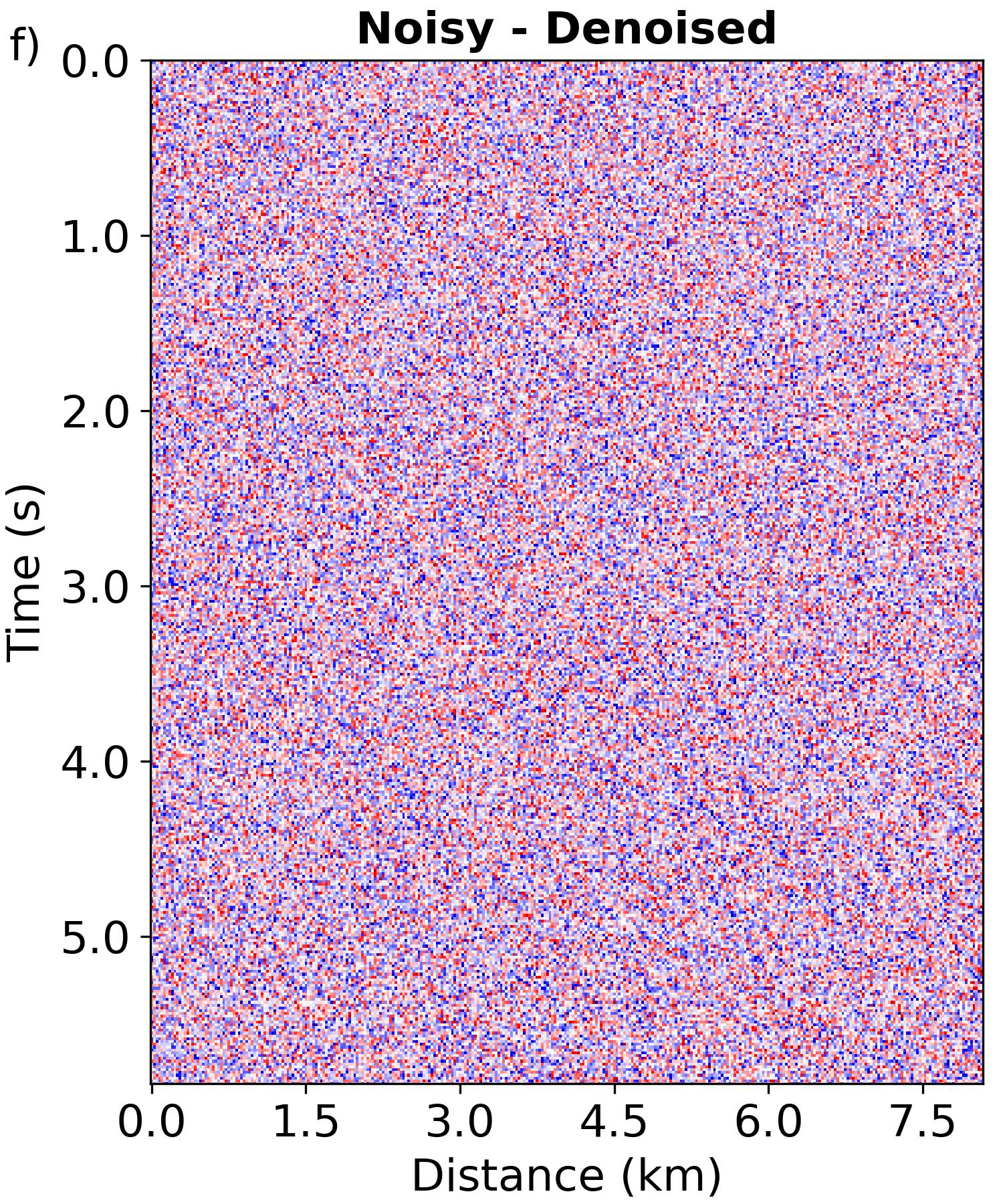} \\
\includegraphics[width=0.3\textwidth]{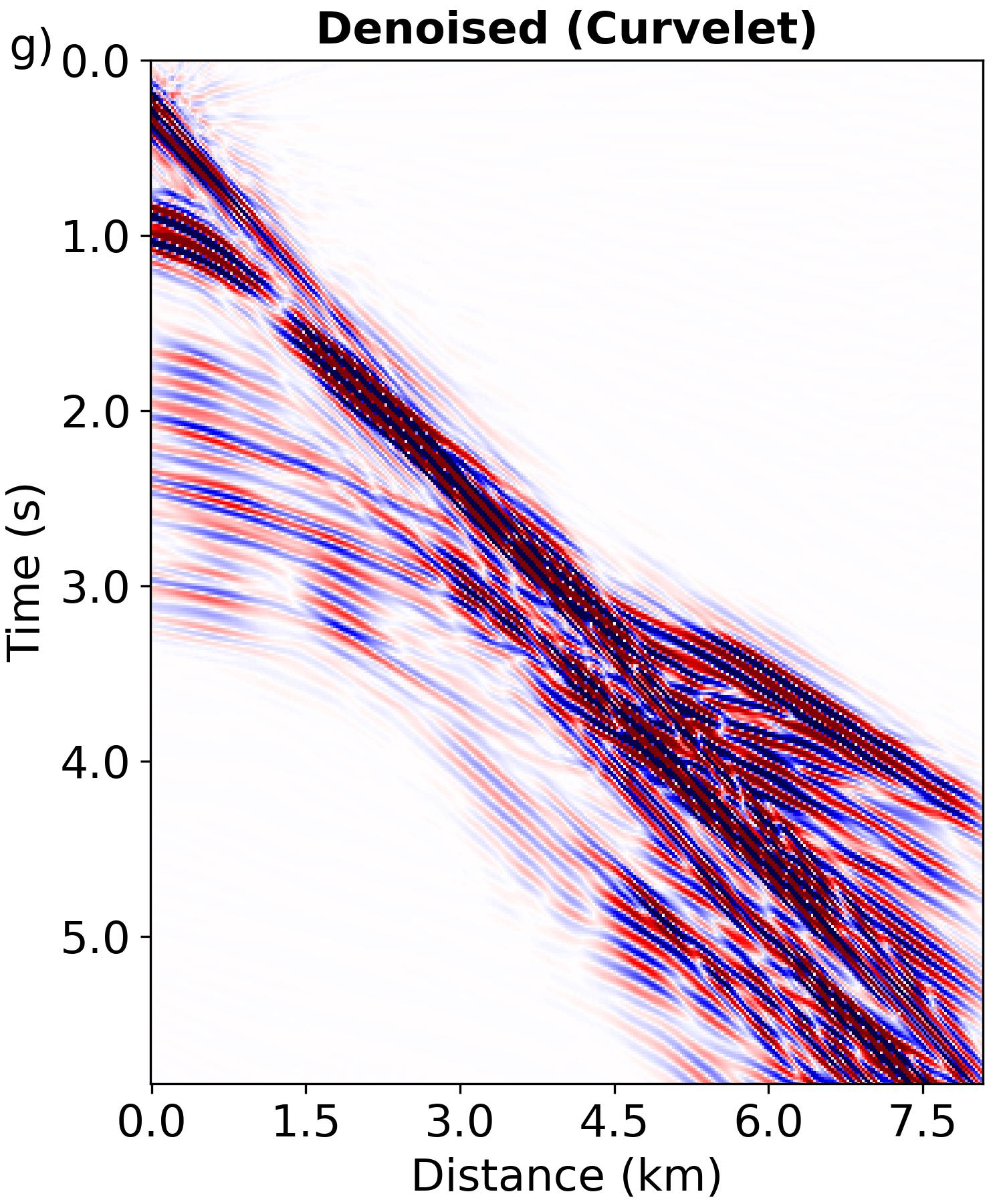} 
\includegraphics[width=0.3\textwidth]{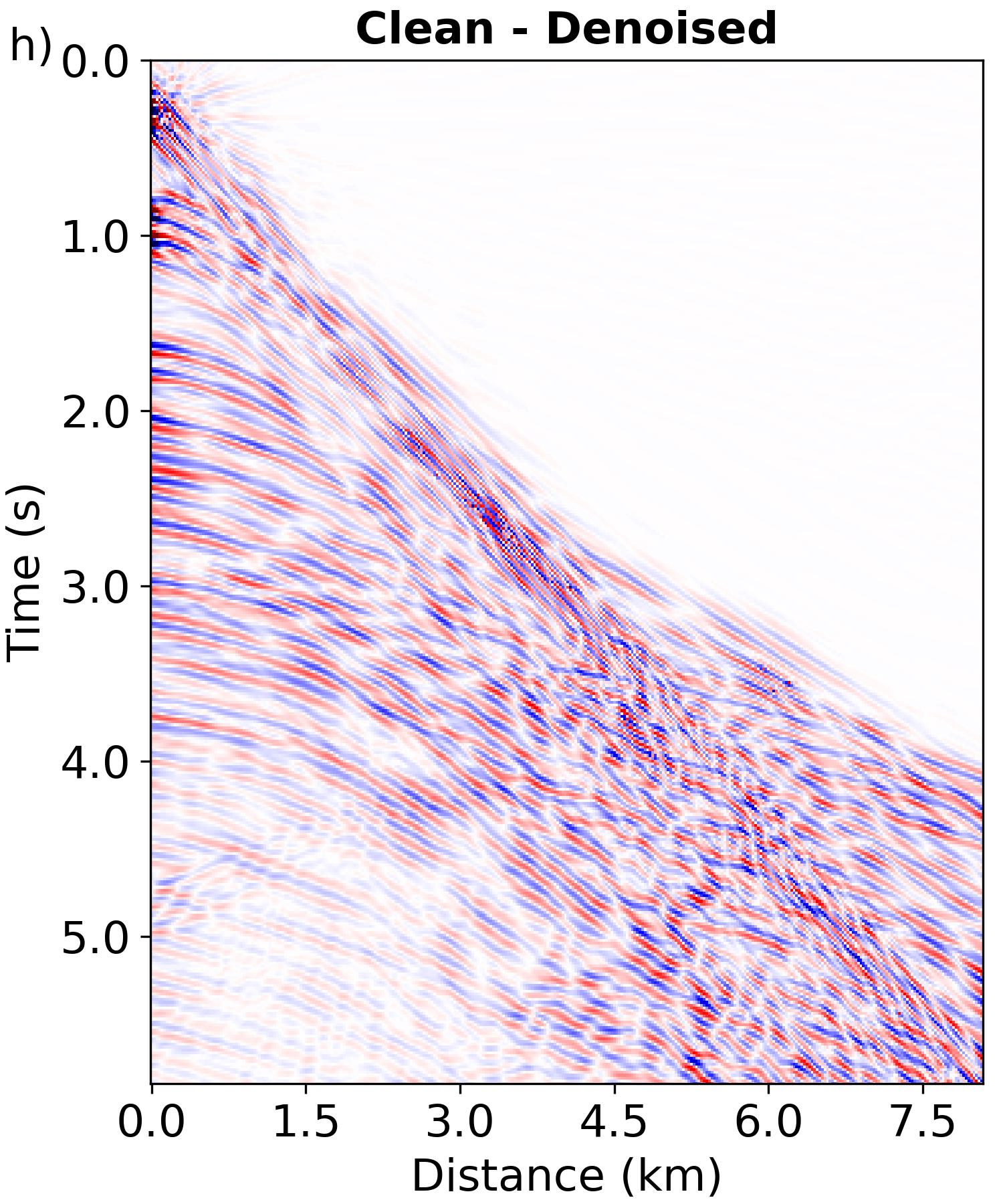}
\includegraphics[width=0.3\textwidth]{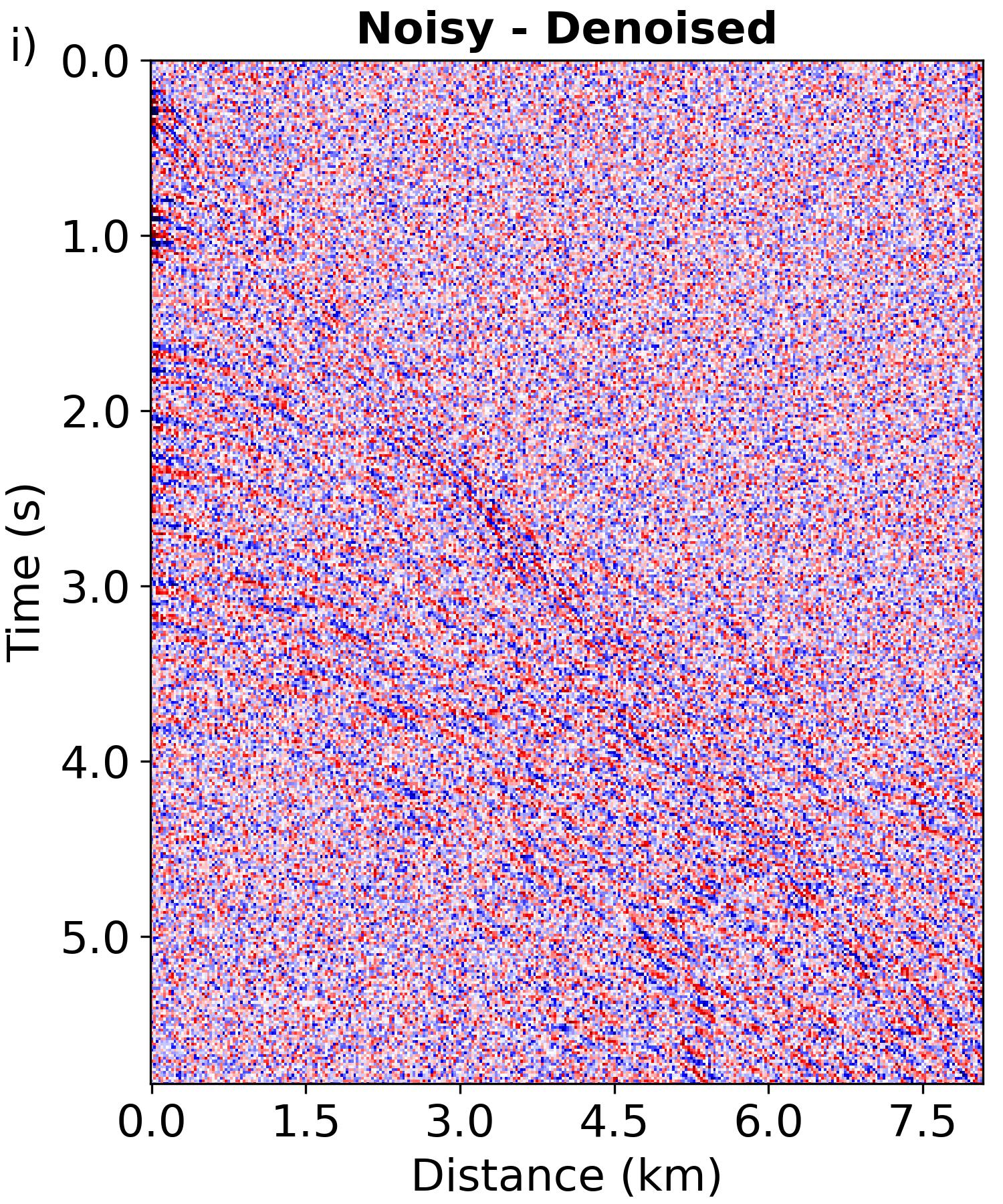}
\caption{Denoising performance comparison on synthetic data contaminated with random noise using our method, SL, and curvelet-based approach, presented from top to bottom respectively. The first column corresponds to denoised results. The second column depicts the difference between the denoised results and the clean data, while the third column shows the difference between the denoised results and the noisy data.}
\label{fig6}
\end{figure*}

Fig. \ref{fig5} shows the noisy data used for testing and the matching clean labels. This noisy data originates from adding random noise with the level of 30 to the clean data. Evidently, the noise level is significant, obscuring deep reflection signals. Fig. \ref{fig6} offers a comparison of the denoising performances achieved by our method, SL, and the curvelet-based approach. We observe that our method appears almost on par with SL in denoising performance. However, there's a slight deviation when addressing weak signals from deeper layers. In contrast, the traditional curvelet denoising technique, while reducing noise, also considerably attenuates the signal, especially for weak reflection signals. Also, it introduces some unwanted artifacts near the direct wave. 

\begin{figure}[htp]
\centering
\includegraphics[width=0.3\textwidth]{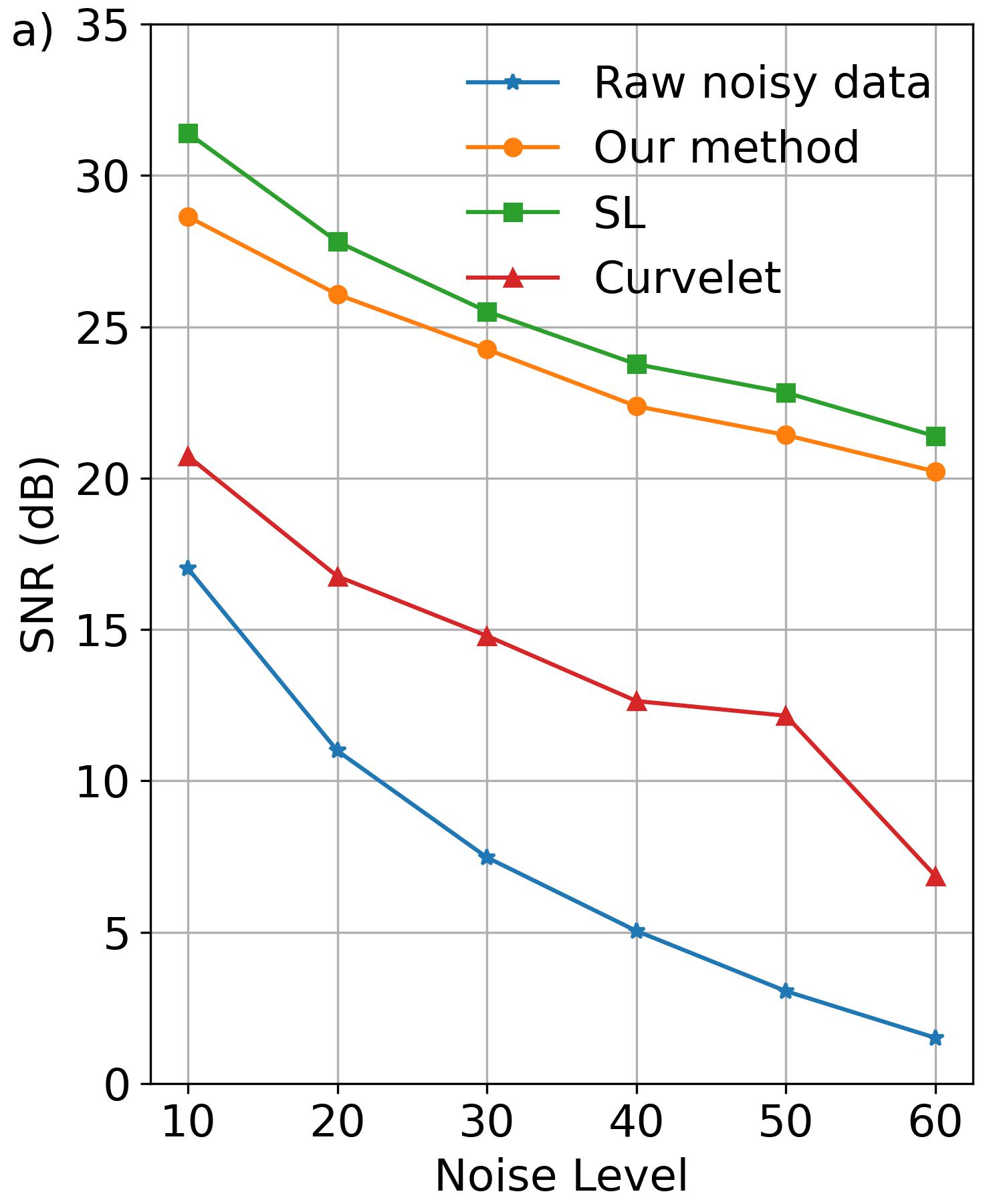}
\hspace{1cm}
\includegraphics[width=0.3\textwidth]{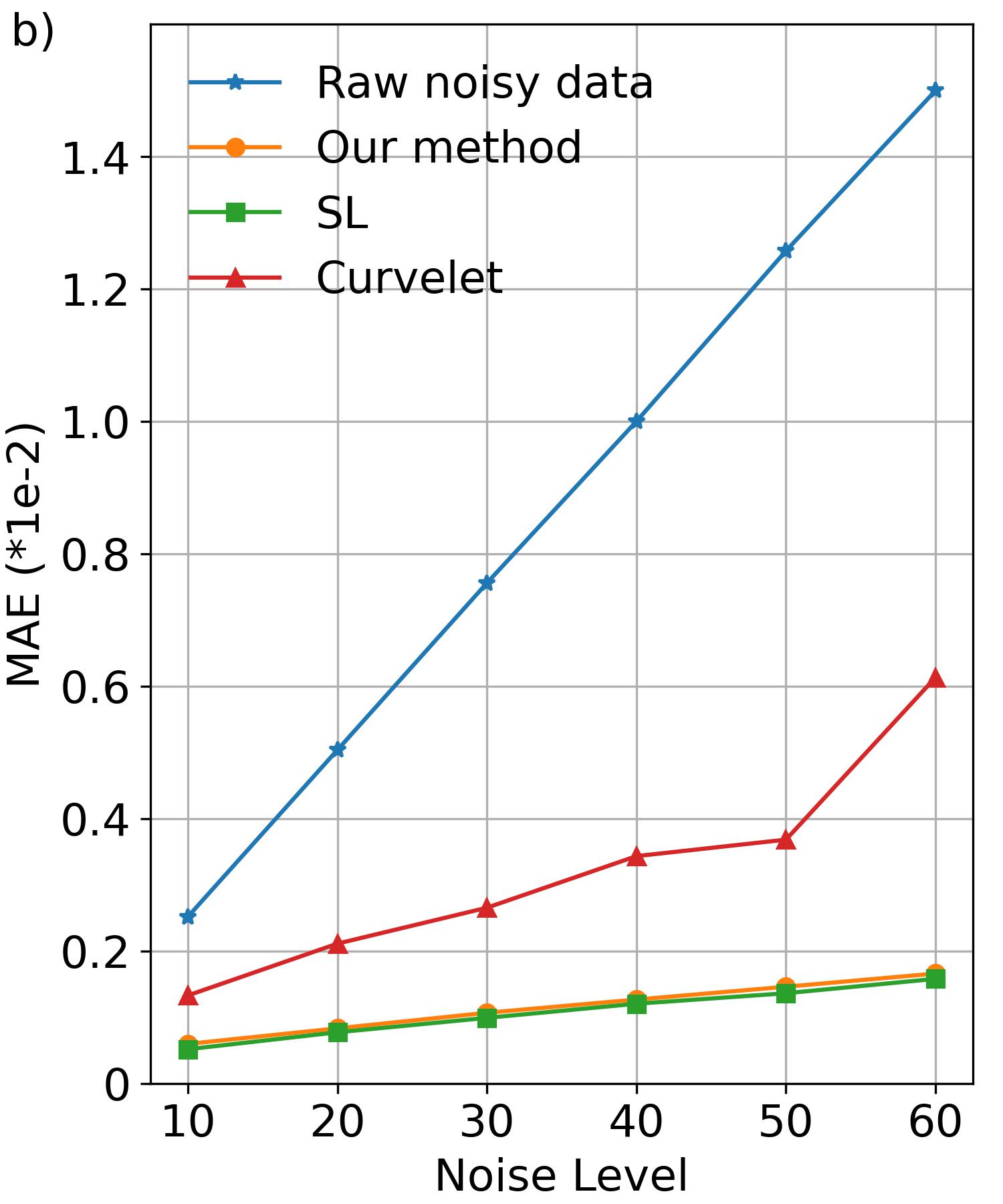}
\caption{The denoising performance comparison of different method (Our method, SL, and curvelet) across six different noise levels, with the blue line representing the raw noisy test data. (a) Denoised results using SNR metric. (b) Denoised results using MAE metric.}
\label{fig7}
\end{figure}

We further evaluate the denoising capabilities of our method to those of the SL and curvelet approaches across varying noise levels, as illustrated in Fig. \ref{fig7}. The results suggest that, irrespective of the noise level, our method's denoising performance remains closely aligned with SL, especially when measured using the MAE metric. Although the curvelet technique offer a certain ability in noise attenuation, its overall enhancement remains relatively modest when compared with both our method and SL. \\

\subsubsection{\textbf{Field data}}
\begin{figure*}[!t]
\centering
\begin{minipage}[t]{0.23\textwidth}
\includegraphics[width=\textwidth]{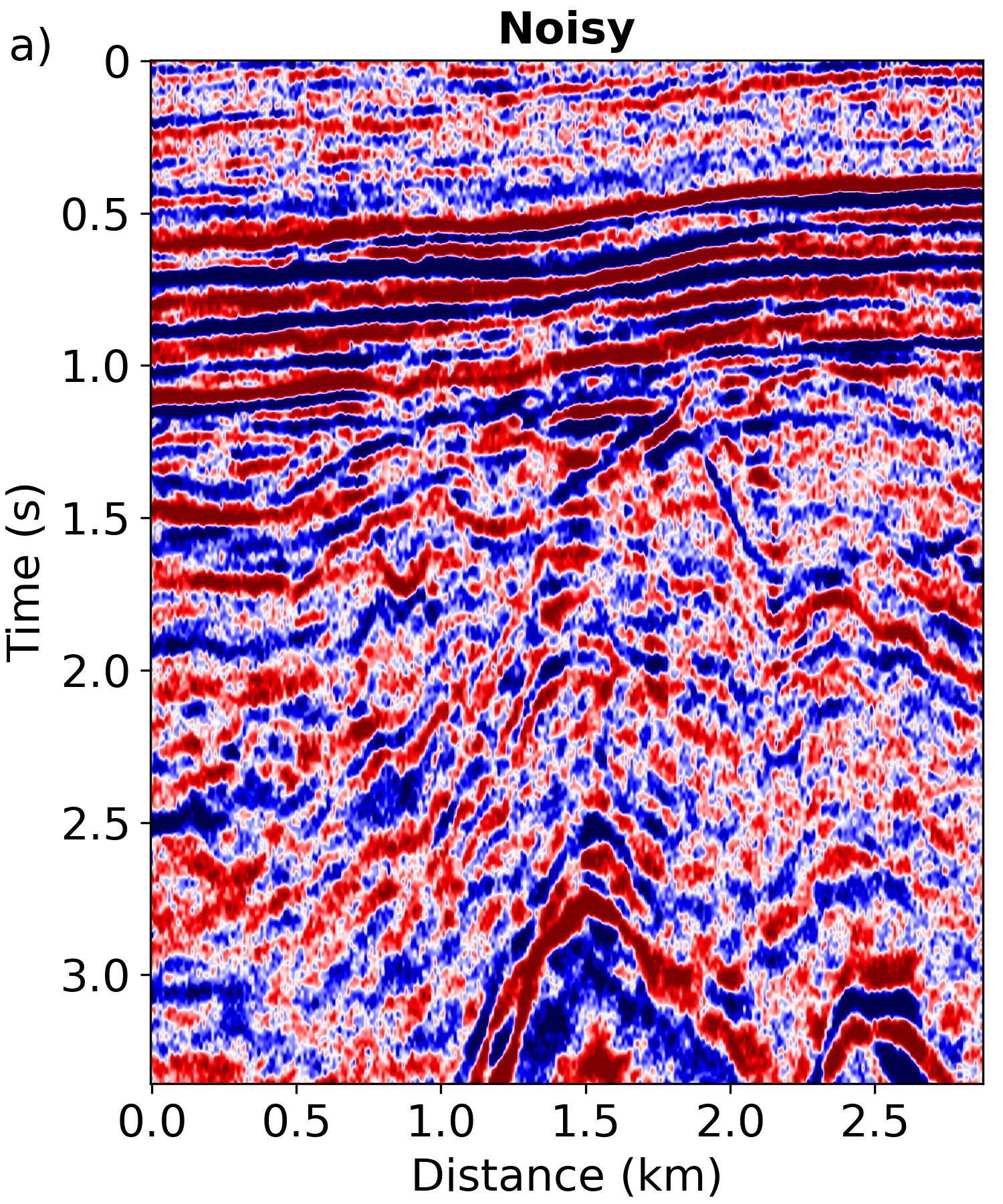} 
\end{minipage}
\hfill
\begin{minipage}[t]{0.23\textwidth}
\includegraphics[width=\textwidth]{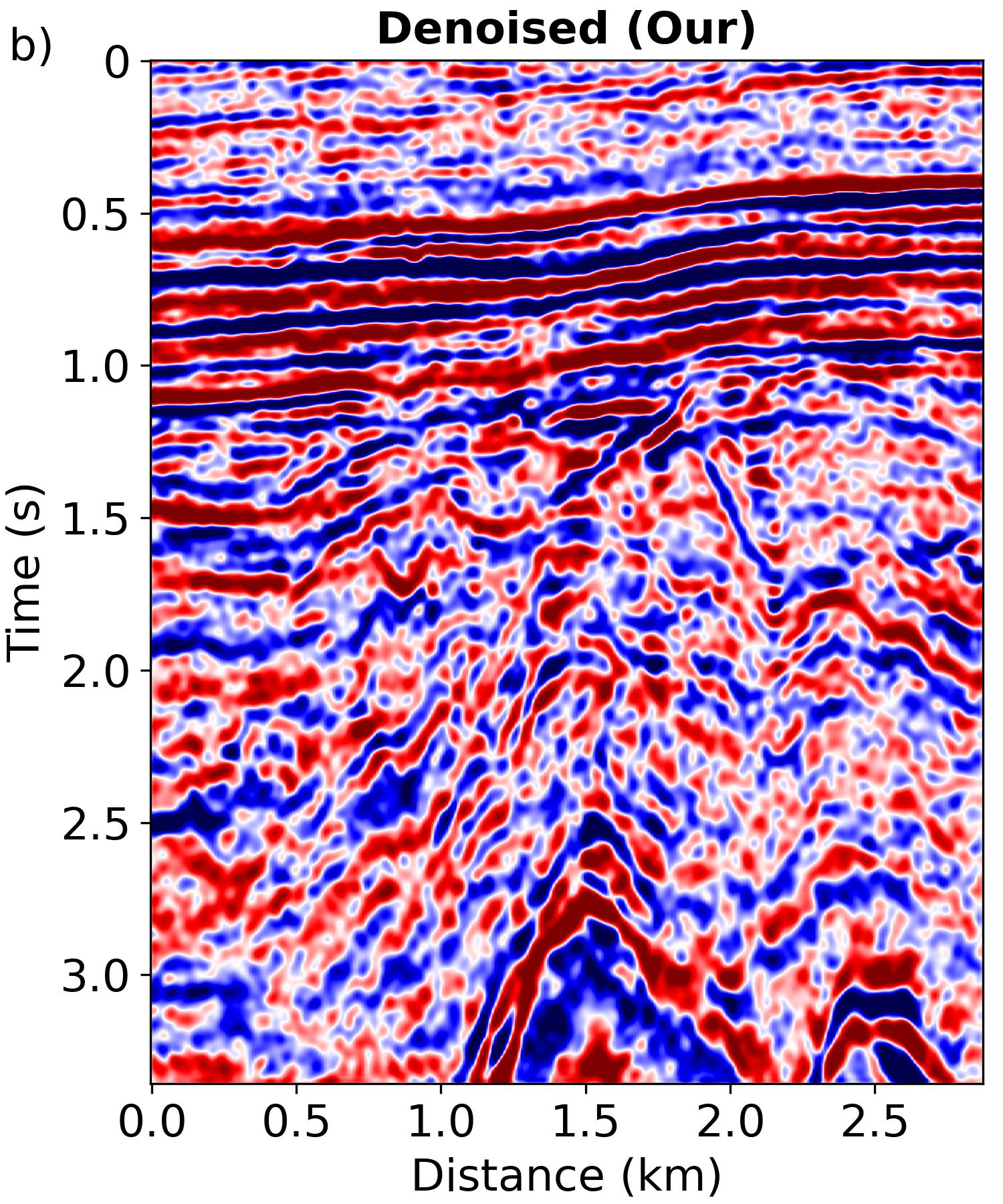} \\
\vfill
\includegraphics[width=\textwidth]{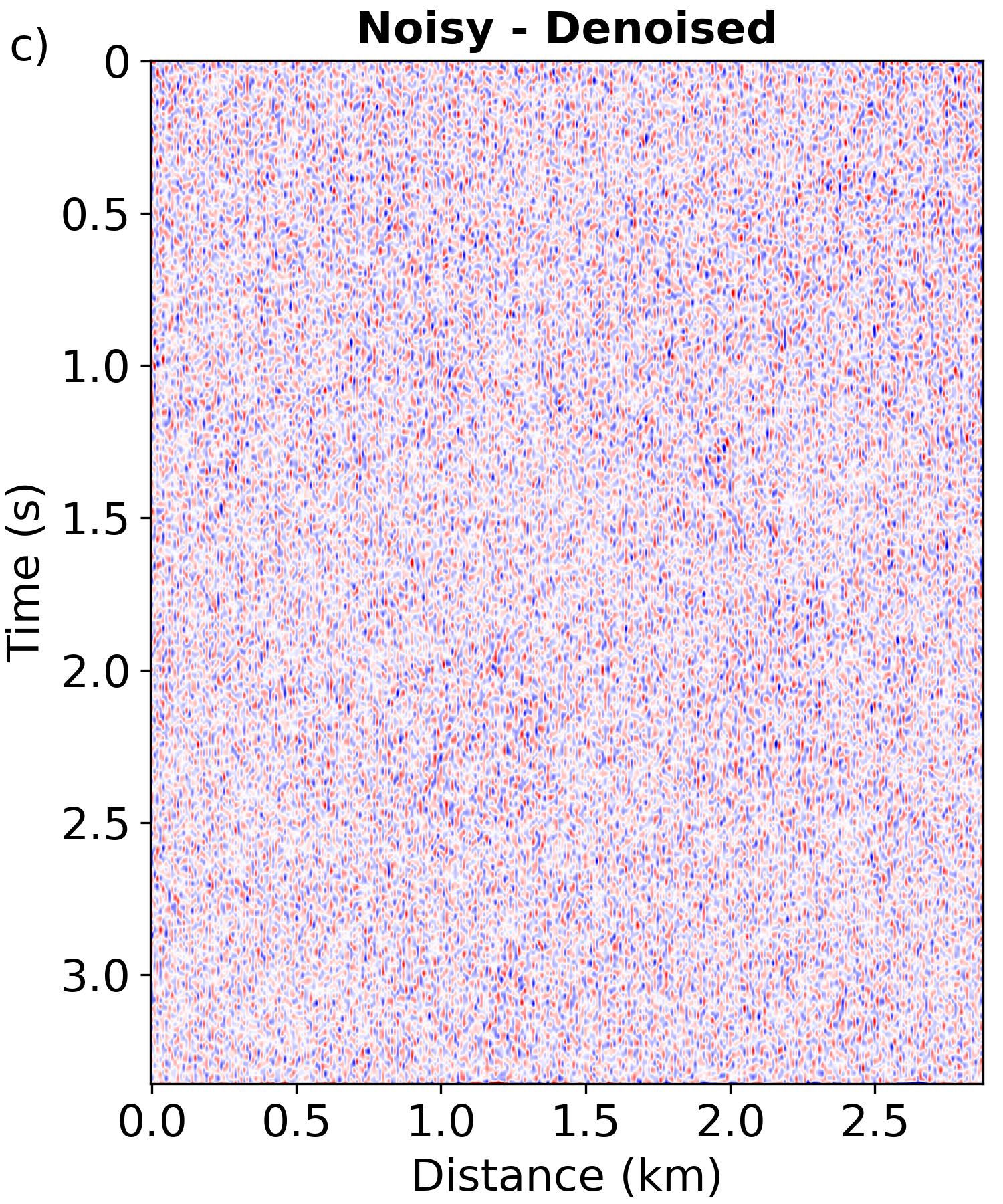} 
\end{minipage}
\hfill
\begin{minipage}[t]{0.23\textwidth}
\includegraphics[width=\textwidth]{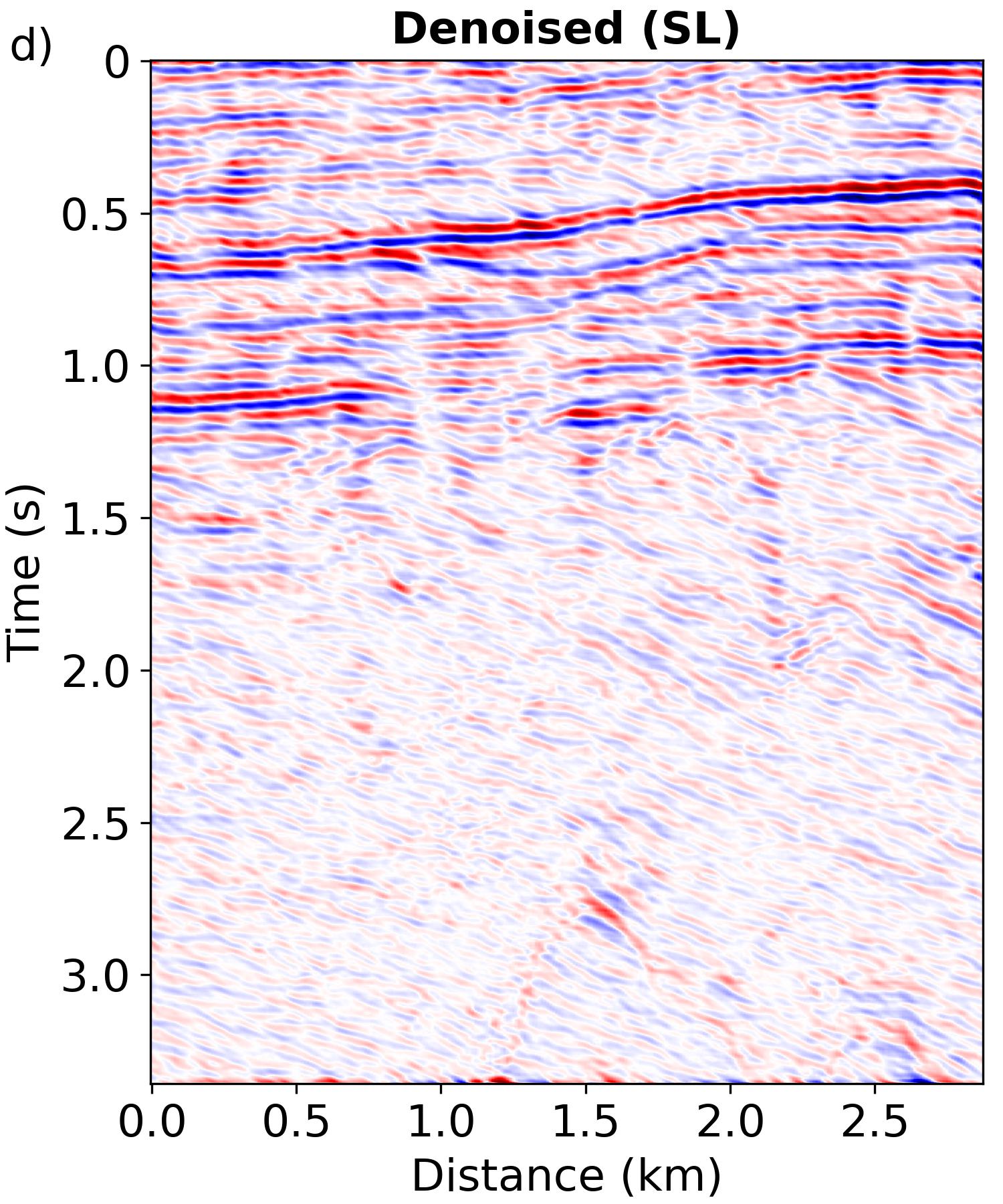} \\
\vfill
\includegraphics[width=\textwidth]{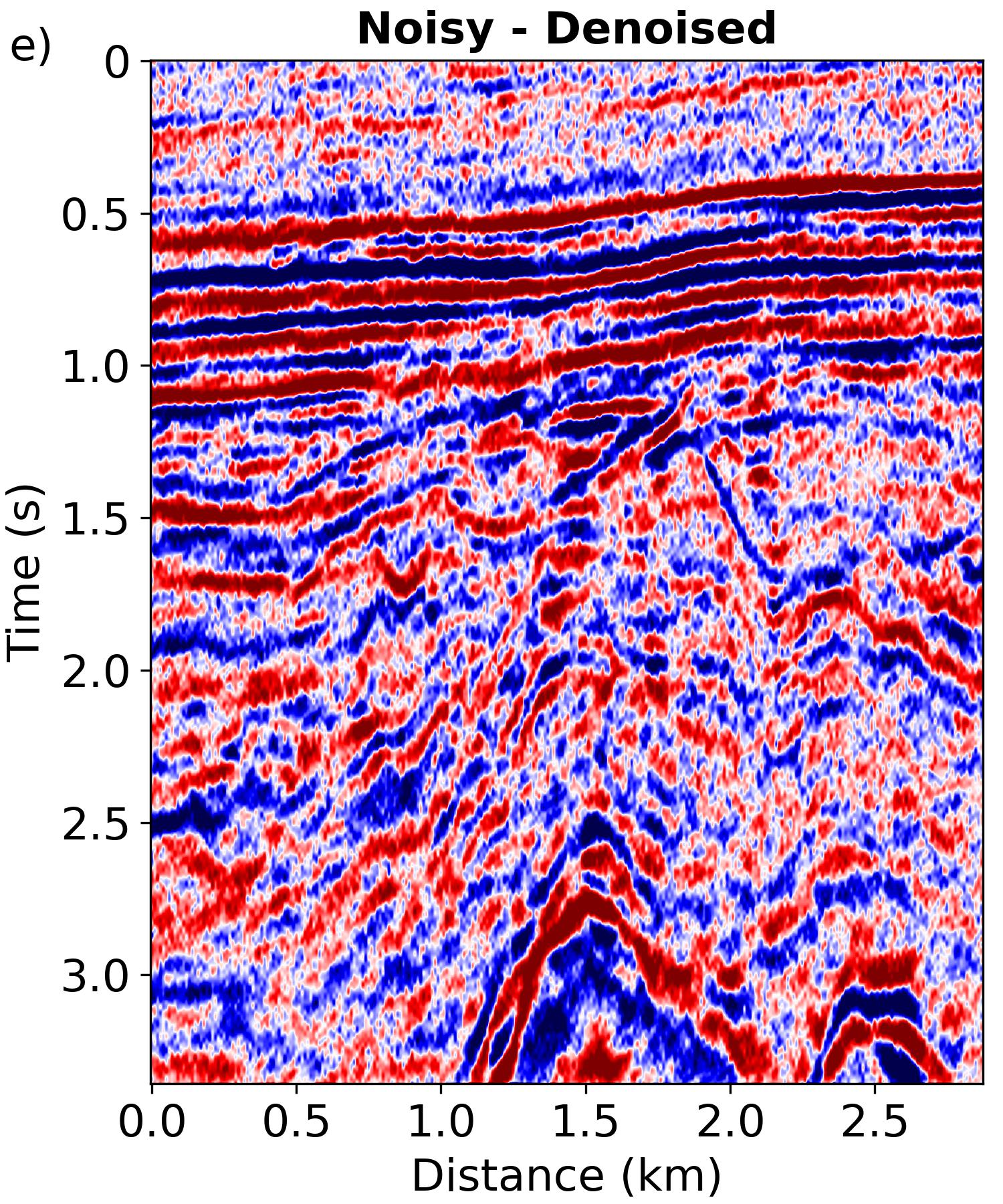} 
\end{minipage}
\hfill
\begin{minipage}[t]{0.23\textwidth}
\includegraphics[width=\textwidth]{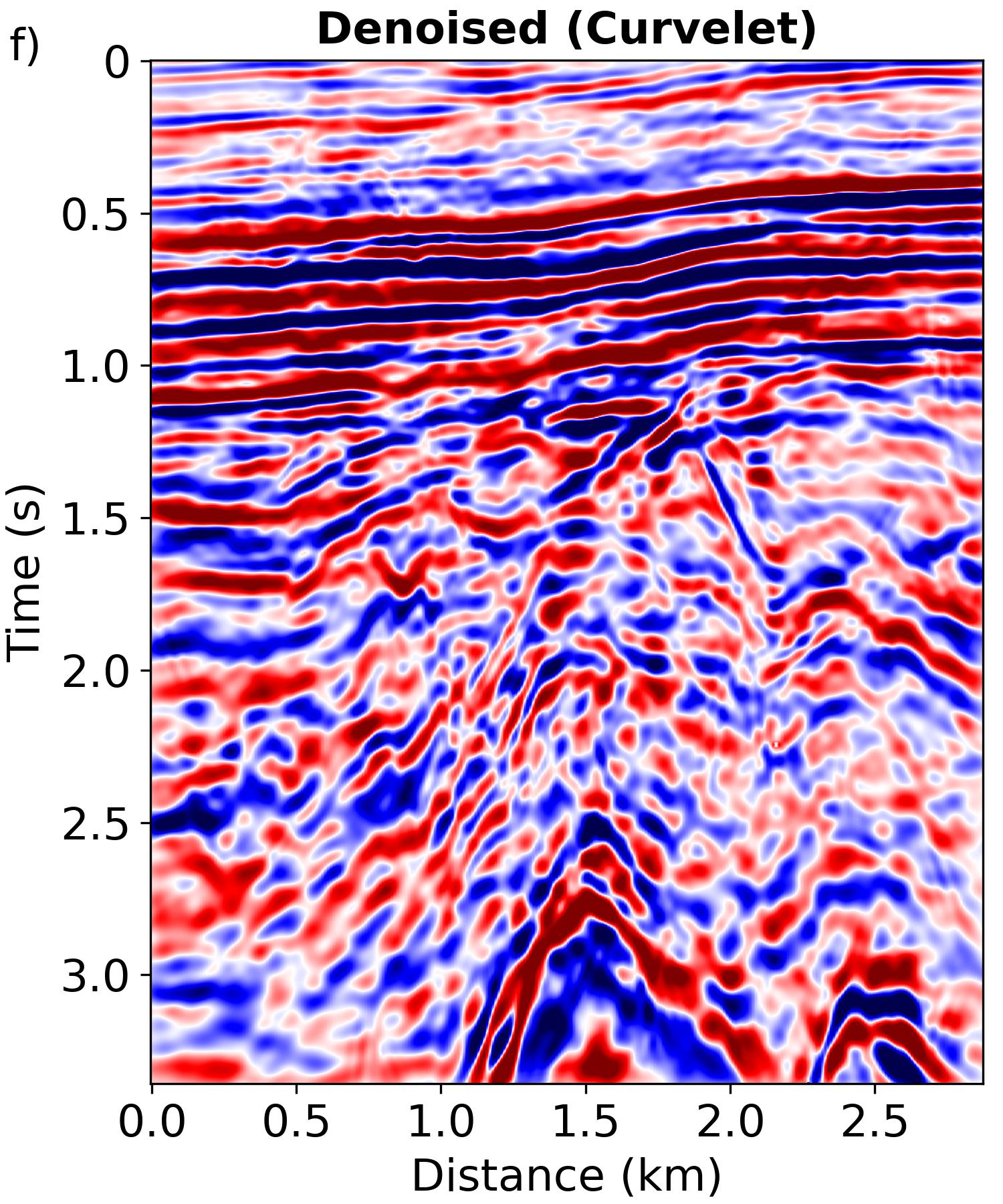} \\
\vfill
\includegraphics[width=\textwidth]{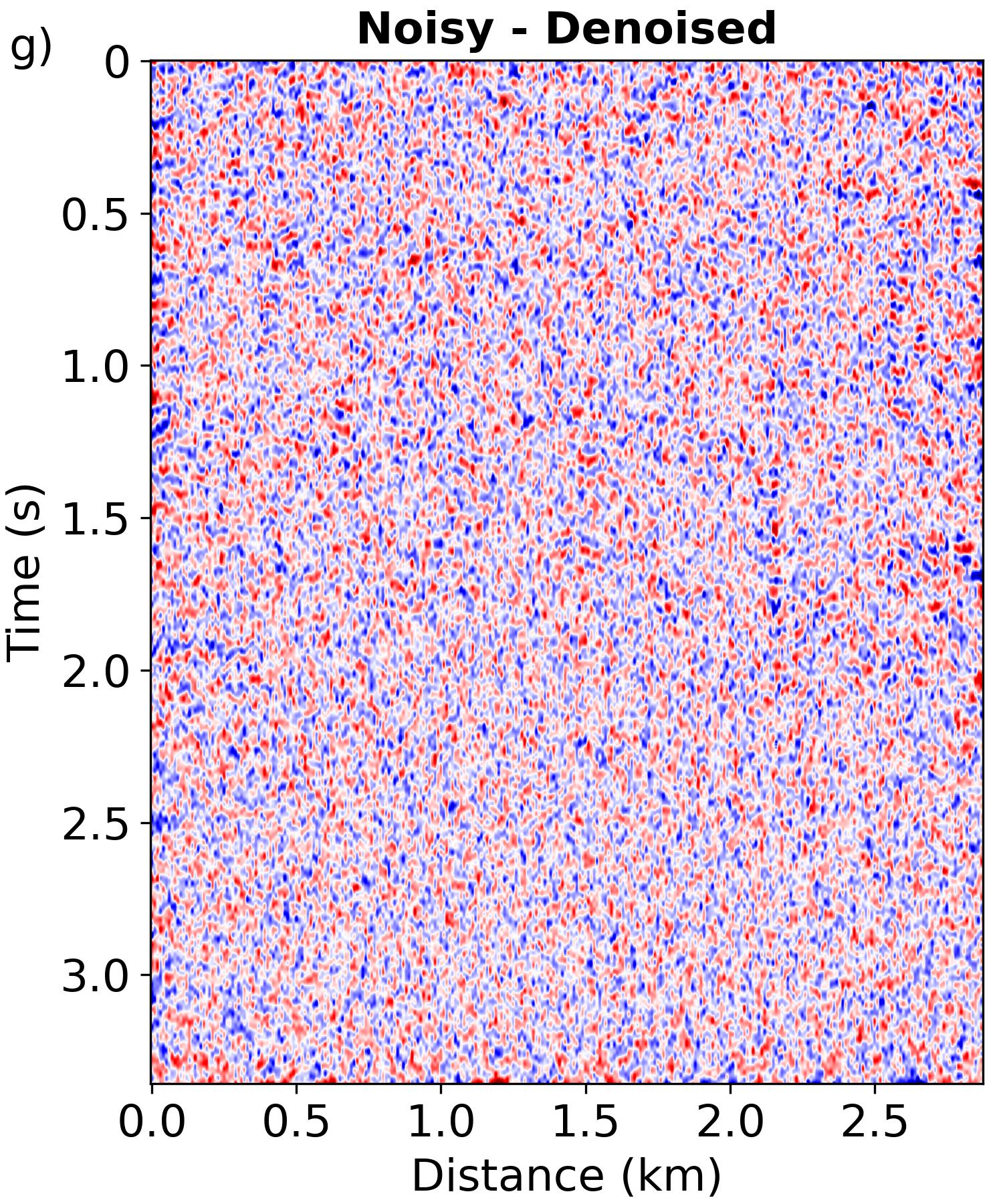} 
\end{minipage}
\caption{Comparison of the denoising performance on a post-stack time-migrated image, which comes from a China land field dataset and is known to be contaminated by random noise. (a) The original noisy data. (b), (d), and (f) represent the denoised results from our method, SL, and curvelet-based method, respectively. The corresponding differences between the original noisy data and the denoised results are depicted on the bottom.}
\label{fig8}
\end{figure*}

We, then, test our method on a post-stack time-migrated image (see Fig. \ref{fig8}a) from a China land field dataset. This image, based on prior knowledge, is known to be contaminated with random noise. To perform training, we extract 528 patches, each sized 128x128, from the original noisy data. The warm-up and IDR phases hold the same noise level $s=[20,100]$. The denoising network undergo a pre-training of 30 epochs in the warm-up phase and then perform 60 epochs of training in the IDR phase. The learning rate remains at 2e-4.

Figs. \ref{fig8}b, \ref{fig8}d, and \ref{fig8}f display the denoising results on the original noisy data by our method, SL, and the curvelet-based approach, respectively. The differences with the original noisy data are illustrated in Figs. \ref{fig8}c, \ref{fig8}e, and \ref{fig8}g, respectively. It's evident that our method achieves superior denoising, effectively reducing noise while preserving the signal. SL applies directly to the field data after training on the synthetic data. Due to the significant feature distribution difference between the synthetic and field data, SL faces a considerable performance drop. This highlights the potential greater applicability of SSL in field data compared to SL. While the traditional curvelet-based method also effectively diminishes noise, it overly smoothes the signal, as reported in some studies \cite{birnie2021potential}. Moreover, we can see that the curvelet-based approach reduces the signals at the boundaries of the noisy data.

\subsection{Backscattered noise attenuation}
\subsubsection{\textbf{Synthetic data}}
In the following, we apply the proposed method in attenuating backscattered noise. We first present the test results on synthetic data. The clean synthetic data used here share the same simulated data from the examples of random noise attenuation. The distinction lies in the backscattered noise extracted from field data that we inject into the clean data. The original noisy data are produced by randomly amplifying the extracted backscattered noise by a factor of 5 to 10 and then adding it to the clean data. The noise levels added during both the warm-up and IDR phase are set as $s=[5,15]$. In our implementation, the denoising network undergoes an initial 30-epoch warm-up training, followed by 90 epochs of training in the IDR phase. The initial learning rate is 2e-4, and a scheduler decreases it by a factor of 0.8 every 20 epochs. SL maintains the same training configuration to ensure a fair comparison. 

\begin{figure}[htp]
\centering
\includegraphics[width=0.3\textwidth]{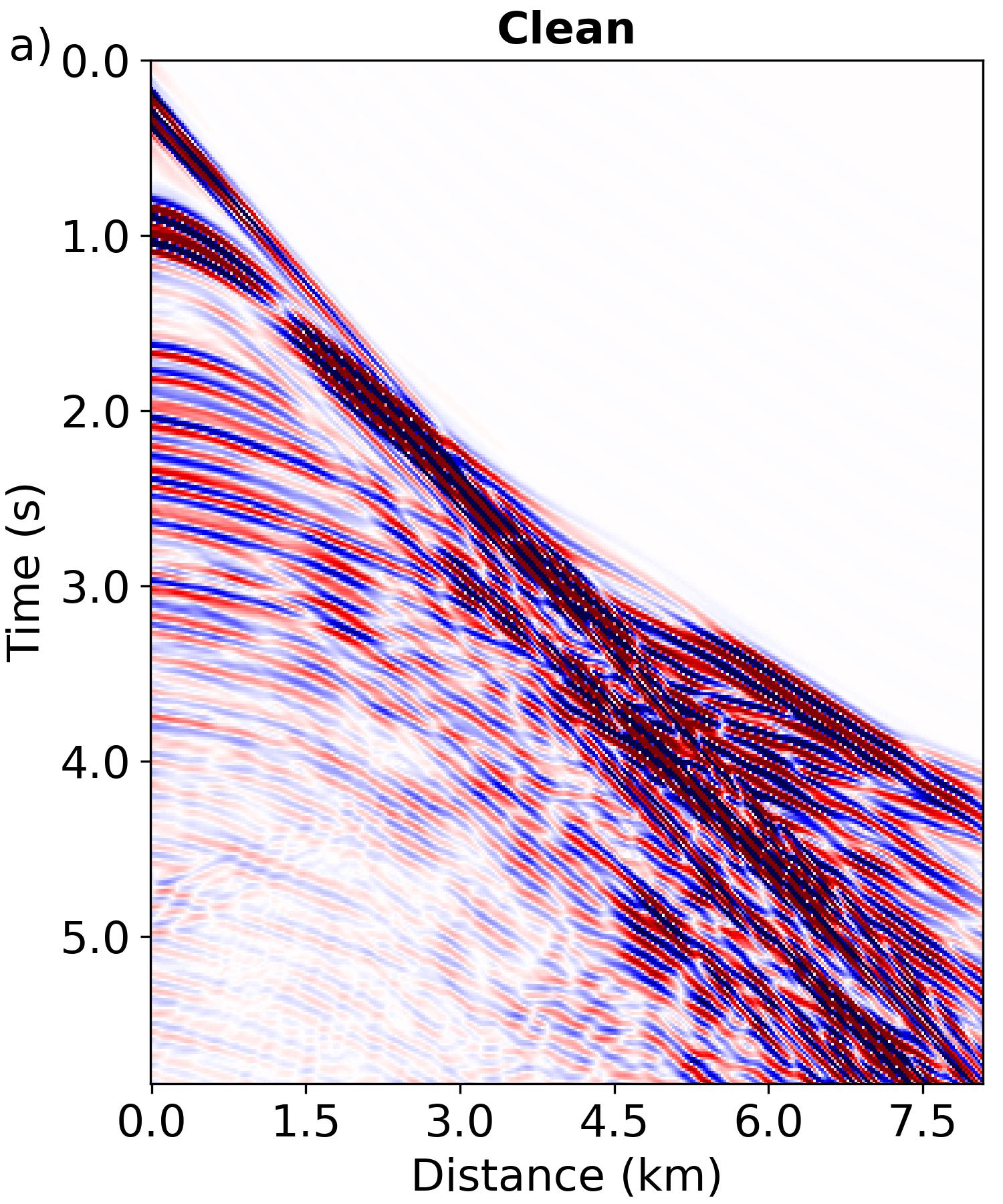}
\hspace{1cm}
\includegraphics[width=0.3\textwidth]{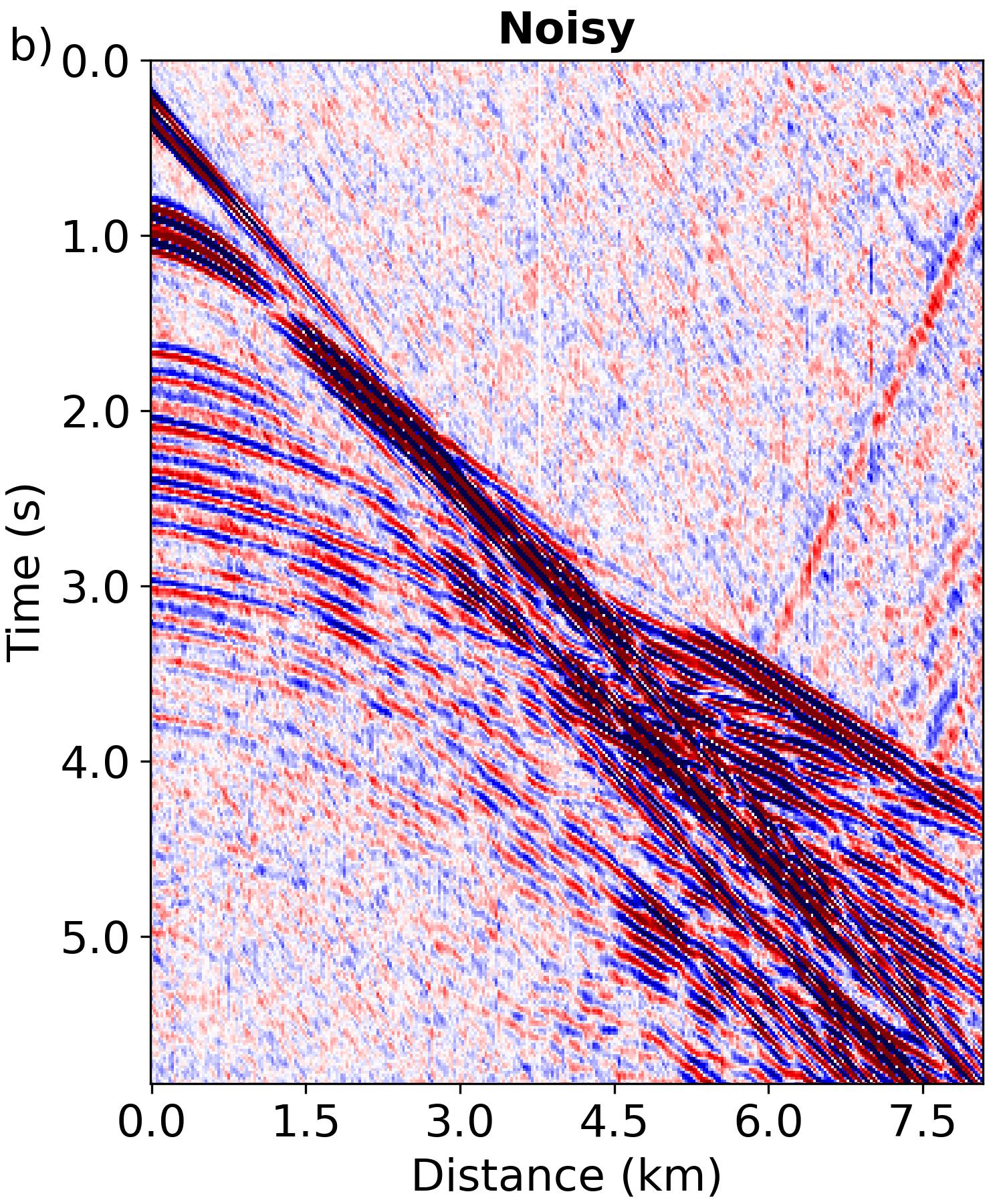}
\caption{The clean (a) and noisy (b) data of the synthetic test data, where the noisy data is contaminated with backscattered noise. The backscattered noise is extracted from the field data and directly inject into the synthetic data with a gain of 10 times to generate the noisy data.}
\label{fig9}
\end{figure}

\begin{figure*}[!t]
\centering
\includegraphics[width=0.3\textwidth]{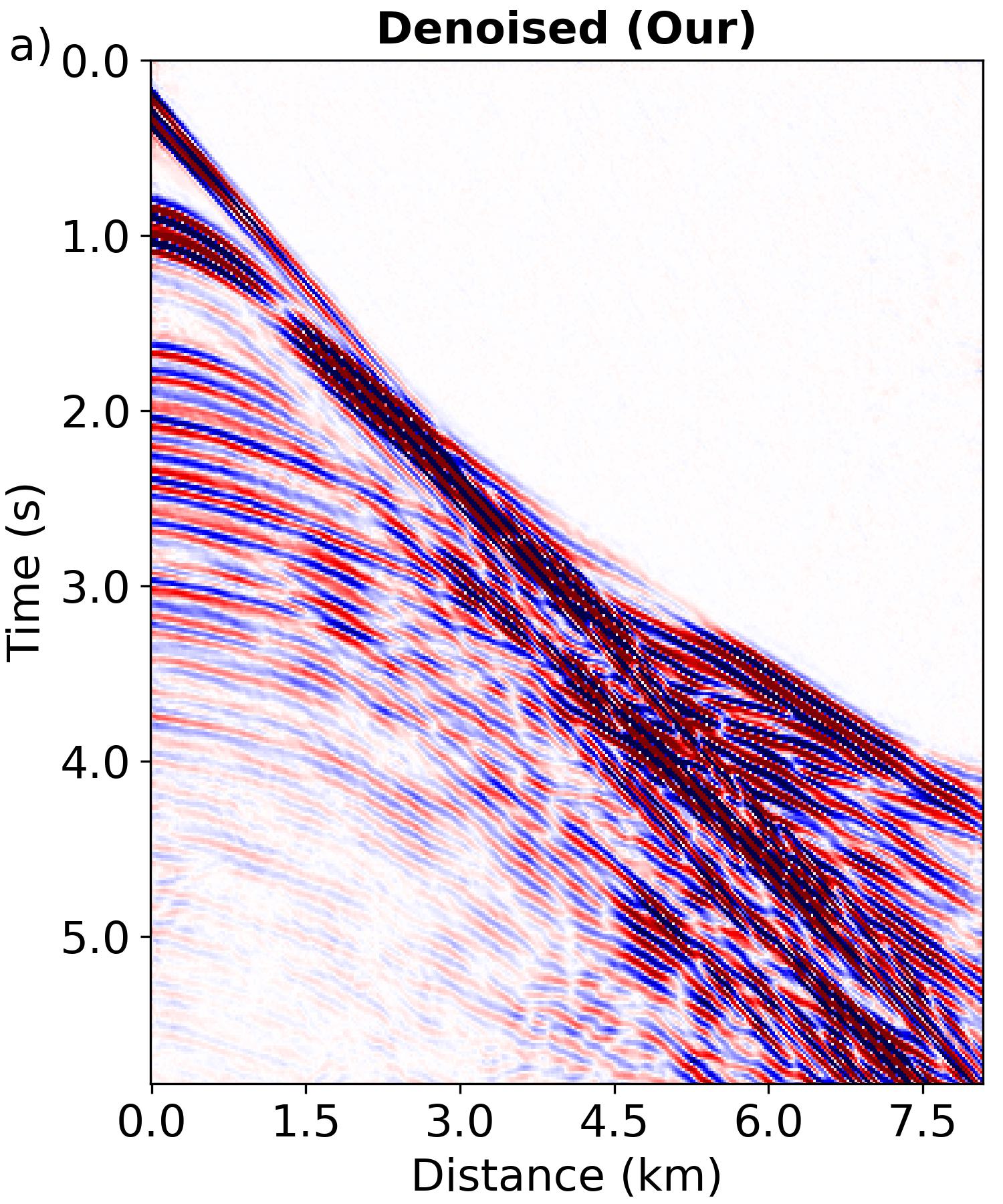} 
\includegraphics[width=0.3\textwidth]{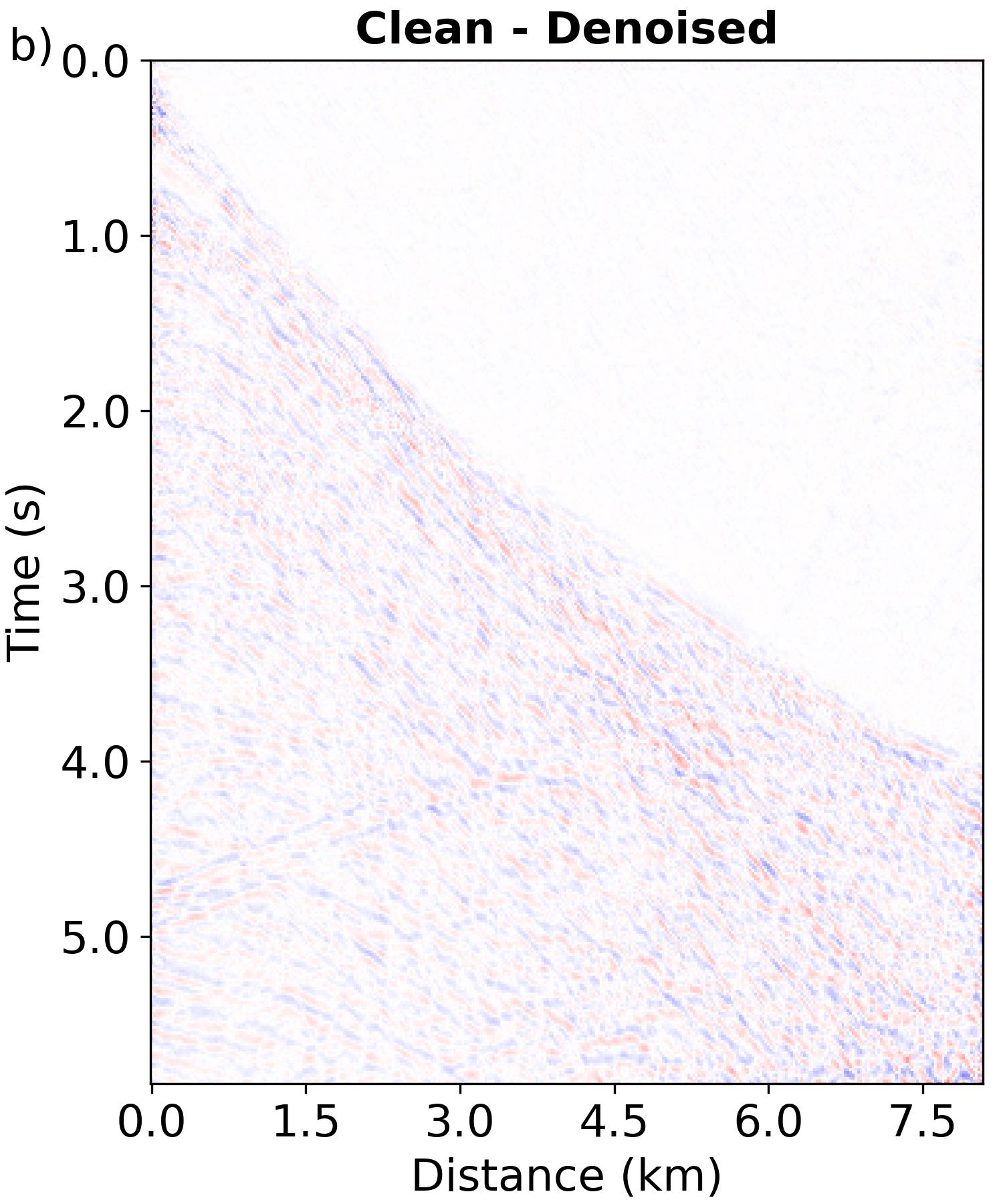}
\includegraphics[width=0.3\textwidth]{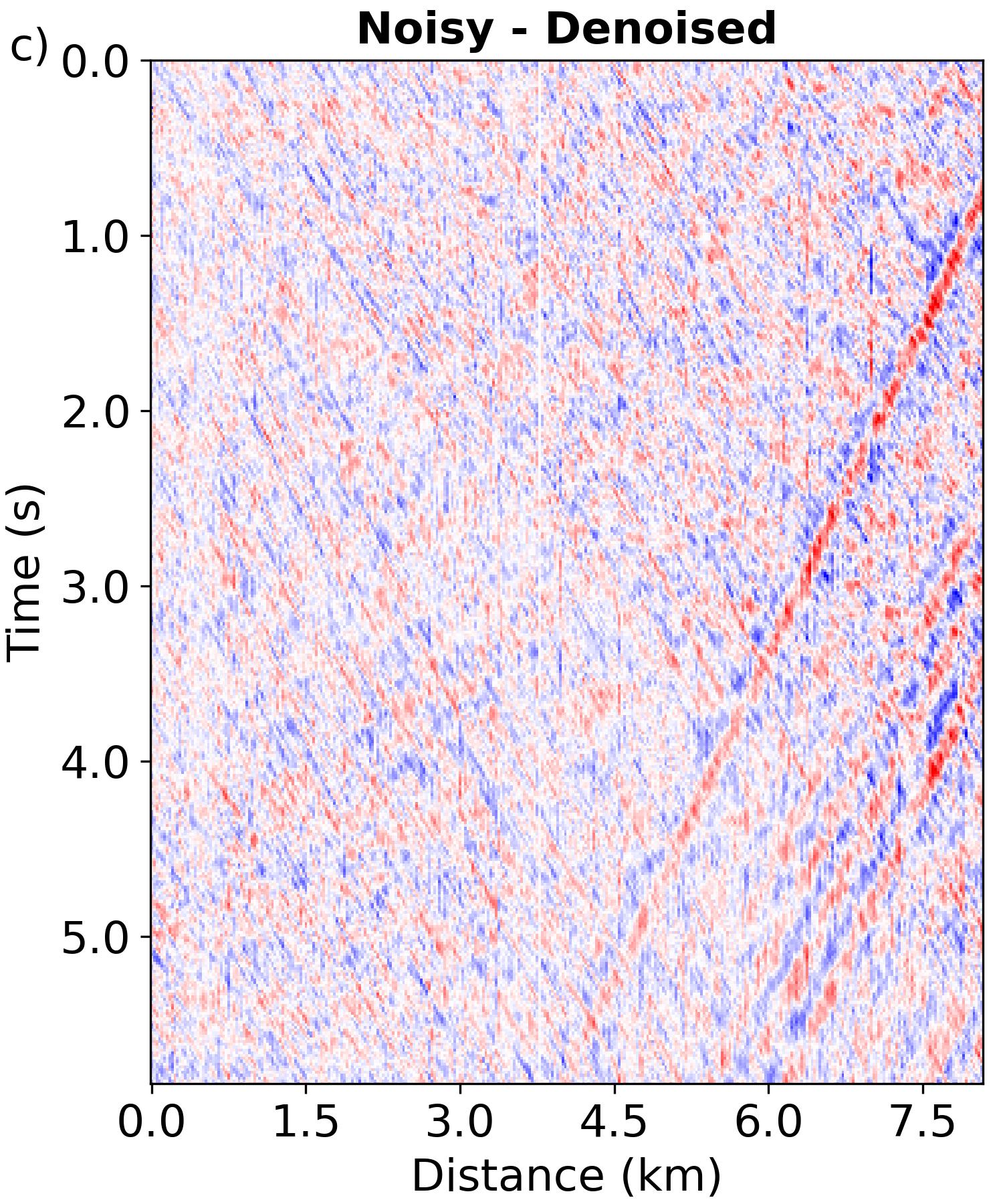} \\
\includegraphics[width=0.3\textwidth]{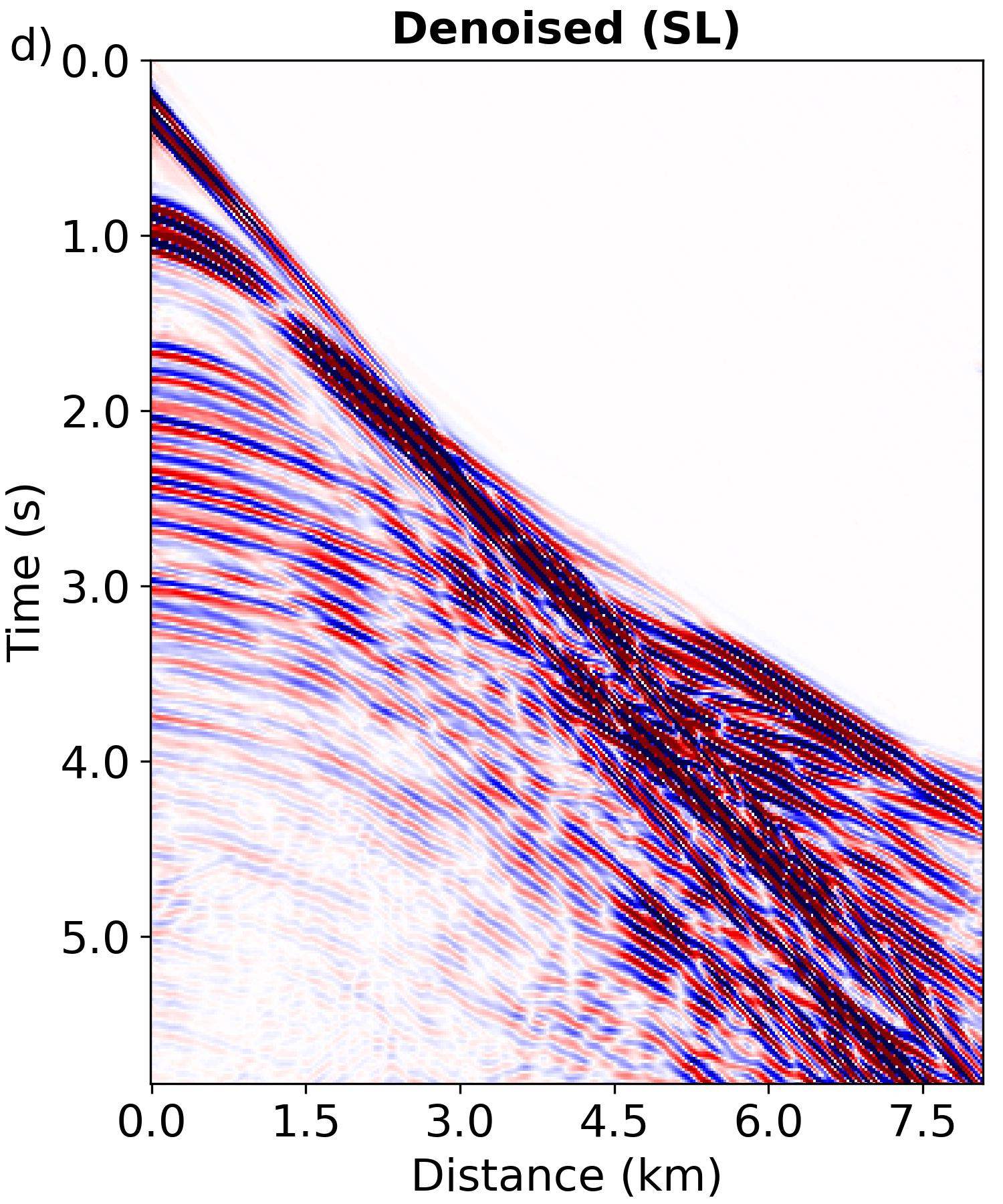} 
\includegraphics[width=0.3\textwidth]{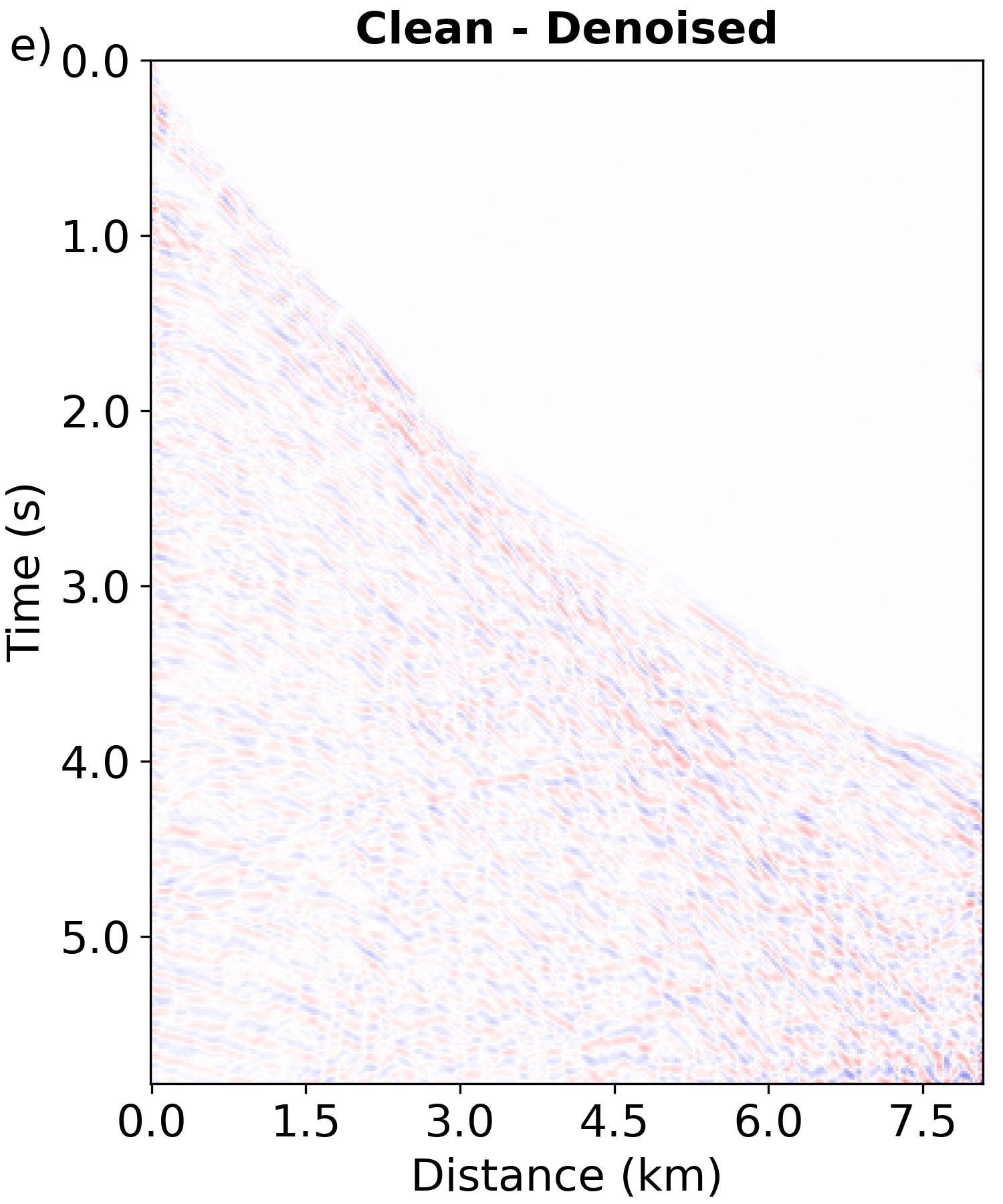} 
\includegraphics[width=0.3\textwidth]{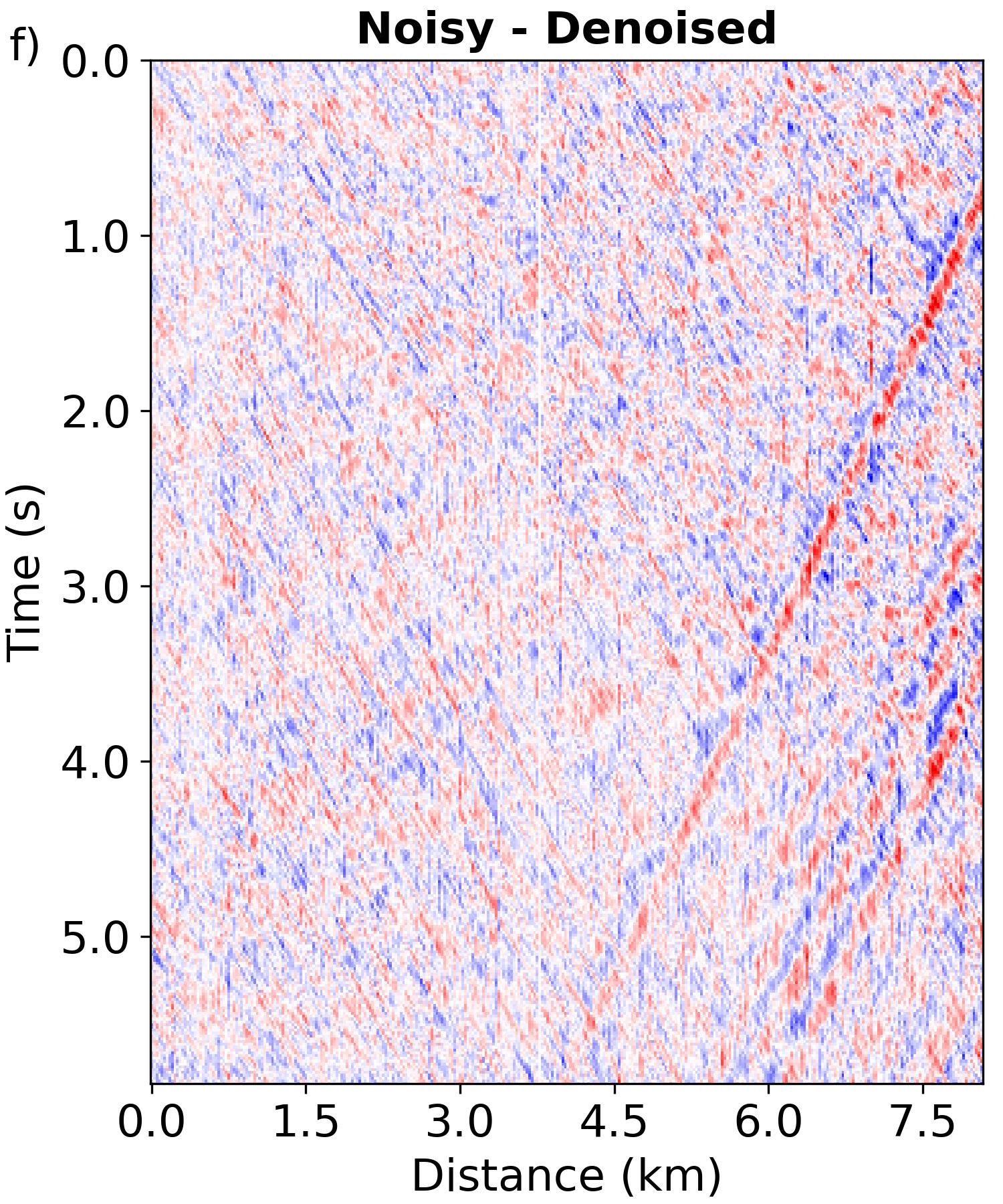} \\
\includegraphics[width=0.3\textwidth]{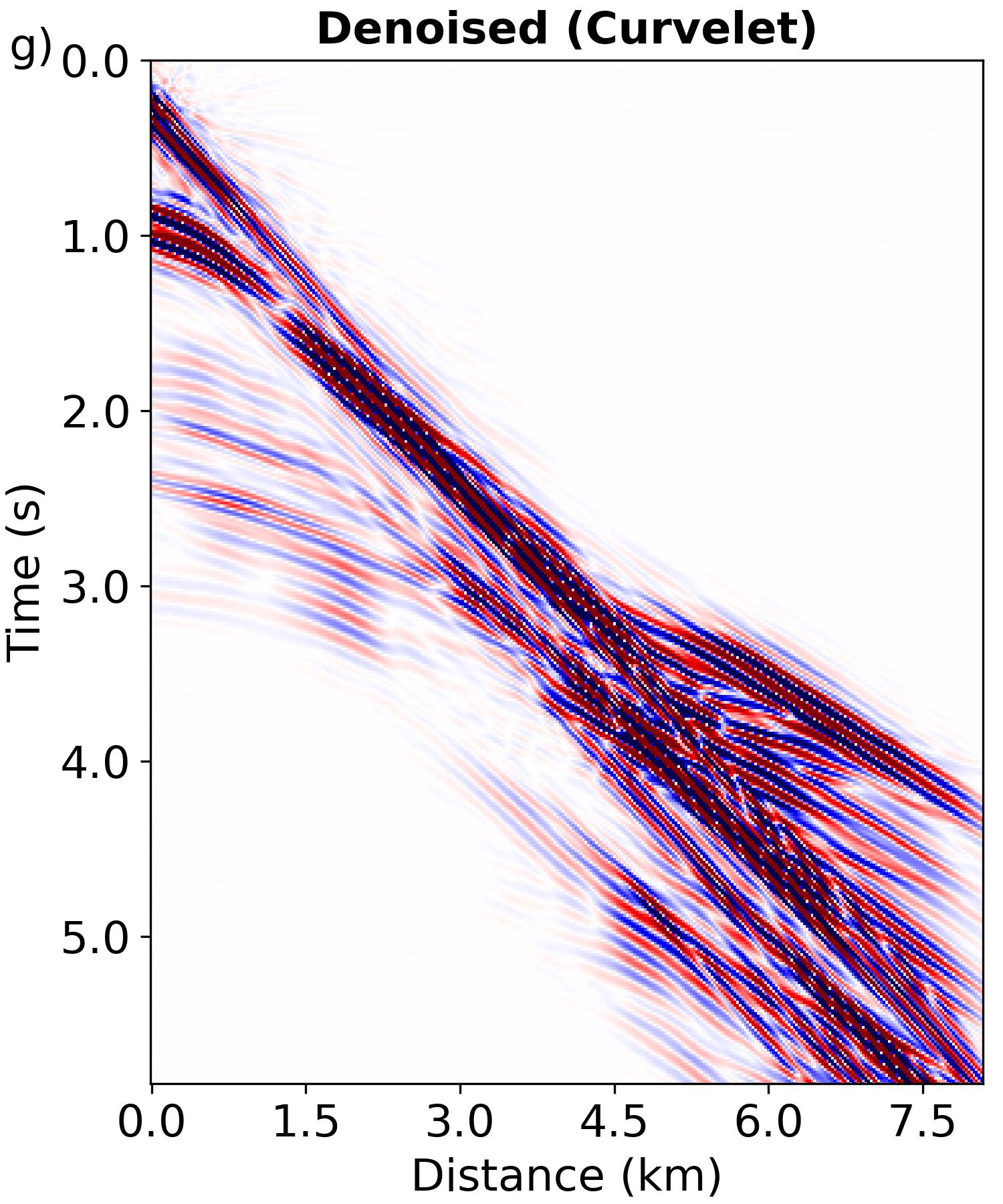} 
\includegraphics[width=0.3\textwidth]{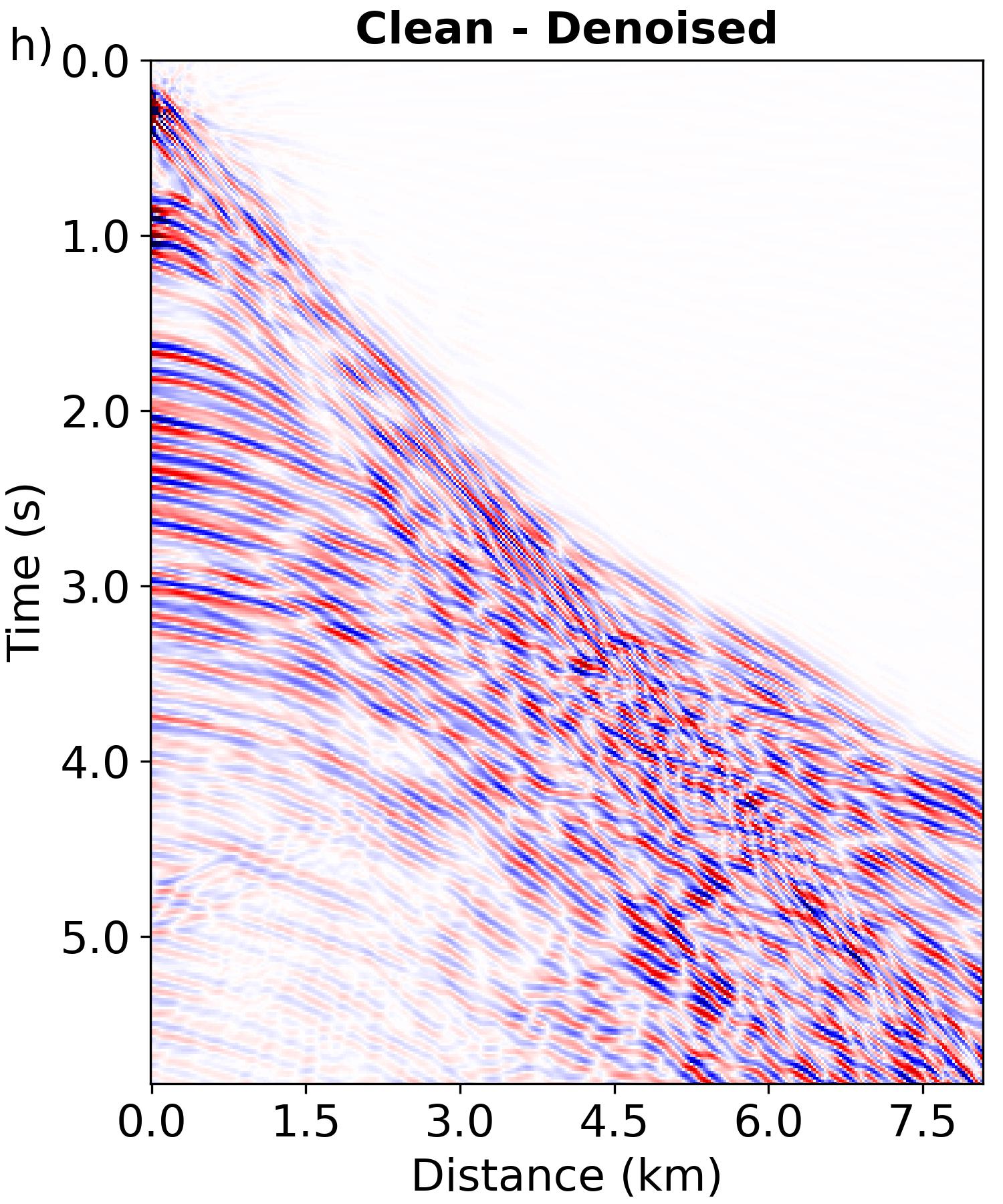}
\includegraphics[width=0.3\textwidth]{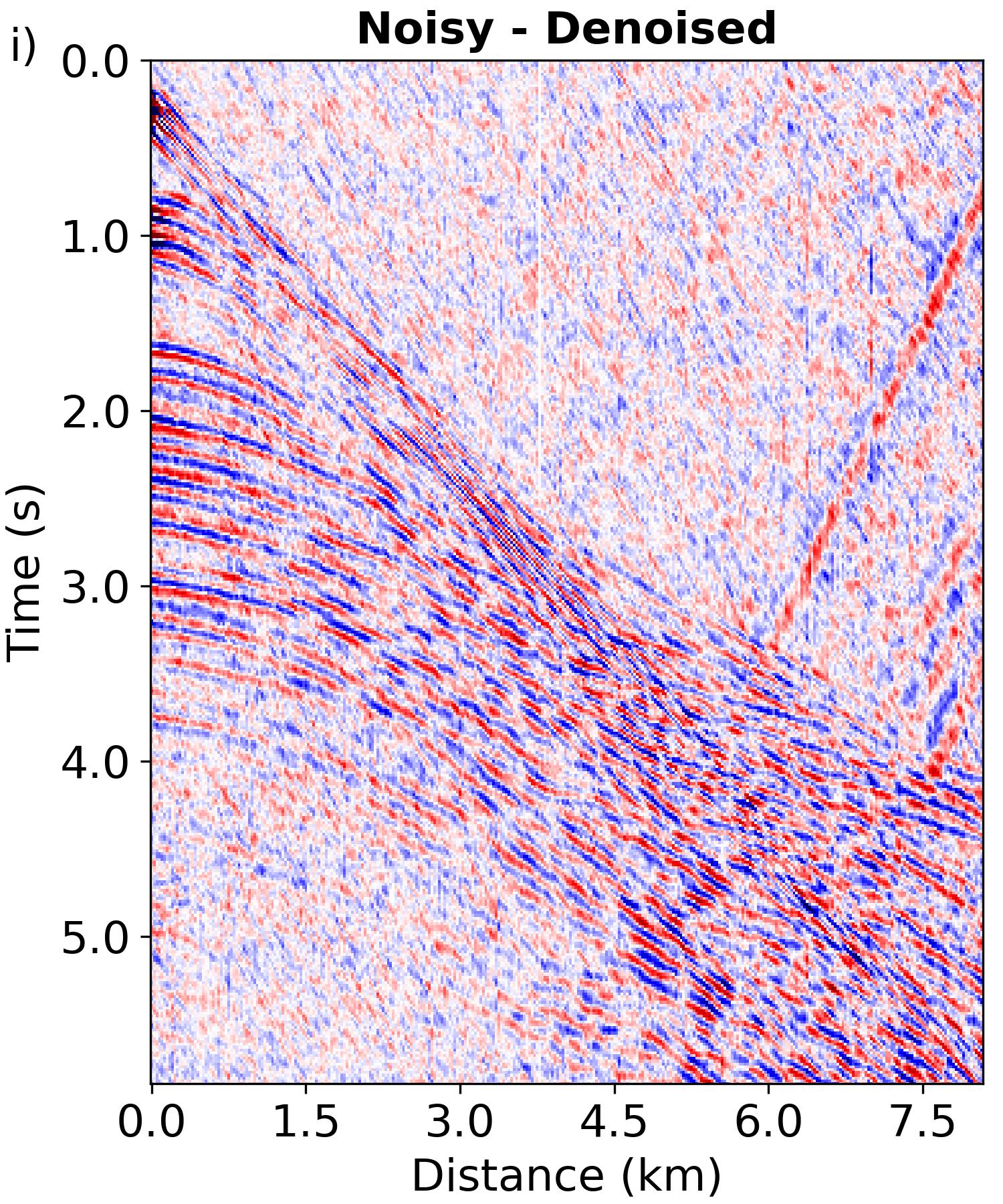}
\caption{Denoising performance comparison on synthetic data contaminated with backscattered noise using our method, SL, and curvelet-based approach, presented from top to bottom respectively. The first column corresponds to denoised results. The second column depicts the difference between the denoised results and the clean data, while the third column shows the difference between the denoised results and the noisy data.}
\label{fig10}
\end{figure*}

Fig. \ref{fig9} displays the clean label data used for testing and its noisy counterpart contaminated with backscattered noise. The noisy data is obtained by injecting the extracted backscattered noise with a gain of 10 into the clean data. It's evident that the characteristics of backscattered noise differ significantly from random noise, particularly in the far-offset backscattered artifacts. Fig. \ref{fig10} shows the denoising products from our approach, SL, and the curvelet-based method. As demonstrated in the case of random noise attenuation, our method still offers competitive denoising performance comparable to SL, nearly eliminating the backscattered noise with only minor impact on the signal. The curvelet method exhibits the same issues seen in the random noise examples, notably reducing the signal energy while decreasing noise and introducing artifacts near the direct wave. To quantitatively compare the denoising performance of the three methods, we computed their respective SNR and MAE metrics, which are shown in Table \ref{tab1}. As observed, both metrics for our method closely align with the SL approach, and their denoising capability significantly outperforms the curvelet method. \\

\begin{table}
\centering
\caption{The comparison of denoising performance of different methods on original noisy data contaminated by backscattered noise, including SNR and MAE metrics.}
\renewcommand\arraystretch{1.5}
\setlength{\tabcolsep}{20pt}
\begin{tabular}{ccc}
    \hline
    \text {~} & \text { SNR } & \text { MAE } \\
    \hline
    \text {Raw noisy data} & $13.89$ & $0.00354$ \\
    \text {Our method} & $26.47$ & $0.00081$ \\
    \text {Supervised} & $28.51$ & $0.00065$ \\
    \text {Curvelet} & $12.80$ & $0.00340$ \\
    \hline
\end{tabular}
\label{tab1}
\end{table}

\subsubsection{\textbf{Field data}}
We further tested our method's capability to attenuate backscattered noise on field data. This data was acquired in North West Australia using a streamer containing 324 hydrophones with a 25 m spacing that recorded 1824 shots. We select 201 shot gathers and extract 10000 patches, each sized 128x128. The levels of the added backscattered noise during the warm-up and IDR phases remain consistent at $s=[0.2,2]$. The initial learning rate and its decay settings inherit the training configuration of synthetic data.

\begin{figure*}[!t]
\centering
\begin{minipage}[t]{0.23\textwidth}
\includegraphics[width=\textwidth]{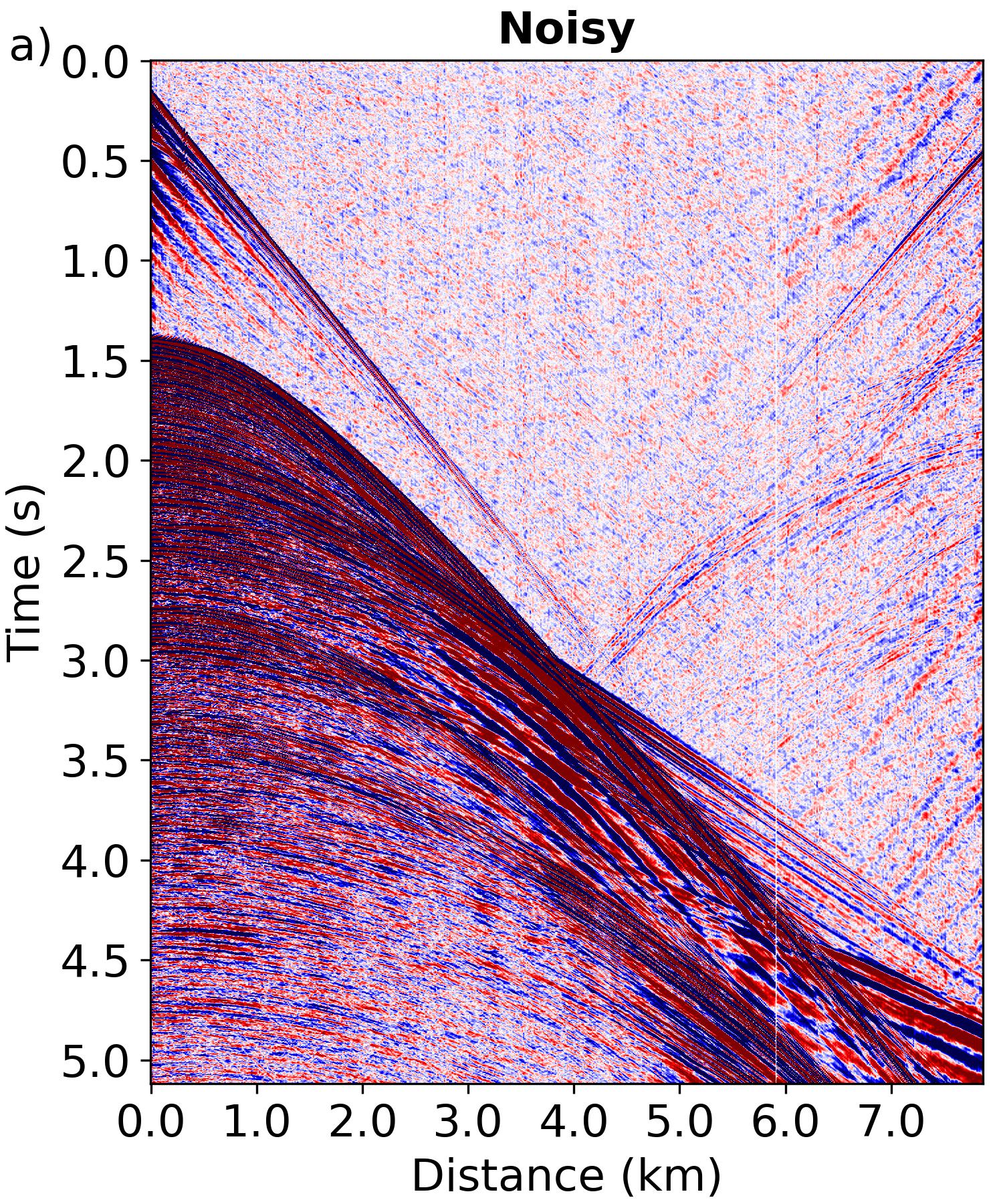} 
\end{minipage}
\hfill
\begin{minipage}[t]{0.23\textwidth}
\includegraphics[width=\textwidth]{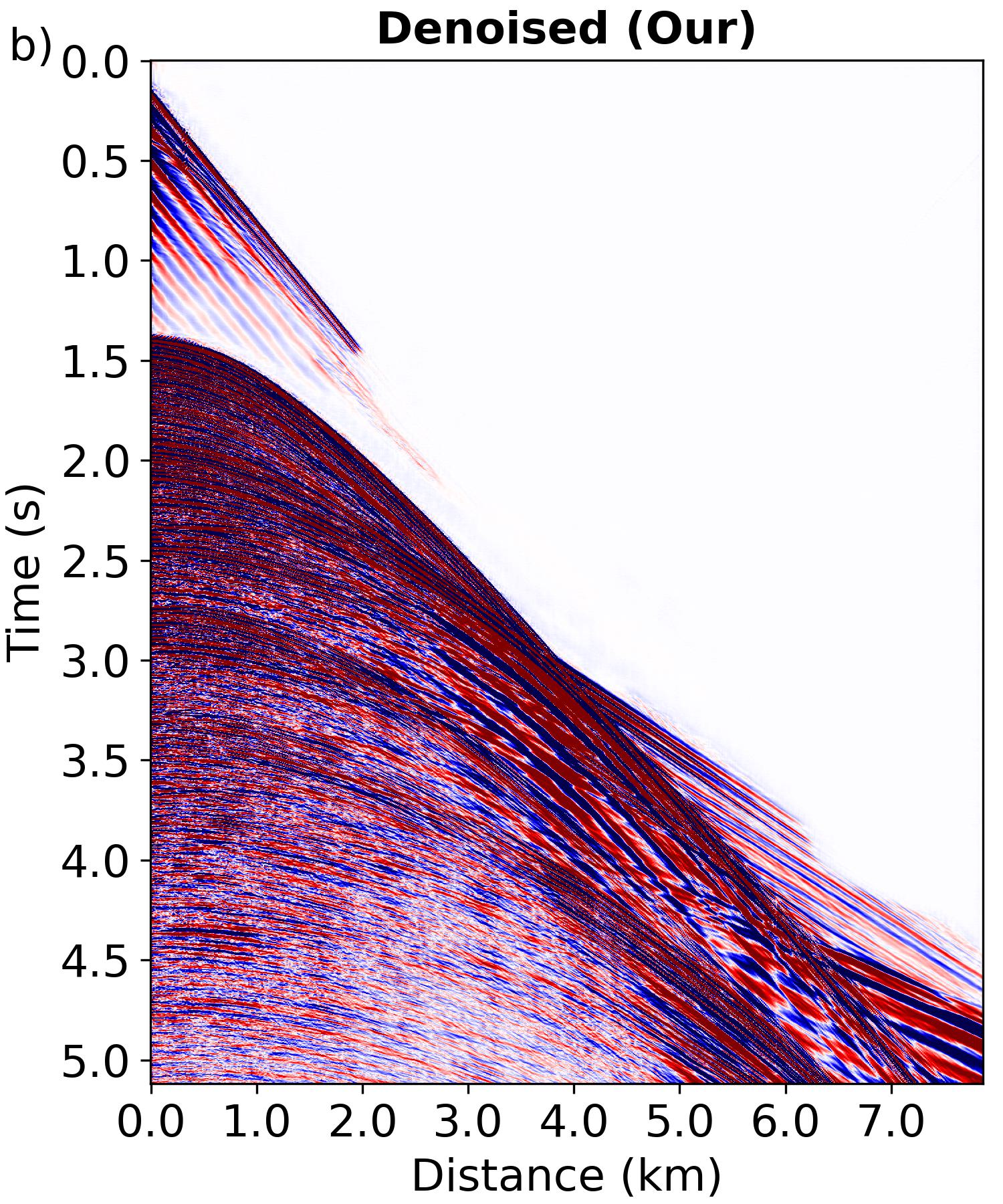} \\
\vfill
\includegraphics[width=\textwidth]{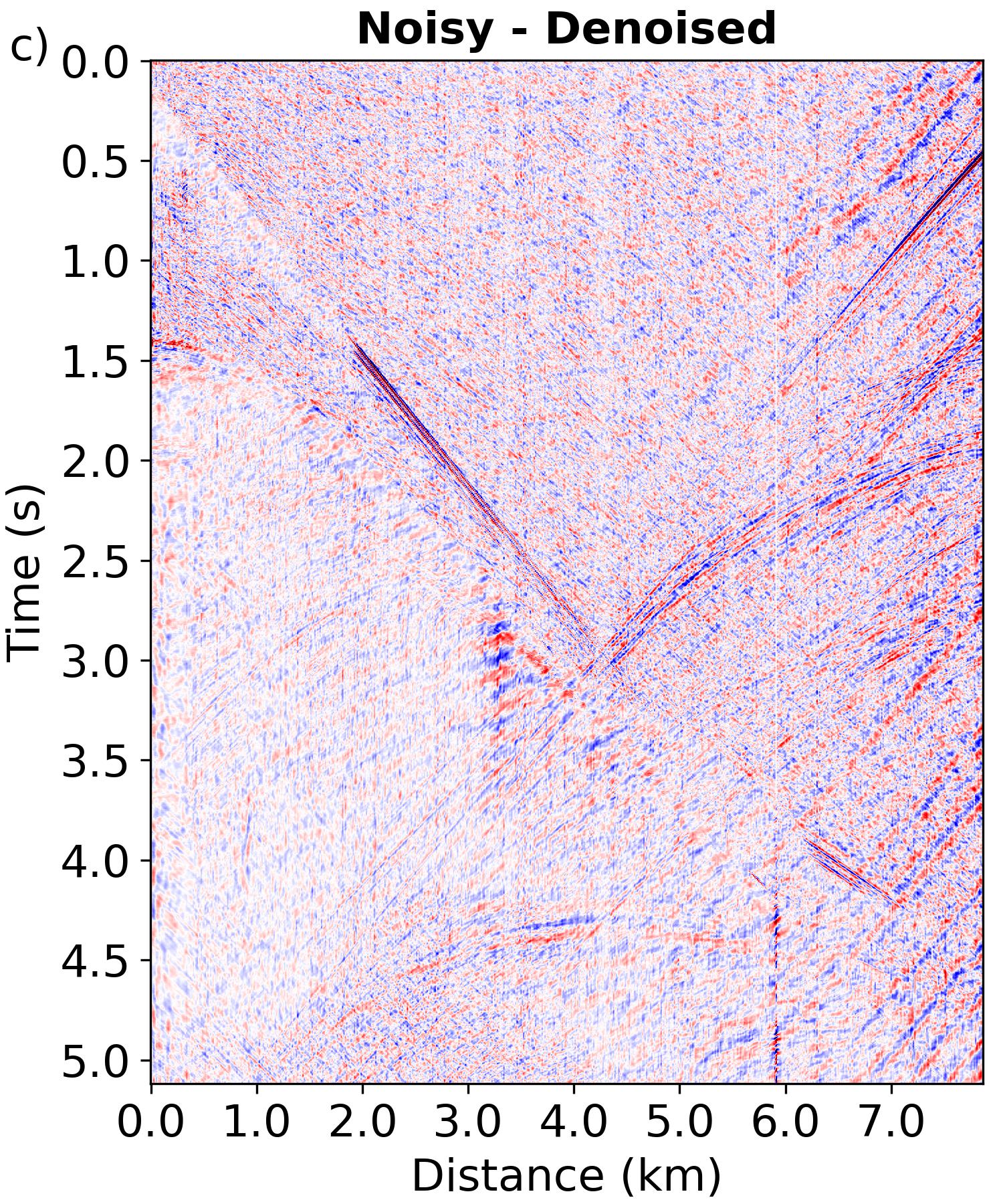} 
\end{minipage}
\hfill
\begin{minipage}[t]{0.23\textwidth}
\includegraphics[width=\textwidth]{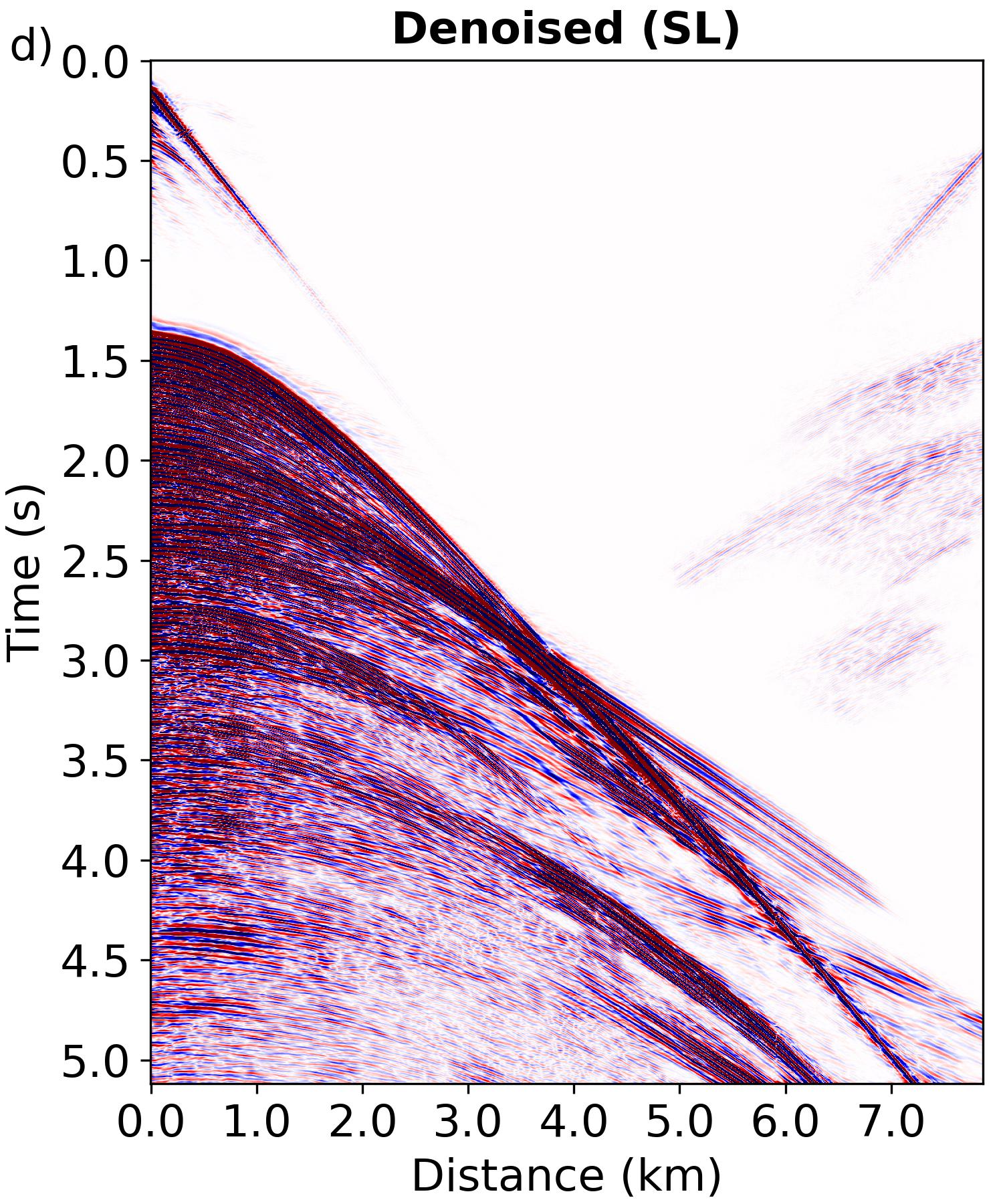} \\
\vfill
\includegraphics[width=\textwidth]{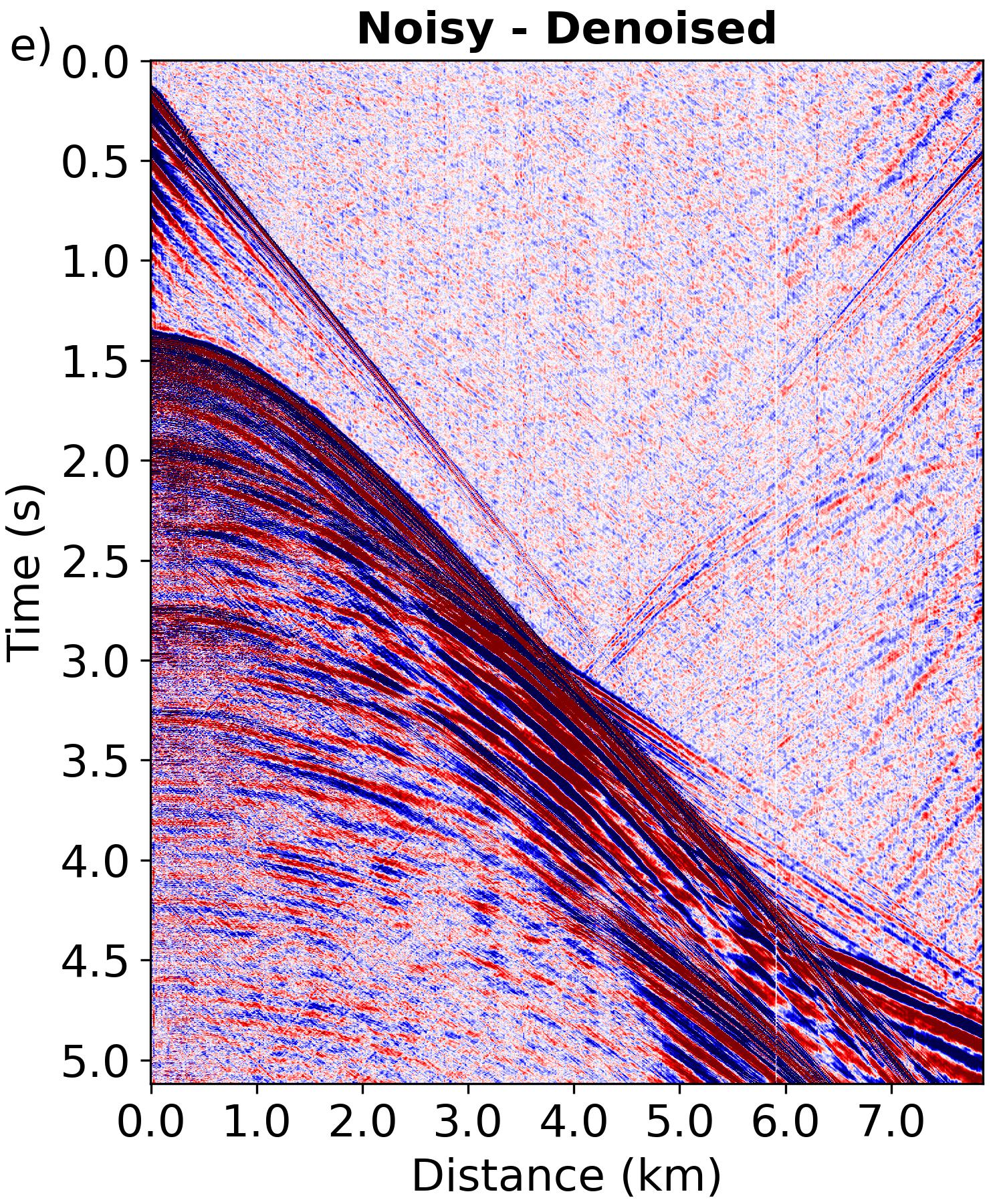} 
\end{minipage}
\hfill
\begin{minipage}[t]{0.23\textwidth}
\includegraphics[width=\textwidth]{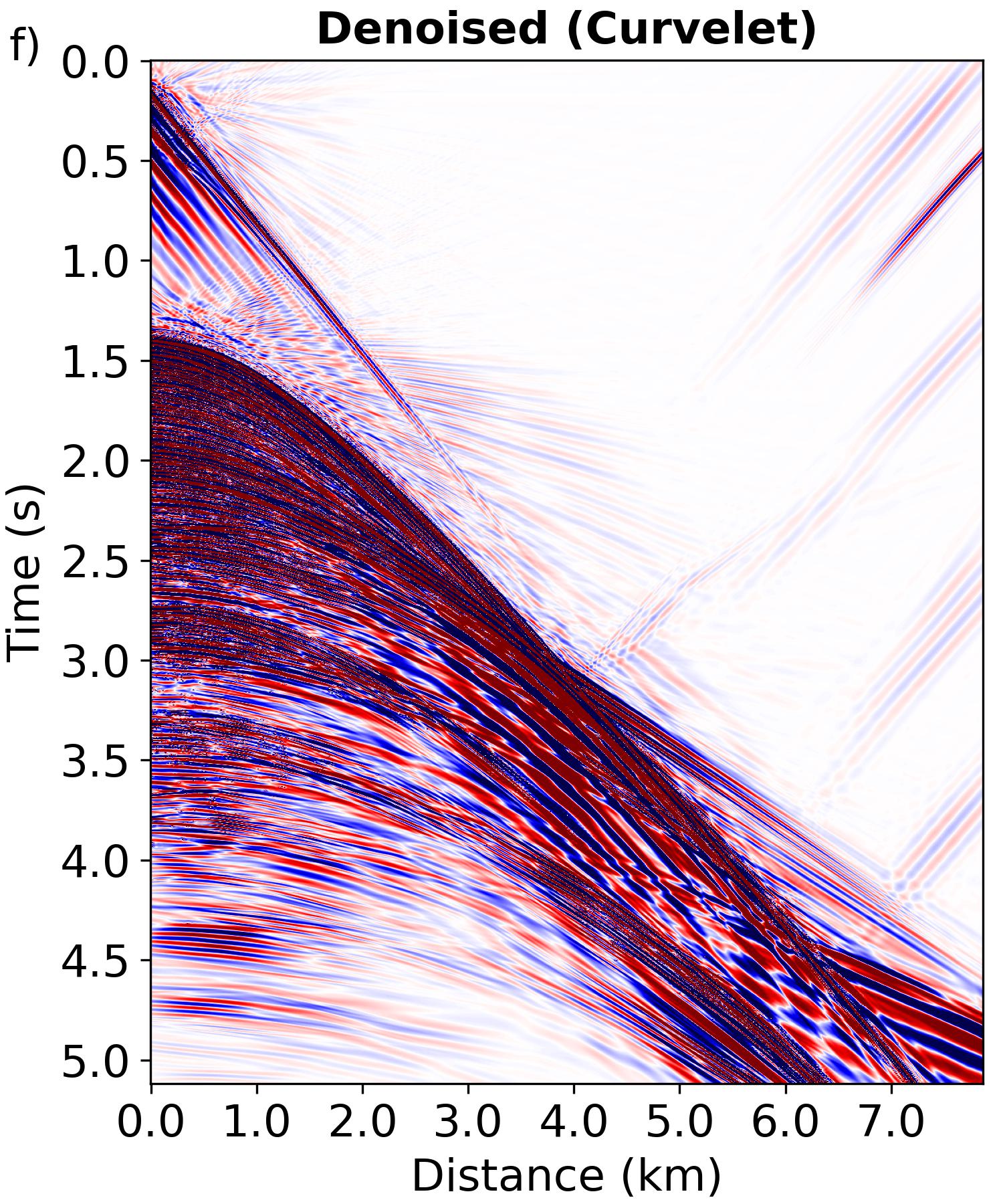} \\
\vfill
\includegraphics[width=\textwidth]{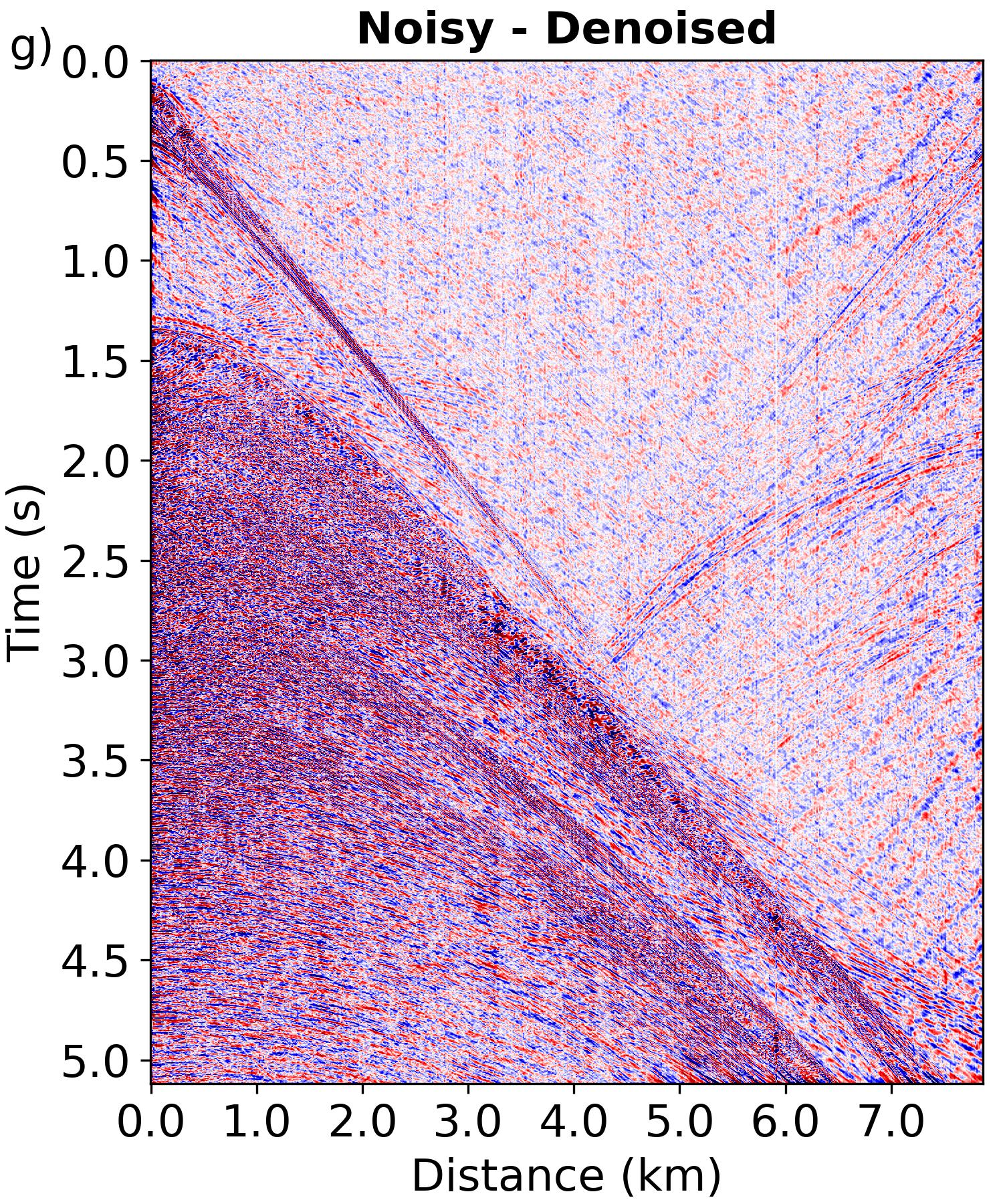} 
\end{minipage}
\caption{Comparison of the denoising performance on a field data, which acquires from North West Australia and is contaminated by backscattered noise. (a) The original noisy data. (b), (d), and (f) represent the denoised results from our method, SL, and curvelet-based method, respectively. The corresponding differences between the original noisy data and the denoised results are shown on the bottom.}
\label{fig11}
\end{figure*}

Fig. \ref{fig11} illustrates the denoising performance comparison of different methods on a noisy shot gather. Fig. \ref{fig11}a shows the original noisy shot. In contrast to what we observe in synthetic noisy data, the backscattered noise here is more intricate, manifesting in reverse arc-like artifacts, which intensifies the denoising challenge. Figs. \ref{fig11}b, \ref{fig11}d, and \ref{fig11}f present the denoising products from our method, SL, and the curvelet approach, respectively, with the associated residuals shown in the bottom row. We can see that, our method once again demonstrates exemplary performance, nearly impeccably mitigating noise contamination, with only a slight attenuation of the direct wave energy. The denoised result from the SL method still retains artifacts from the backscattered noise and notably diminishes the signal's low-frequency components. The curvelet approach still yield a relatively inferior denoising result, evident in the residual backscattered noise artifacts, attenuation of high-frequency signals, and the introduction of artifacts near the direct wave.

\subsection{Blending noise attenuation}
\subsubsection{\textbf{Synthetic data}}
Finally, we evaluate the capability of our method to mitigate the blending noise. Compared to the random noise and the backscattered noise, the blending noise is a type of coherent noise with strong energy, causing it more challenging to handle. Our examination begin with synthetic data. The clean synthetic data is simulated based on the SEAM model. From 14 pseudo-deblended shot gathers, we extract 10000 patches, each with the size of 128x128, to establish our original noisy dataset. The corresponding blending noise is extracted outside of the first arrivals from the pseudo-deblended shot gathers. The noise levels introduced during the warm-up and IDR phases are set at $s=[1,2]$. The warm-up phase performs 30 epochs of training, followed by 300 epochs of training in the IDR stage. The initial learning rate is 2e-4, which is reduced by a factor of 0.8 at the 20, 40, 60, 80, and 100 epochs. For the SL, it also has the same number of training patches as our method, except we exclude the test data. Furthermore, it retains the same training epochs and configurations as our approach.

\begin{figure}[htp]
\centering
\includegraphics[width=0.3\textwidth]{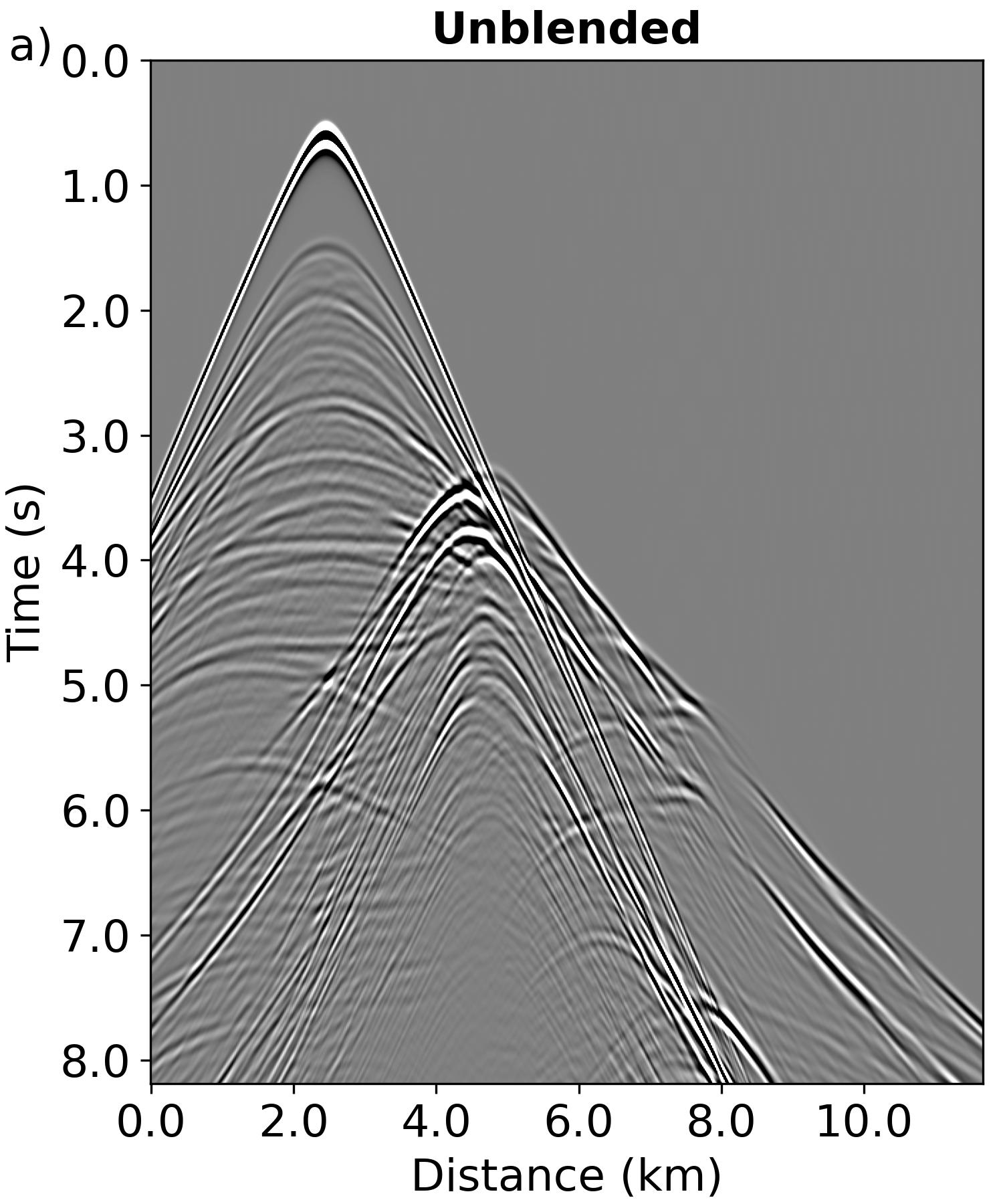}
\hspace{1cm}
\includegraphics[width=0.3\textwidth]{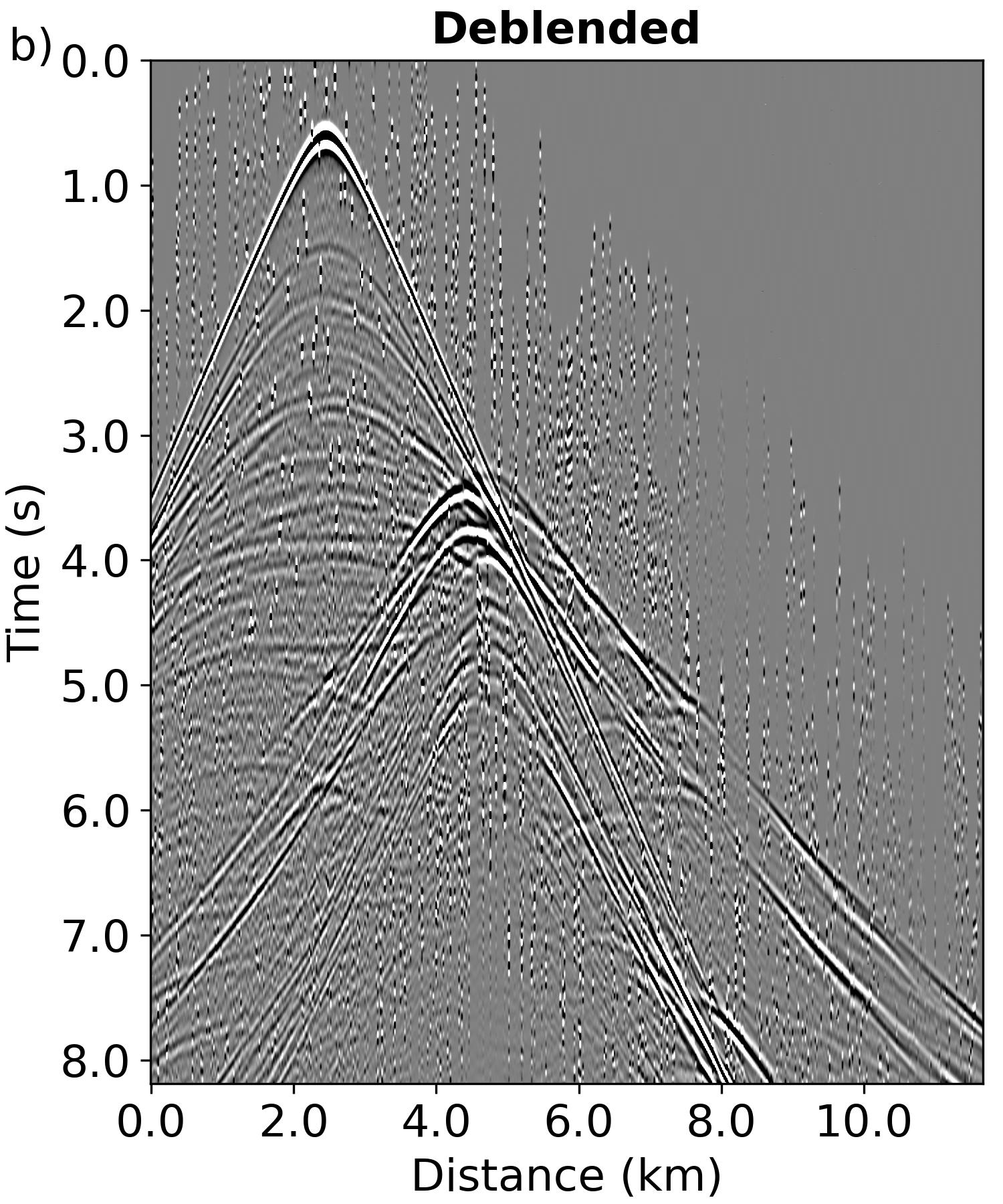}
\caption{The raw unblended shot (a) and the pseudo-deblended (b) shot from the SEAM model. }
\label{fig12}
\end{figure}

\begin{figure*}[!t]
\centering
\includegraphics[width=0.3\textwidth]{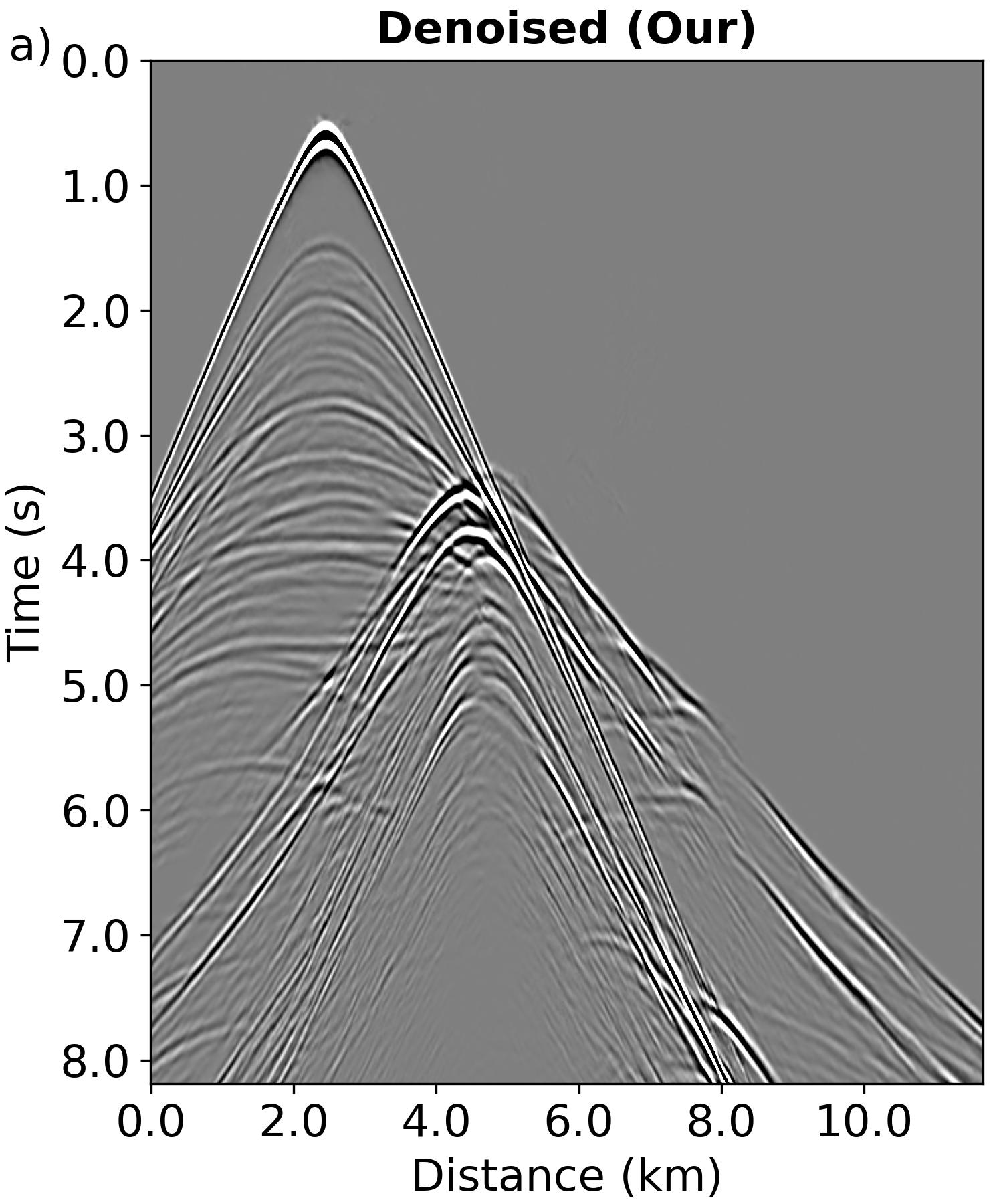} 
\includegraphics[width=0.3\textwidth]{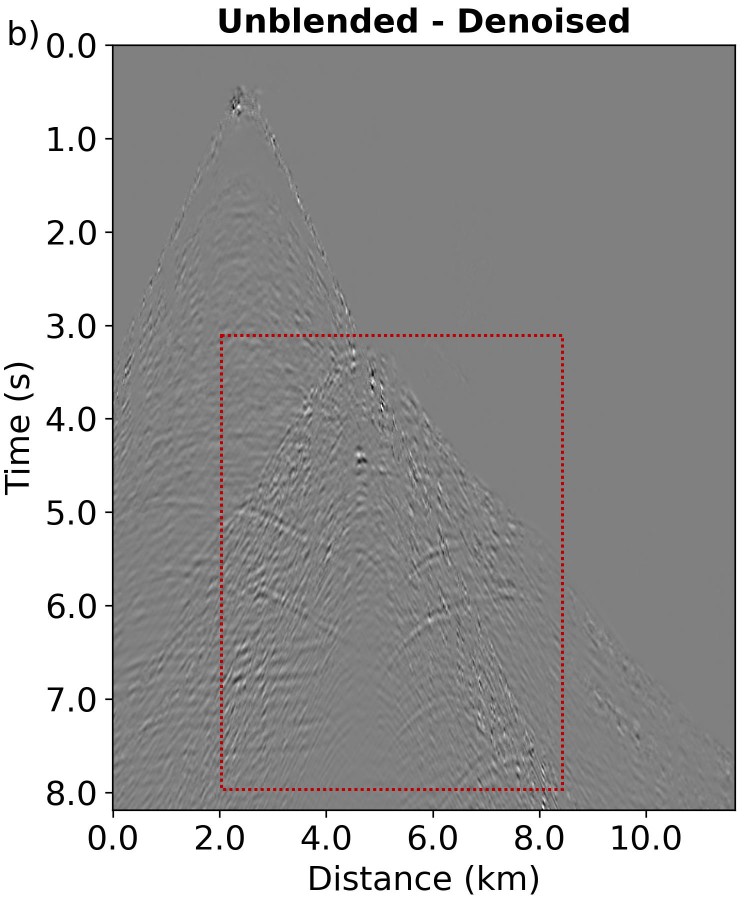}
\includegraphics[width=0.3\textwidth]{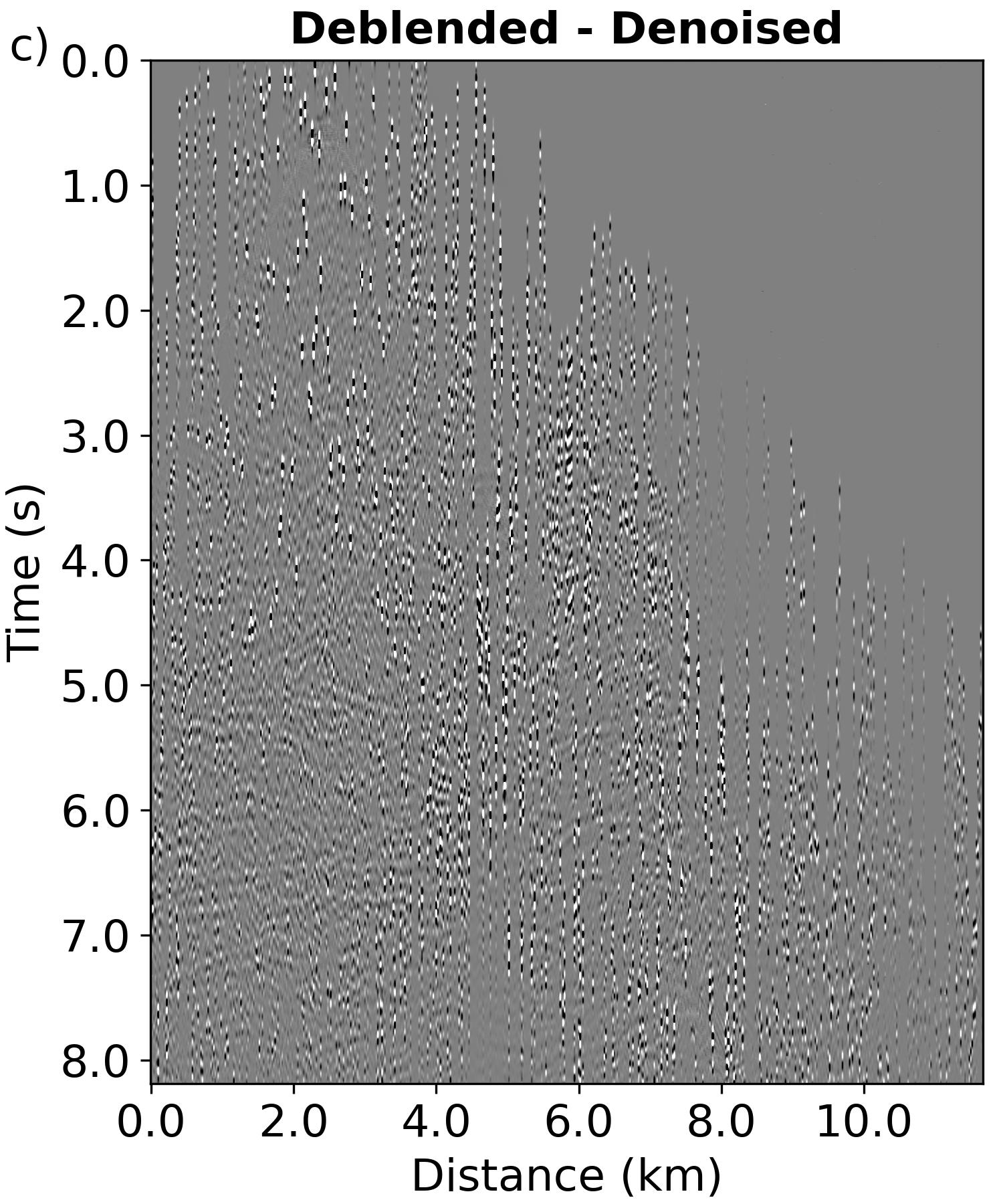} \\
\includegraphics[width=0.3\textwidth]{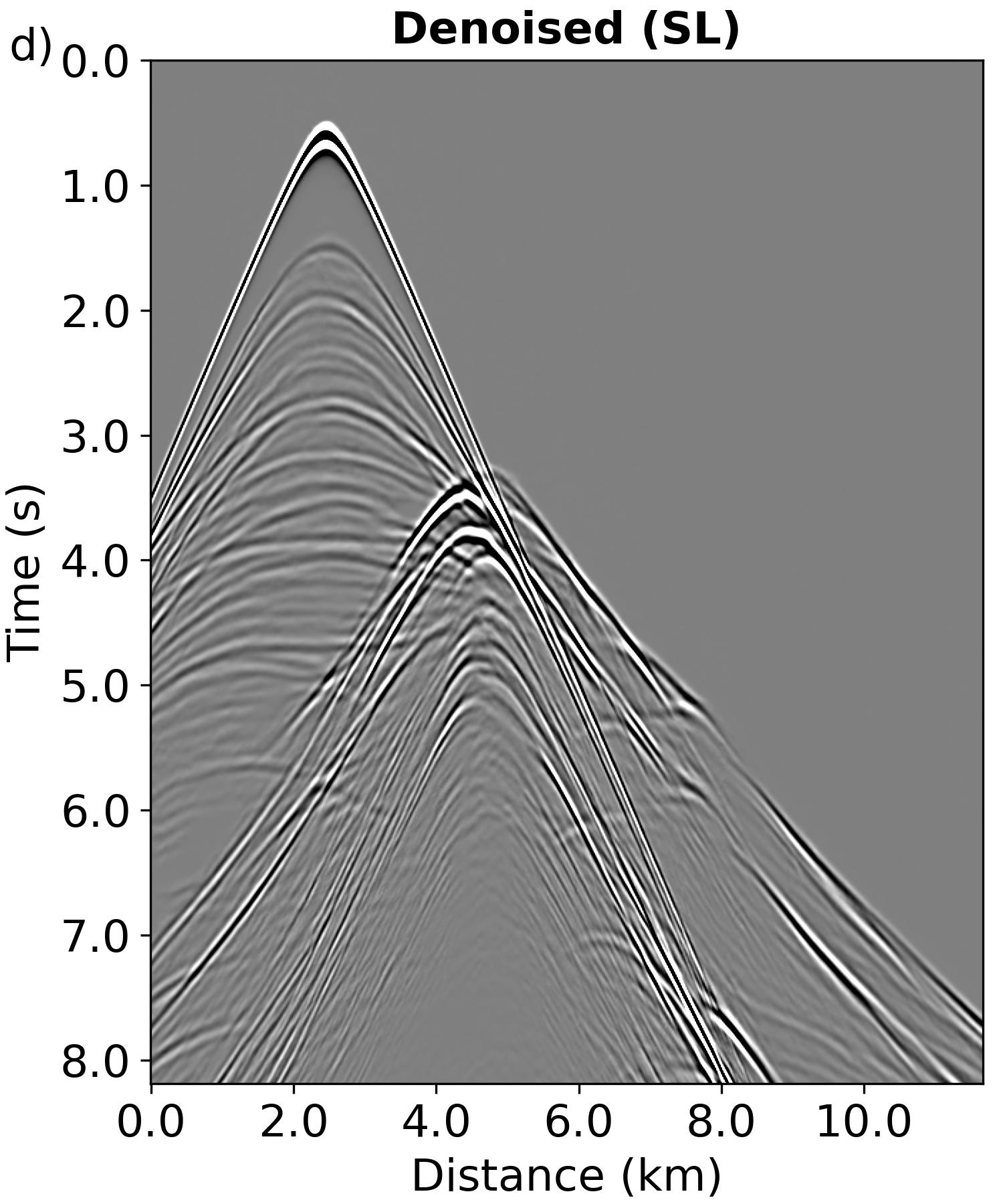} 
\includegraphics[width=0.3\textwidth]{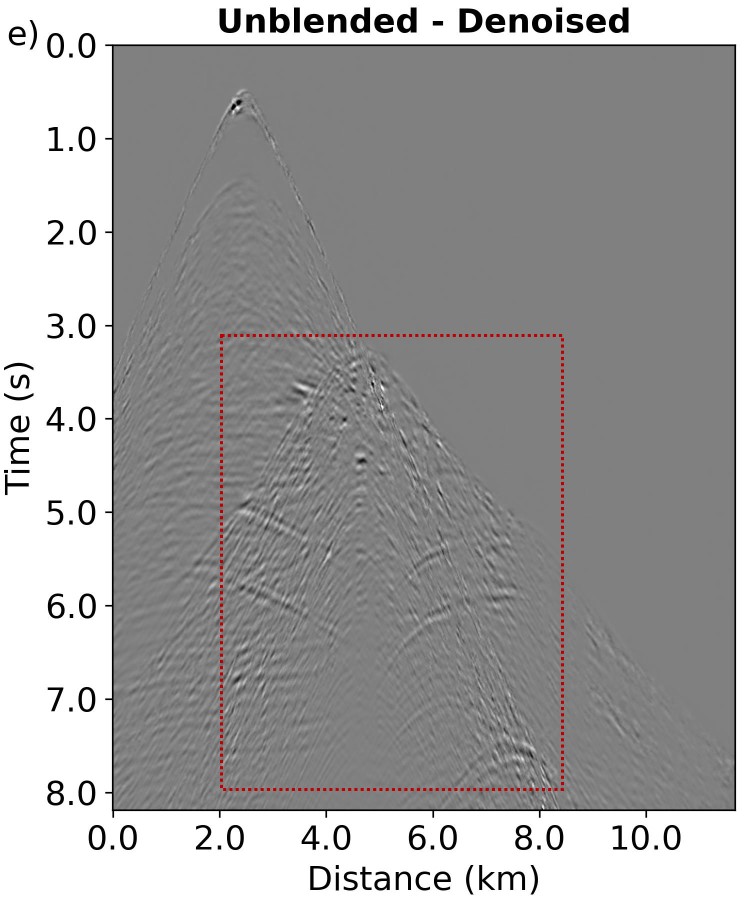} 
\includegraphics[width=0.3\textwidth]{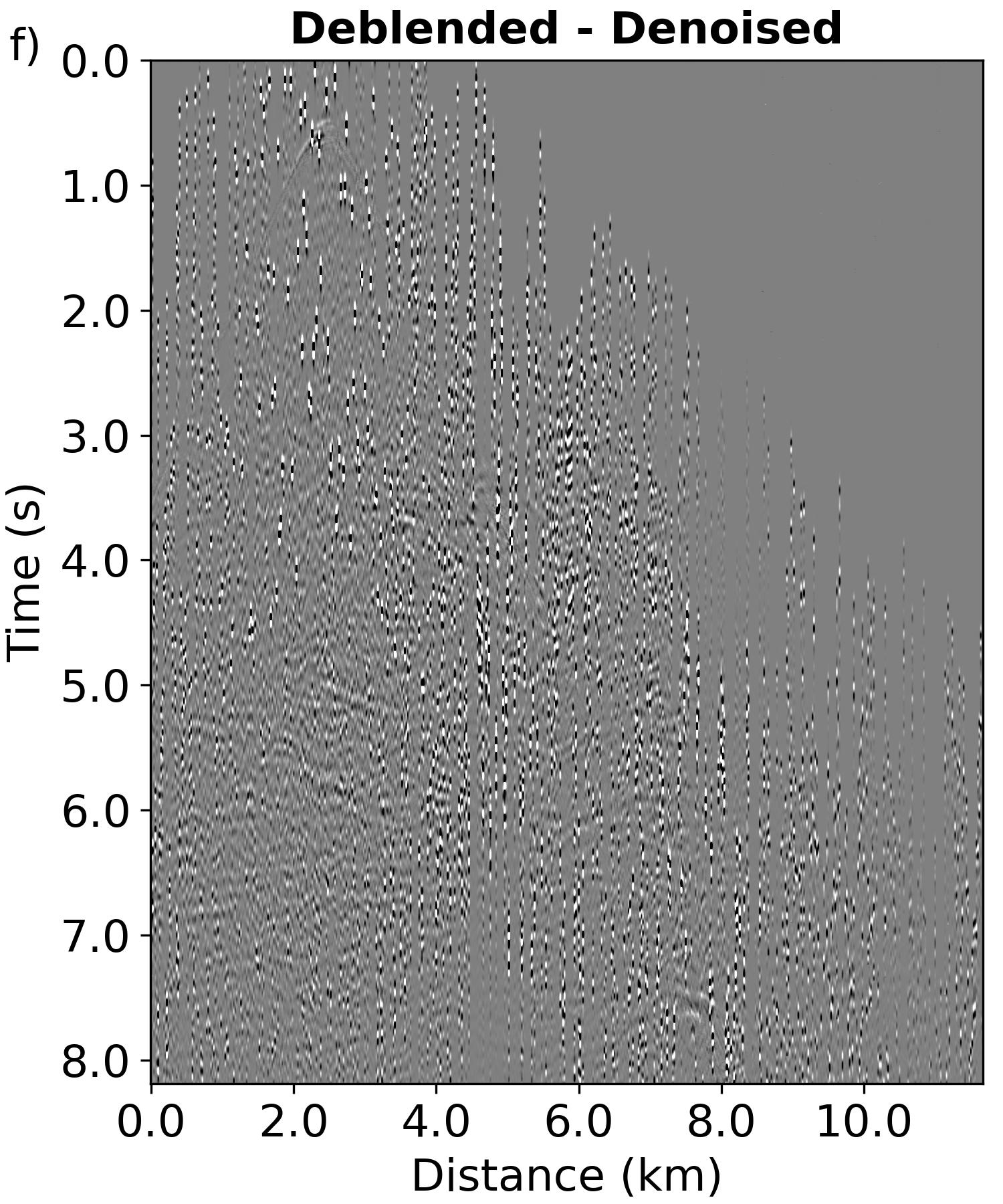} \\
\includegraphics[width=0.3\textwidth]{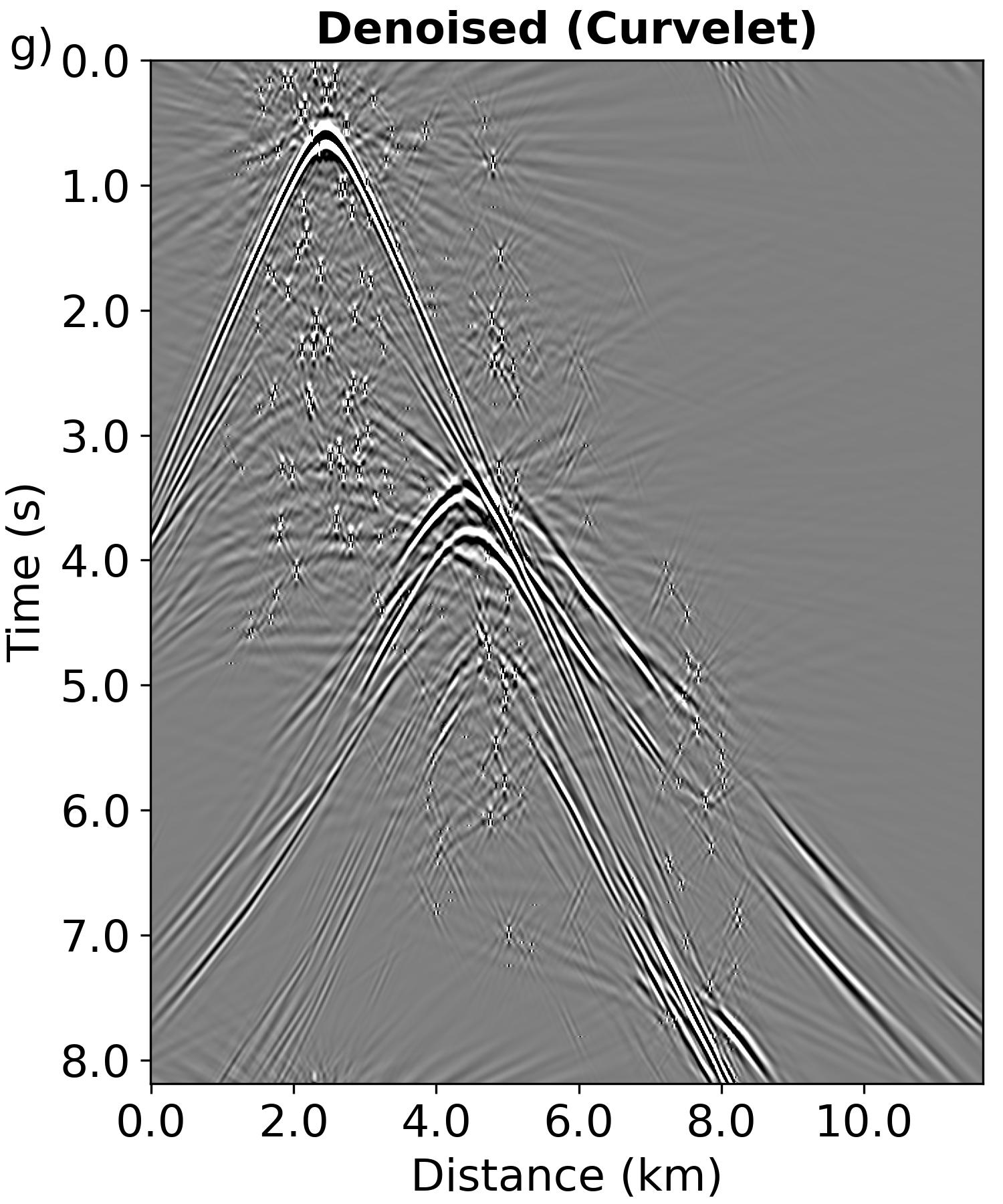} 
\includegraphics[width=0.3\textwidth]{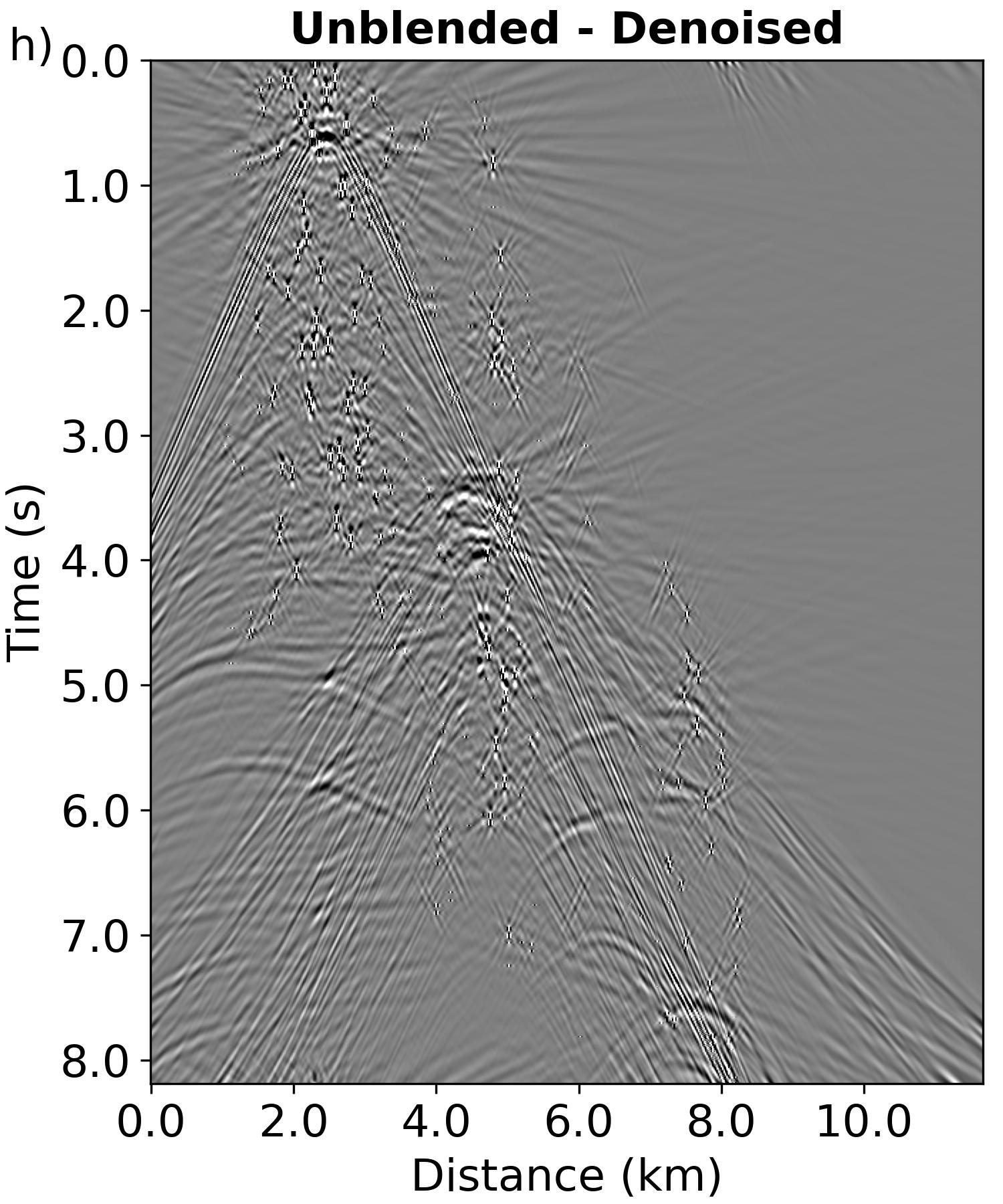}
\includegraphics[width=0.3\textwidth]{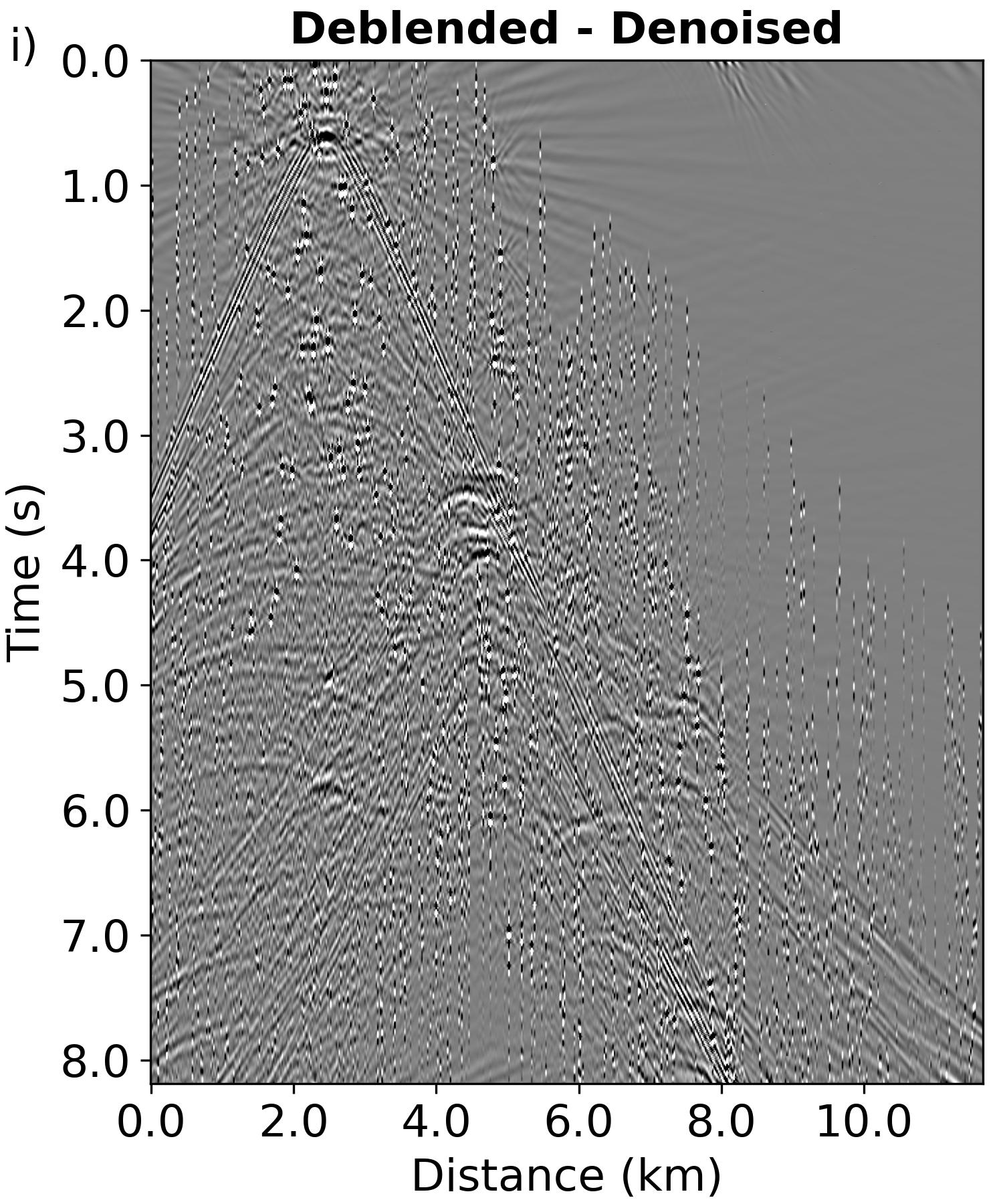}
\caption{Denoising performance comparison on synthetic data contaminated with blending noise using our method, SL, and curvelet-based approach, presented from top to bottom respectively. The first column corresponds to denoised results. The second column depicts the difference between the denoised results and the raw unblended data, while the third column shows the difference between the denoised results and the deblended data.}
\label{fig13}
\end{figure*}

We select one shot gather for testing from the 14 pseudo-deblended shot gathers involved in our method training, which is depicted in Fig. \ref{fig12}b, while the corresponding raw unblended data is shown in panel a. The denoised results of our method, SL, and the curvelet method are presented in Fig. \ref{fig13}. It's evident that our method achieves outstanding denoising performance, holding its own even when compared with the SL method. Notably, as displayed within the red boxes in panels b and e of Fig. \ref{fig13}, our method even demonstrates more minimal signal leakage, proving its superior performance over the SL method in this instance, which is quite encouraging. The curvelet method displays very limited capability in mitigating blending noise, not only significantly damaging the signal but also introducing strong artifacts. The denoising metrics of the three methods on the test data are exhibited in Table \ref{tab2}. It's observable that our method's SNR metric is almost same with the SL method, while our approach outperforms the SL in the MAE metric. The curvelet method does not demonstrate effective denoising ability, thus its SNR and MAE metrics are only slightly better than the raw noisy data. \\

\begin{table}
\centering
\caption{The comparison of denoising performance of different methods on original noisy data contaminated by blending noise, including SNR and MAE metrics.}
\renewcommand\arraystretch{1.5}
\setlength{\tabcolsep}{20pt}
\begin{tabular}{ccc}
    \hline
    \text {~} & \text { SNR } & \text { MAE } \\
    \hline
    \text {Raw noisy data} & $-3.84$ & $0.00759$ \\
    \text {Our method} & $17.83$ & $0.0014$ \\
    \text {Supervised} & $18.01$ & $0.0015$ \\
    \text {Curvelet} & $-0.01$ & $0.00679$ \\
    \hline
\end{tabular}
\label{tab2}
\end{table}

\subsubsection{\textbf{Field data}}
The field data collected from the southern North Sea serves to further validate the capability of our approach to attenuate blending noise in real scenarios. From this collection, we select 8 pseudo-deblended shot gathers, extracting 3000 patches, each of dimensions 120x120. In a manner analogous to our examples with synthetic data, the blending noise is extracted from regions beyond the direct wave of the pseudo-deblended shot gathers. The noise levels introduced during the warm-up and IDR stages remain consistent at $s=[1,2]$. The warm-up stage employs 30 epochs for pre-training, which is subsequently extended with an additional 570 epochs in the IDR stage. The learning rate starts with 2e-4, and a scheduler reduces it by a factor of 0.8 at the 60, 120, 180, 240, and 300 epochs.

\begin{figure}[htp]
\centering
\includegraphics[width=0.3\textwidth]{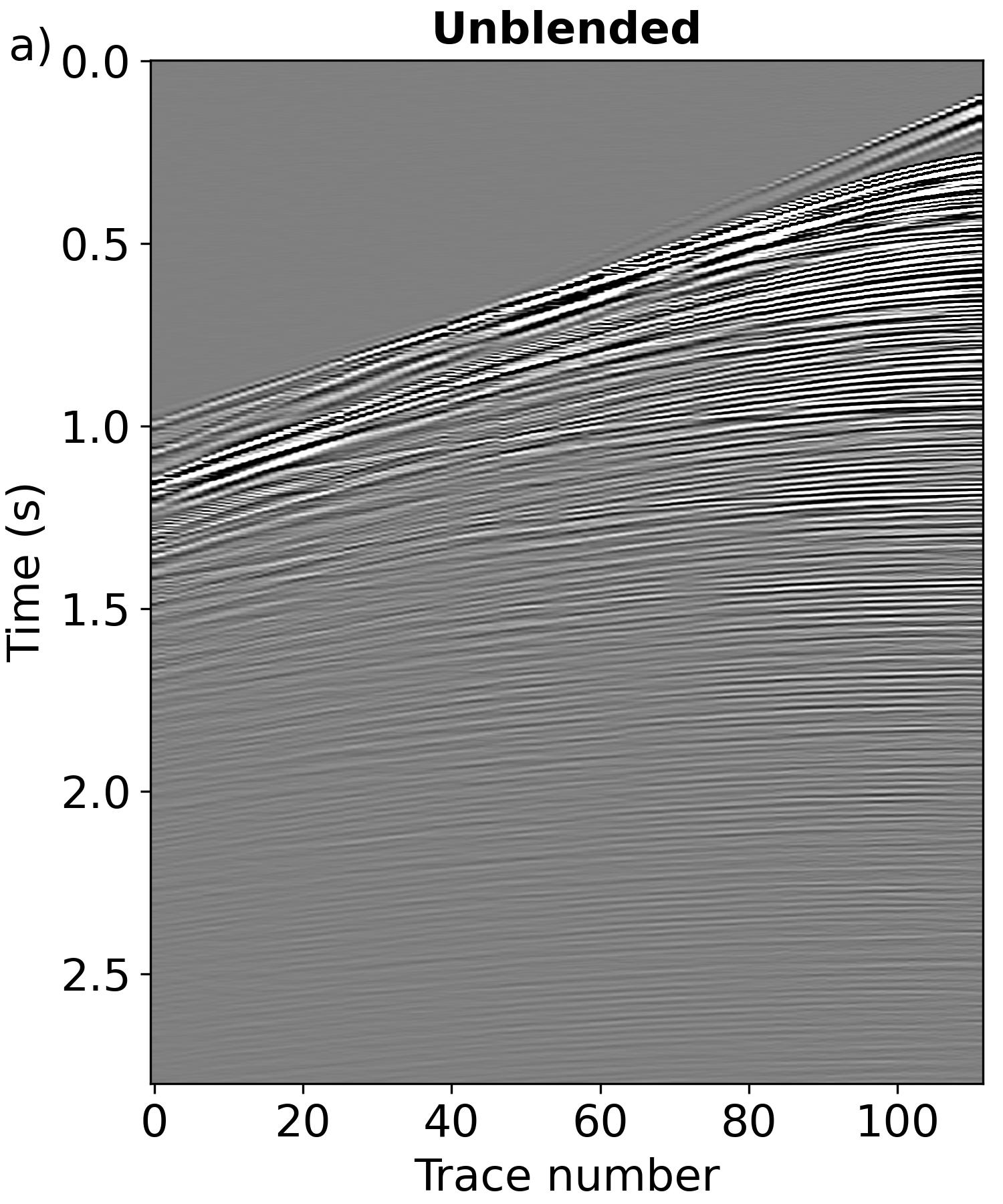}
\hspace{1cm}
\includegraphics[width=0.3\textwidth]{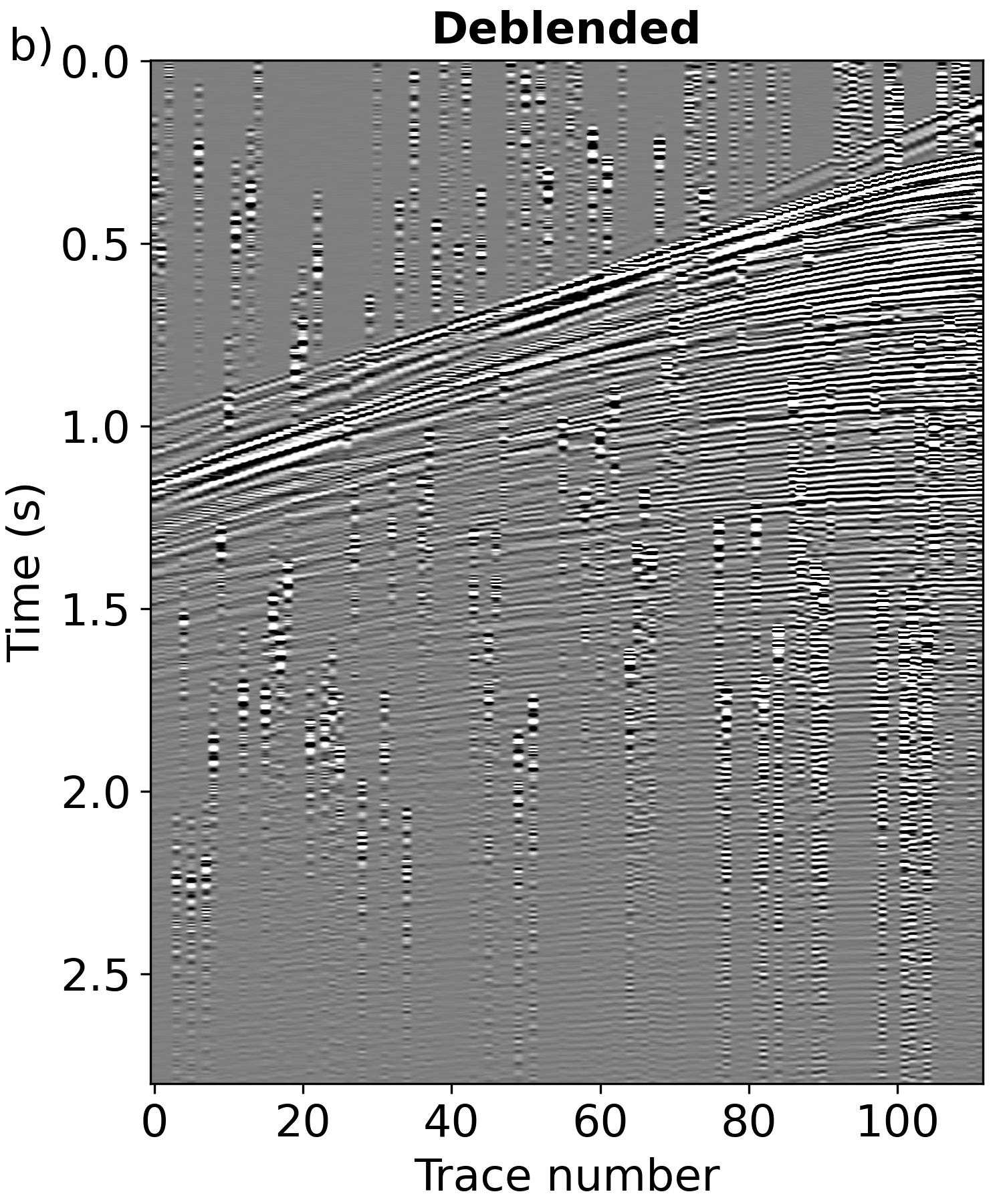}
\caption{The raw unblended shot (a) and the pseudo-deblended (b) shot acquired from southern North Sea. }
\label{fig14}
\end{figure}

\begin{figure*}[!t]
\centering
\includegraphics[width=0.3\textwidth]{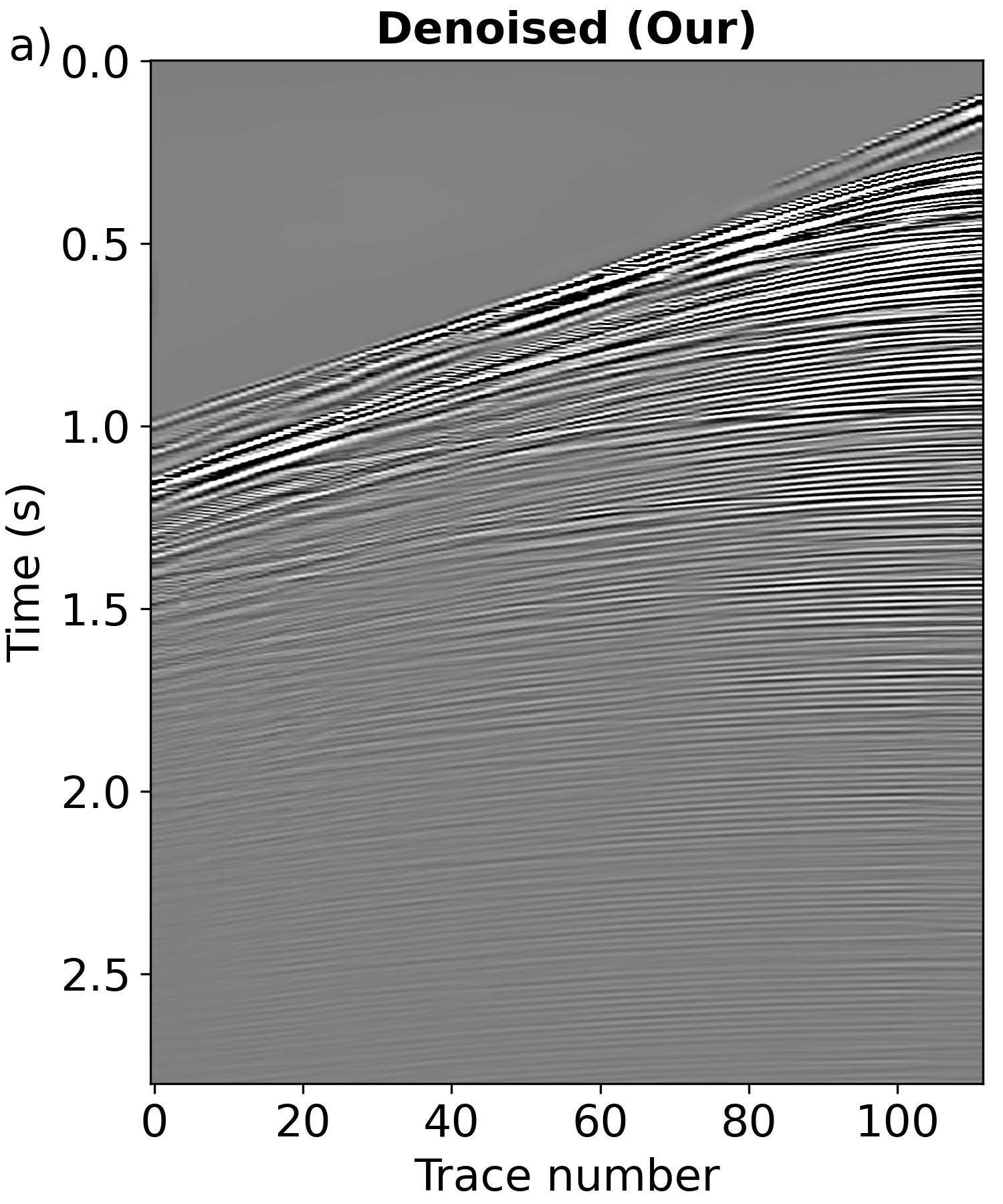} 
\includegraphics[width=0.3\textwidth]{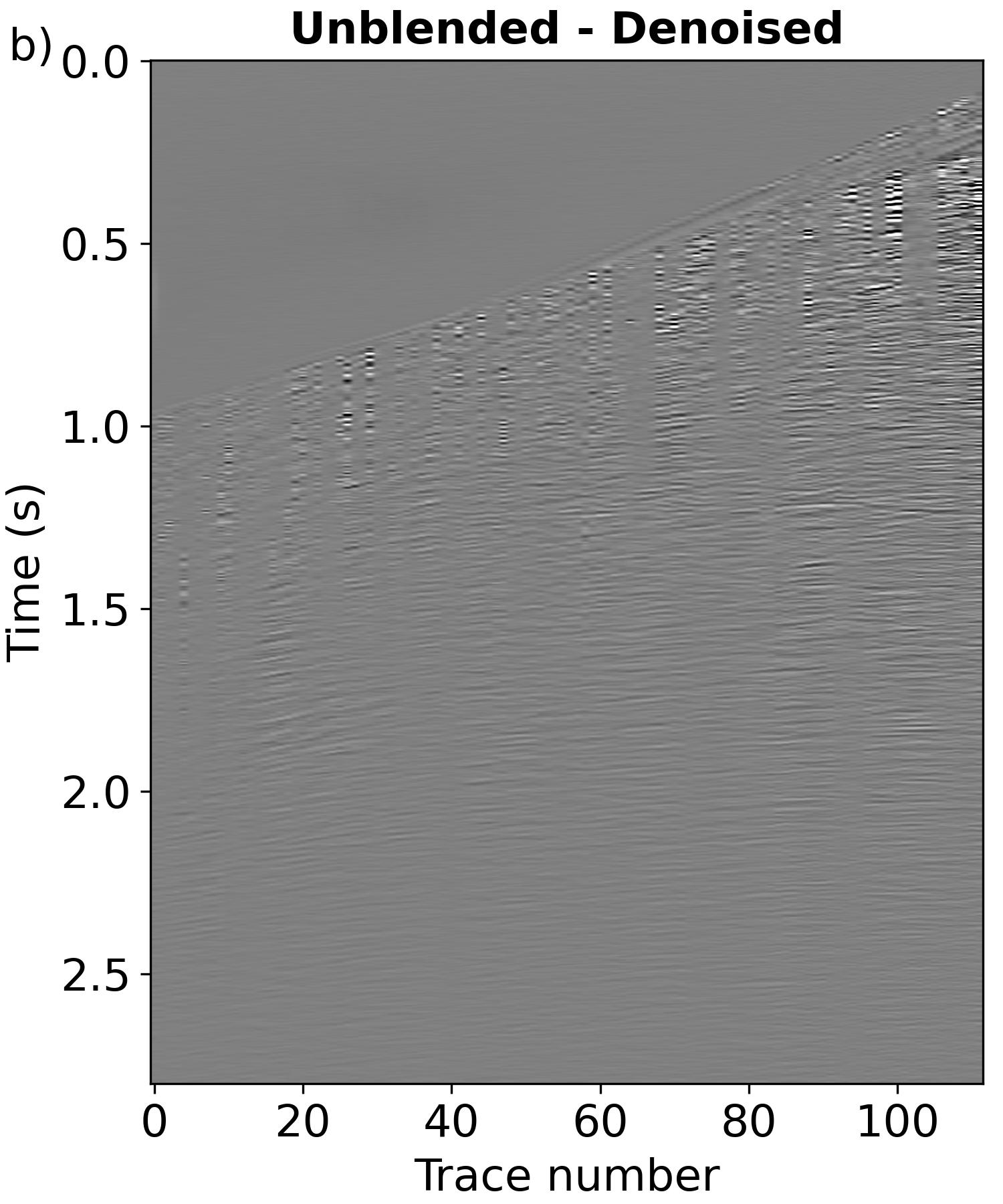}
\includegraphics[width=0.3\textwidth]{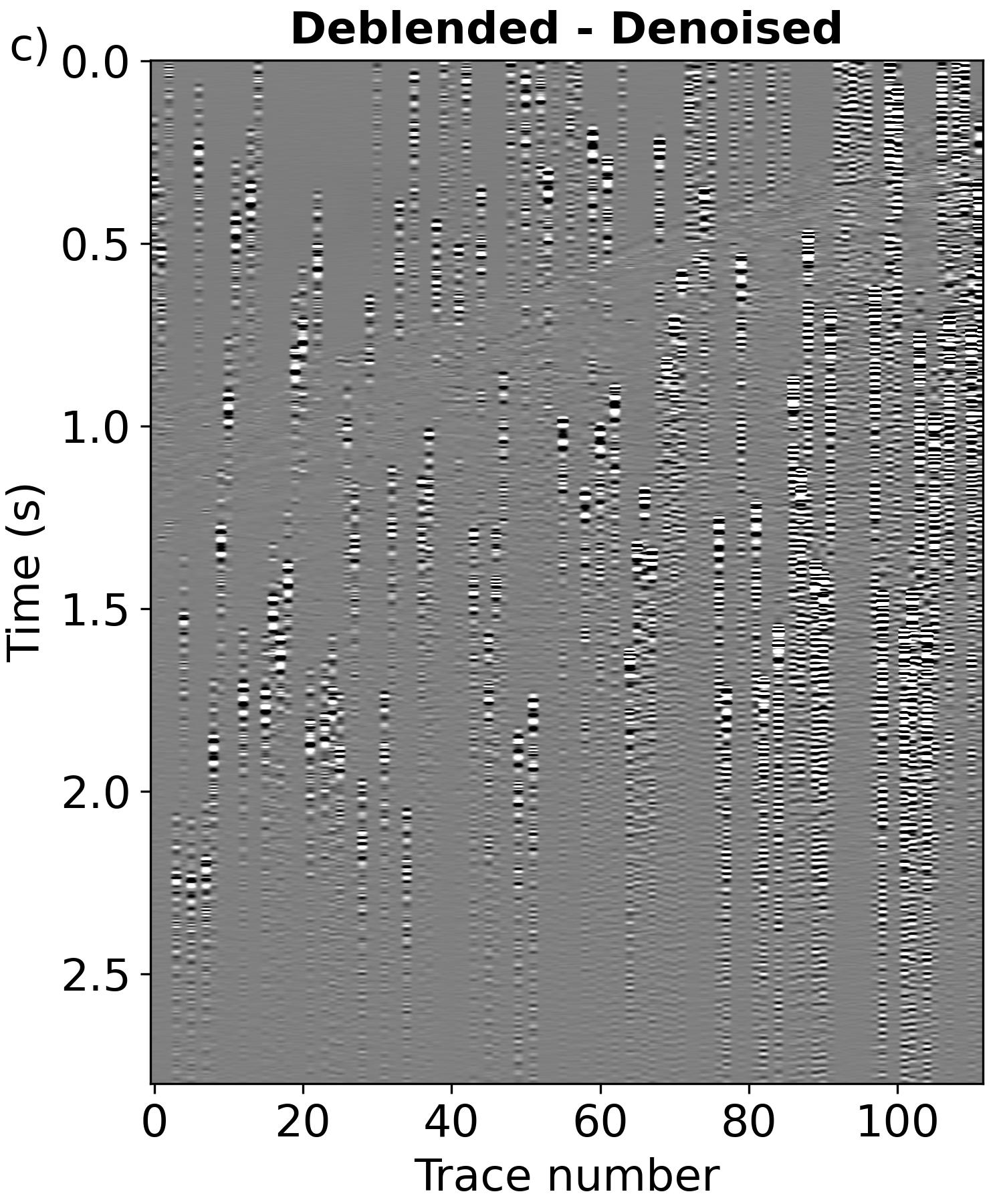} \\
\includegraphics[width=0.3\textwidth]{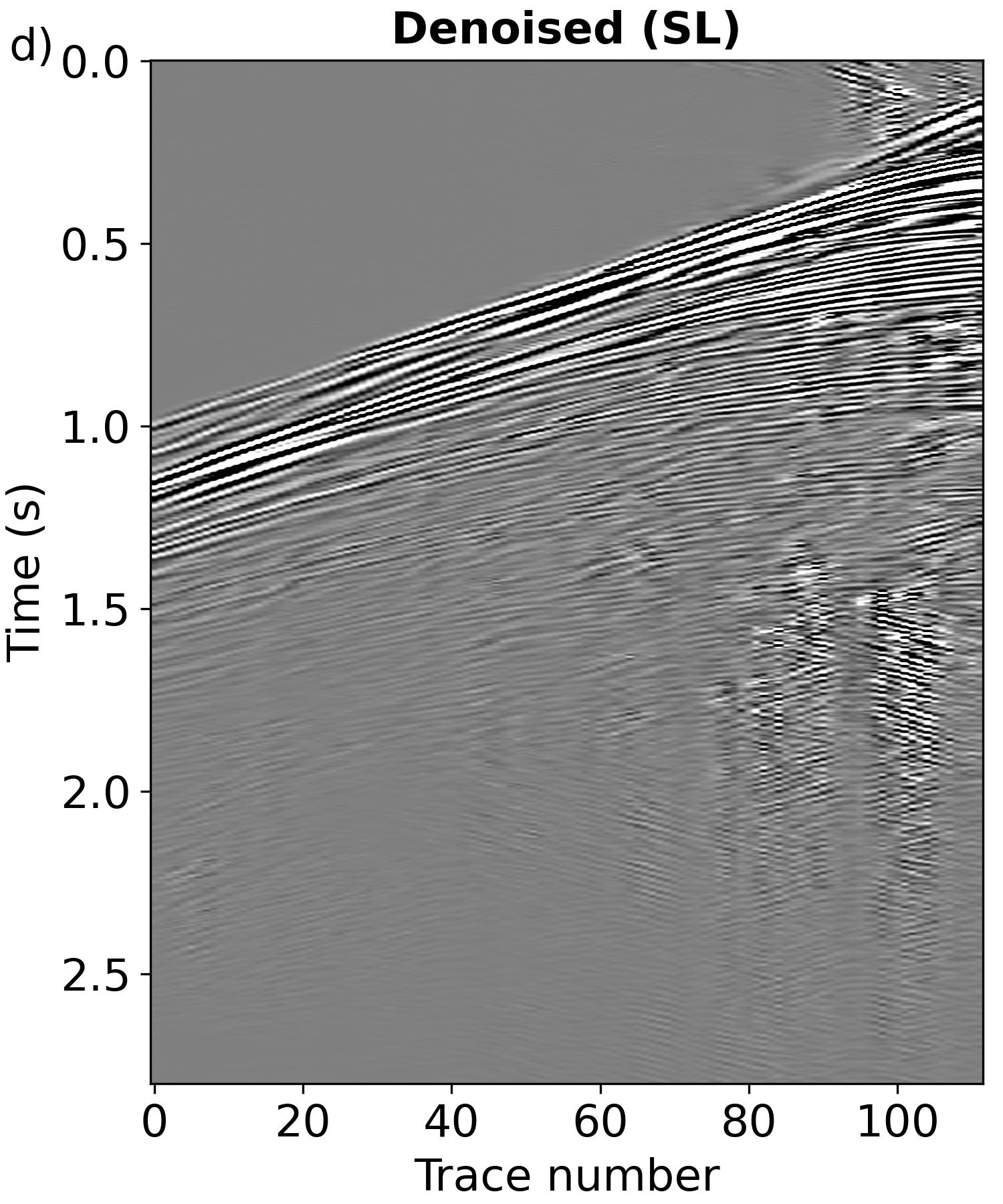} 
\includegraphics[width=0.3\textwidth]{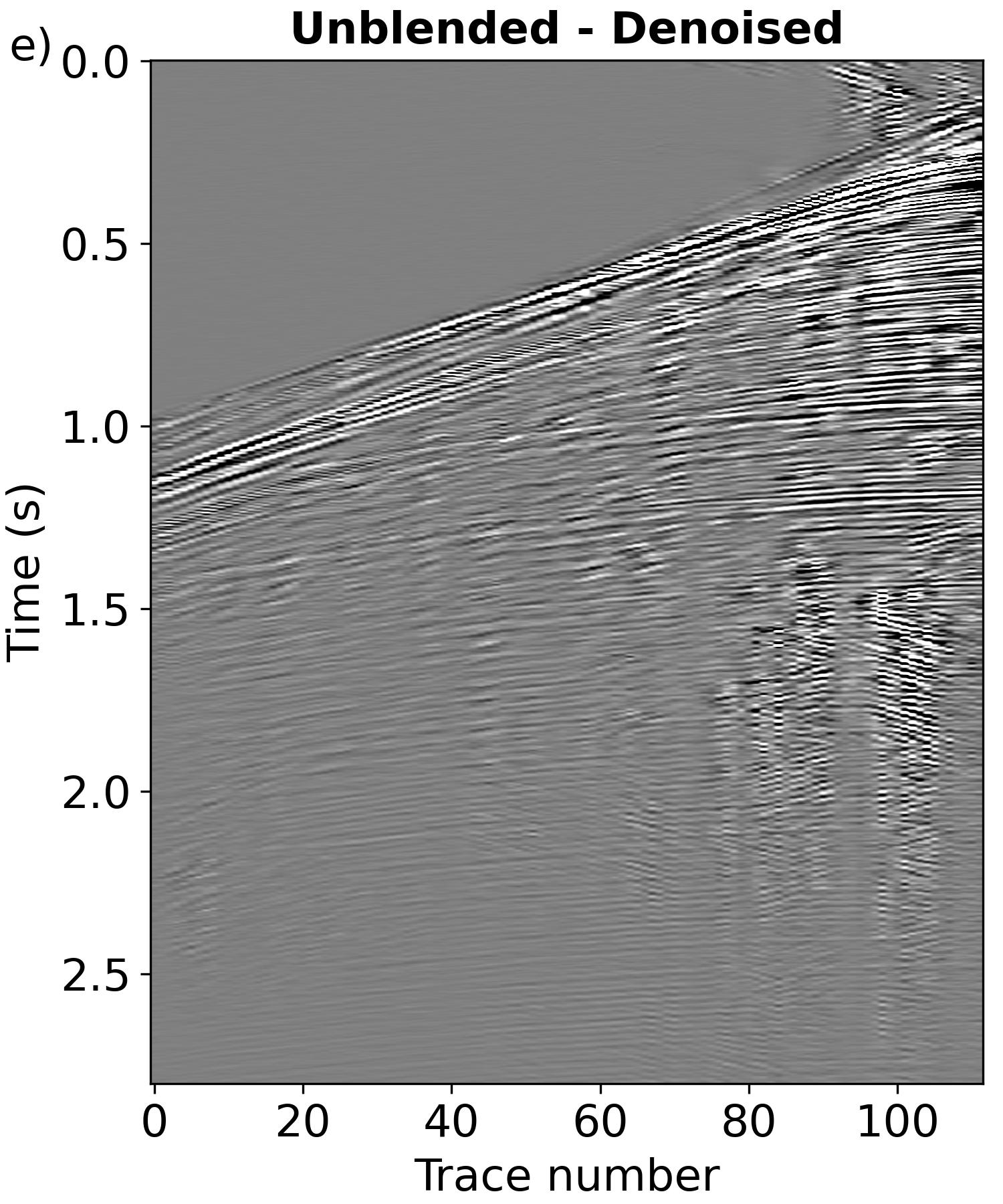} 
\includegraphics[width=0.3\textwidth]{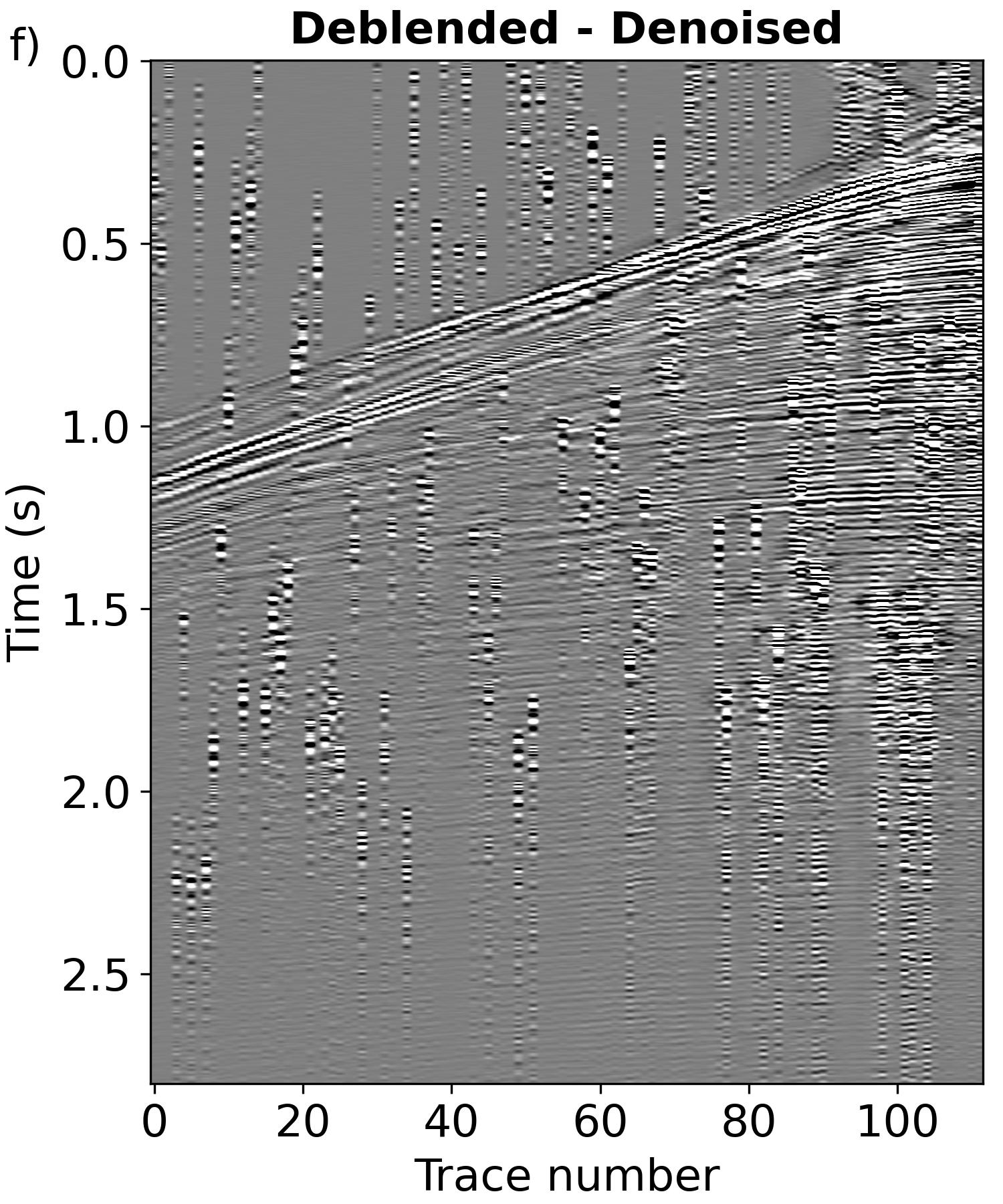} \\
\includegraphics[width=0.3\textwidth]{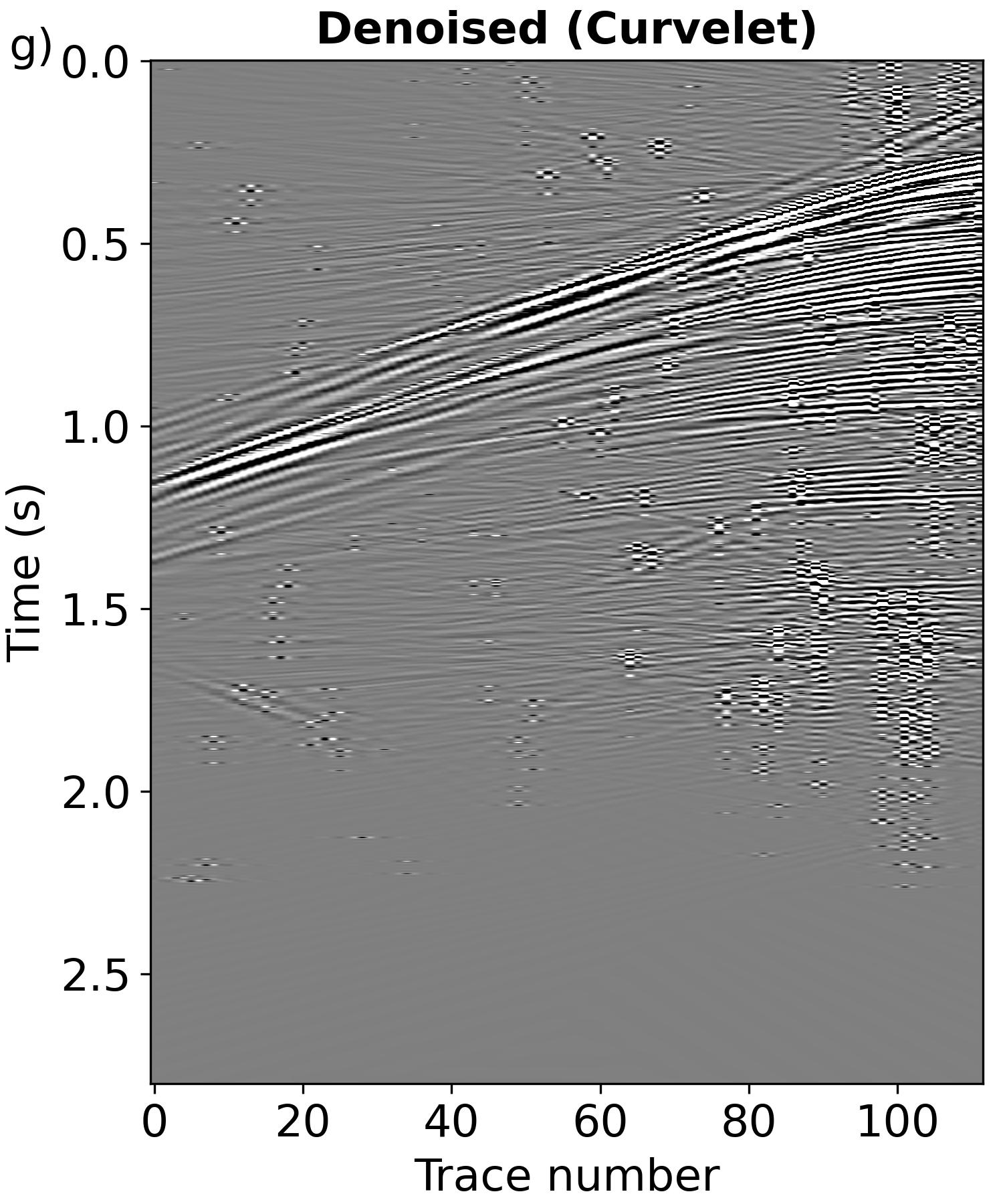} 
\includegraphics[width=0.3\textwidth]{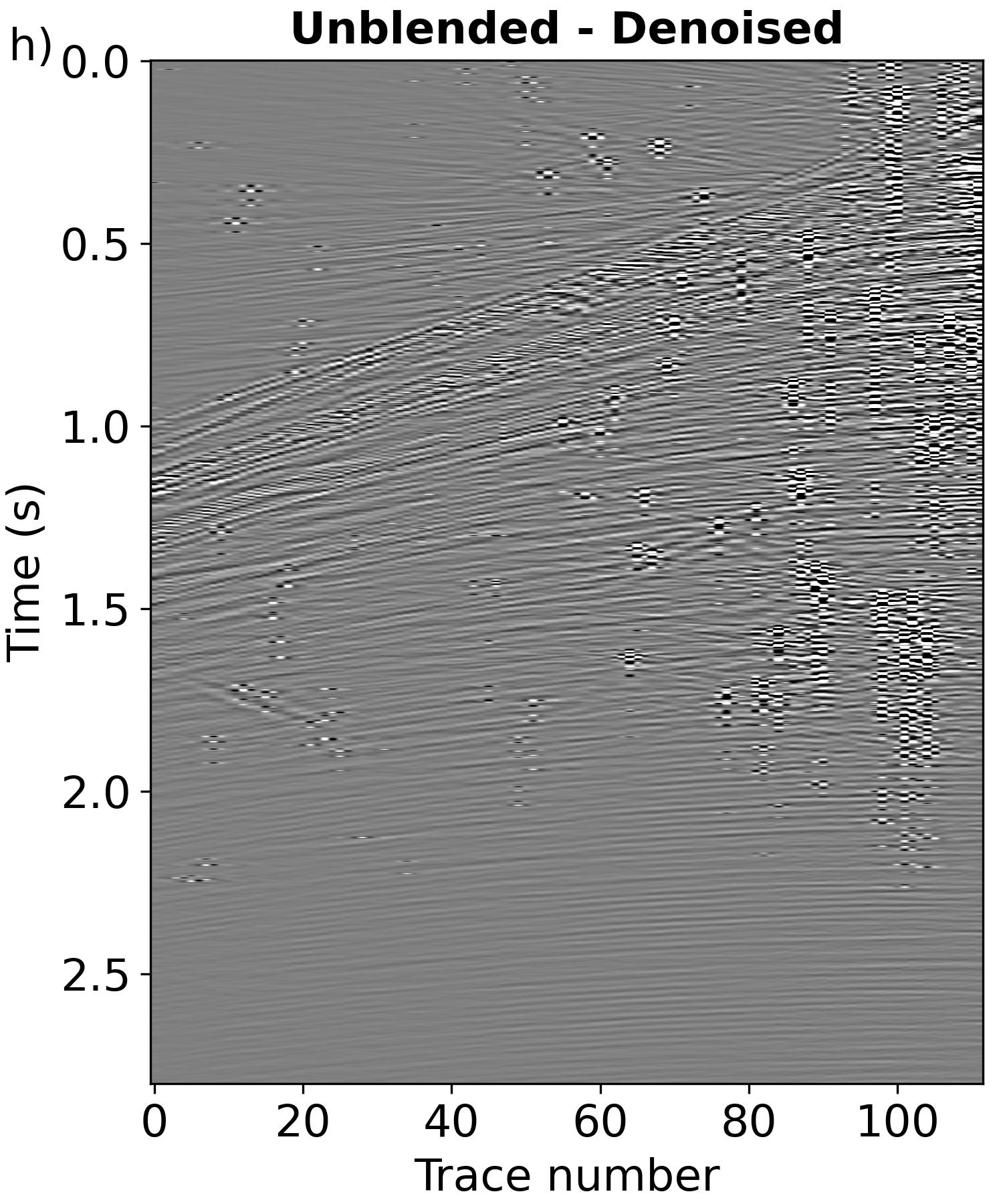}
\includegraphics[width=0.3\textwidth]{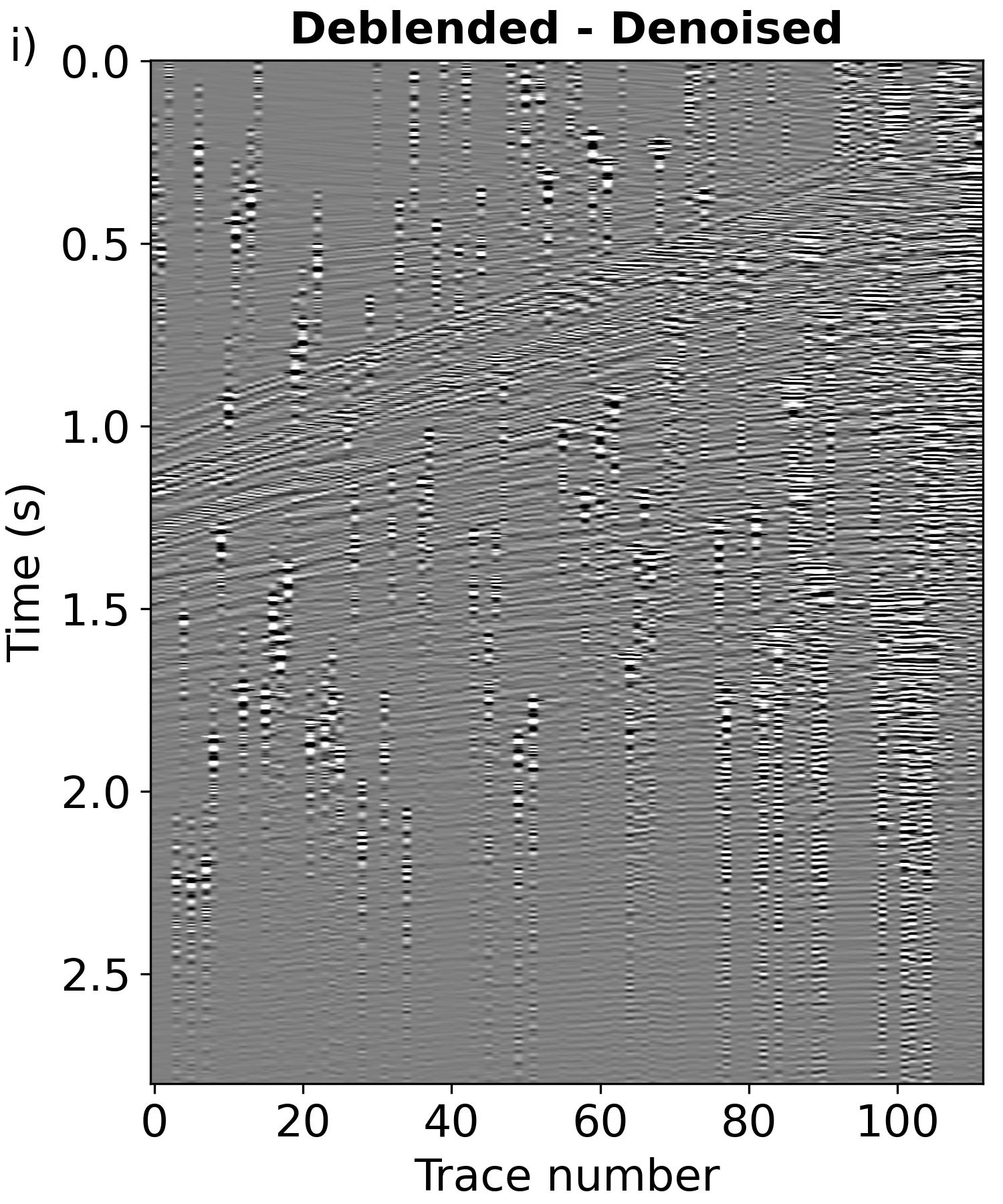}
\caption{Comparison of the denoising performance on a field pseudo-deblended shot gather, which acquires from southern North Sea and is contaminated by blending noise. The first column corresponds to denoised results. The second column shows the difference between the denoised results and the raw unblended data, while the third column shows the difference between the denoised results and the deblended data.}
\label{fig15}
\end{figure*}

Fig. \ref{fig14} presents a pseudo-deblended shot gather used for testing (panel b) and the original undeblended shot gather (panel a). The denoising products of three methods are depicted in Fig. \ref{fig15}. It can be seen that our approach results in some signal leakage in the shallower regions, likely due to the strong direct wave energy at this region. However, in the deeper parts, our method preserves the effective signals almost impeccably. In contrast, since the SL method is trained solely on synthetic data and directly applied to field data, it predictably delivers an inferior denoising product, producing pronounced artifacts in both the direct wave and deeper regions. The curvelet method also fails to provide satisfactory denoising results, clearly struggling with the challenge of handling strong coherent noise. These comparisons further validate the proficiency of our method in attenuating complex blending noise.

\section{Discussion}
\subsection{Why warm-up is needed}
In the results we present, it is evident that, compared to training solely on a fixed noisier-noisy dataset (as shown in Fig. \ref{fig2}), IDR phase offers a notable enhancement in denoising performance. However, the significance of the warm-up phase within our SSL denoising approach cannot be overlooked. In the early stages of training, the denoising network model hasn't yet adapted to the noisy data nor effectively captured the characteristics of the noise, rendering it less robust. Utilizing the network to directly generate pseudo-labels at this stage might introduce an excessive amount of noise. If we integrate these noisy labels into our training regimen, the denoising network model might diverge in a counterproductive direction. Consequently, our denoising network may struggle to converge to a more favorable optimization point. To testify this perspective, we next present an example of attenuating blending noise from synthetic pseudo-deblended data.

We further train three additional denoising networks. These networks share the same training data and configurations as our synthetic blending noise attenuation numerical example. The only difference is that during the warm-up phase, these three denoising networks undergo 1, 10, and 20 epochs of training, respectively, referred to as warm-up 1, warm-up 10, and warm-up 20. In contrast, the previous blending attenuation example adopted a warm-up of 30 epochs (denoted as warm-up 30). The performance metrics for the denoised results of the test data (shown in Fig. \ref{fig12}) by the four trained networks are presented in Table \ref{tab3}.

We can see that with just one warm-up training, the corresponding denoising result's SNR has only a slight improvement, and the MAE metric, surprisingly, decreased compared to the original noisy data, suggesting a suboptimal denoising product. As the number of epochs in the warm-up phase increases, there is a noticeable enhancement in the SNR of the denoised results, with the MAE steadily declining. This confirms the indispensability of the warm-up phase, and that a lengthier warm-up training progressively augments the network's denoising capability. However, an extended warm-up training also increases computational costs. Hence, to strike a balance between the training duration and the denoising performance, we chose to perform 30 epochs of training during the warm-up phase in our implementation.

\begin{table}
\centering
\caption{The comparison of denoising performance of our methods across different warm-up training epochs, including SNR and MAE metrics.}
\renewcommand\arraystretch{1.5}
\setlength{\tabcolsep}{20pt}
\begin{tabular}{ccc}
    \hline
    \text {~} & \text { SNR } & \text { MAE } \\
    \hline
    \text {Raw noisy data} & $-3.84$ & $0.00759$ \\
    \text {Warm-up 1}  & $-1.87$ & $0.00935$ \\
    \text {Warm-up 10} & $17.16$ & $0.00152$ \\
    \text {Warm-up 20} & $17.25$ & $0.00154$ \\
    \text {Warm-up 30} & $17.83$ & $0.0014$ \\
    \hline
\end{tabular}
\label{tab3}
\end{table}

\subsection{How to set the noise level}
In the three distinct examples where we attenuated seismic noise, the level of noise introduced when constructing the noisier-noisy dataset varied. So, when confronted with different data and diverse noise categories for denoising, how does one ascertain this noise level? In the following, we will share our experiences on determining the appropriate noise level.

\begin{figure}[htp]
\centering
\includegraphics[width=0.3\textwidth]{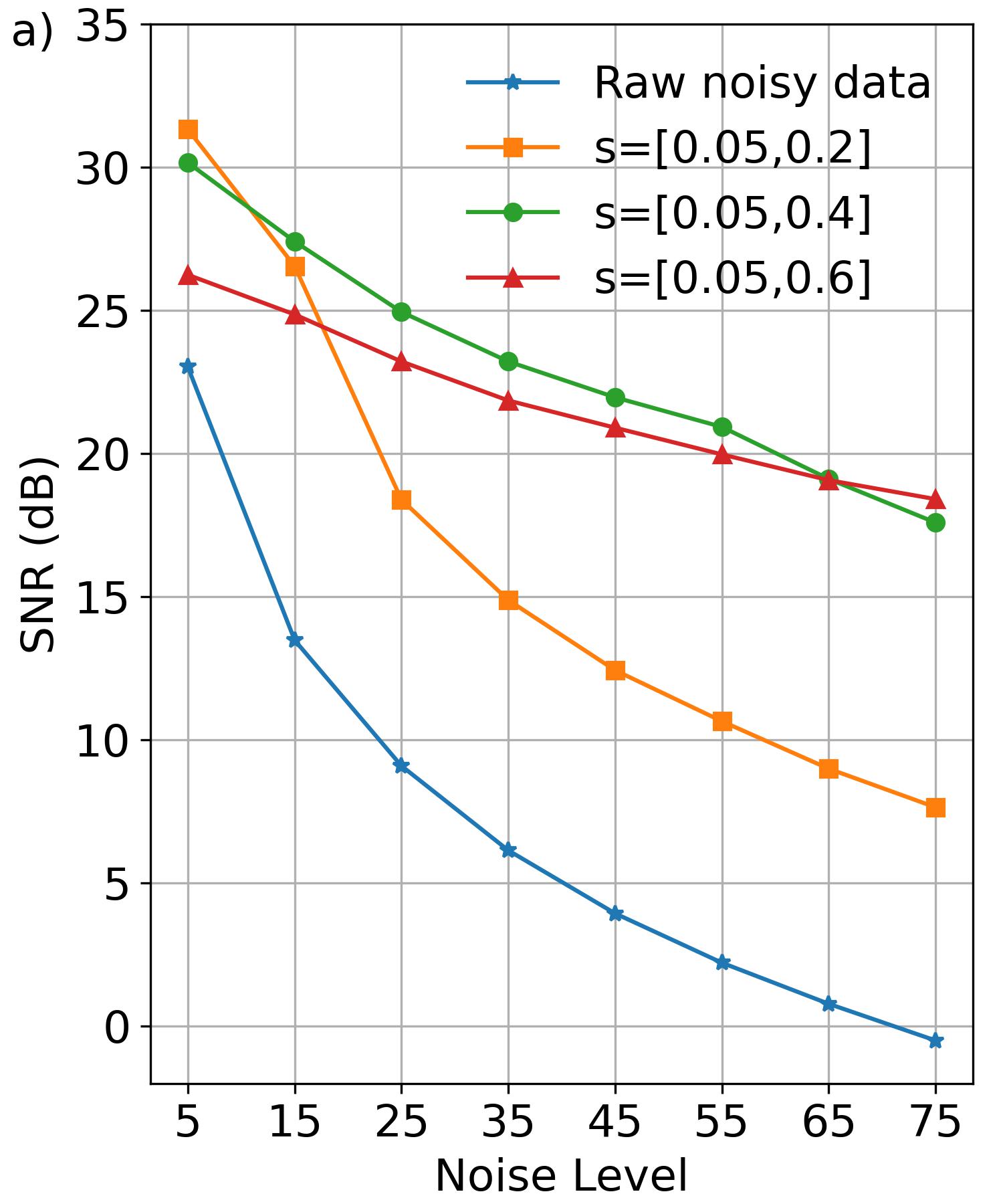}
\hspace{1cm}
\includegraphics[width=0.3\textwidth]{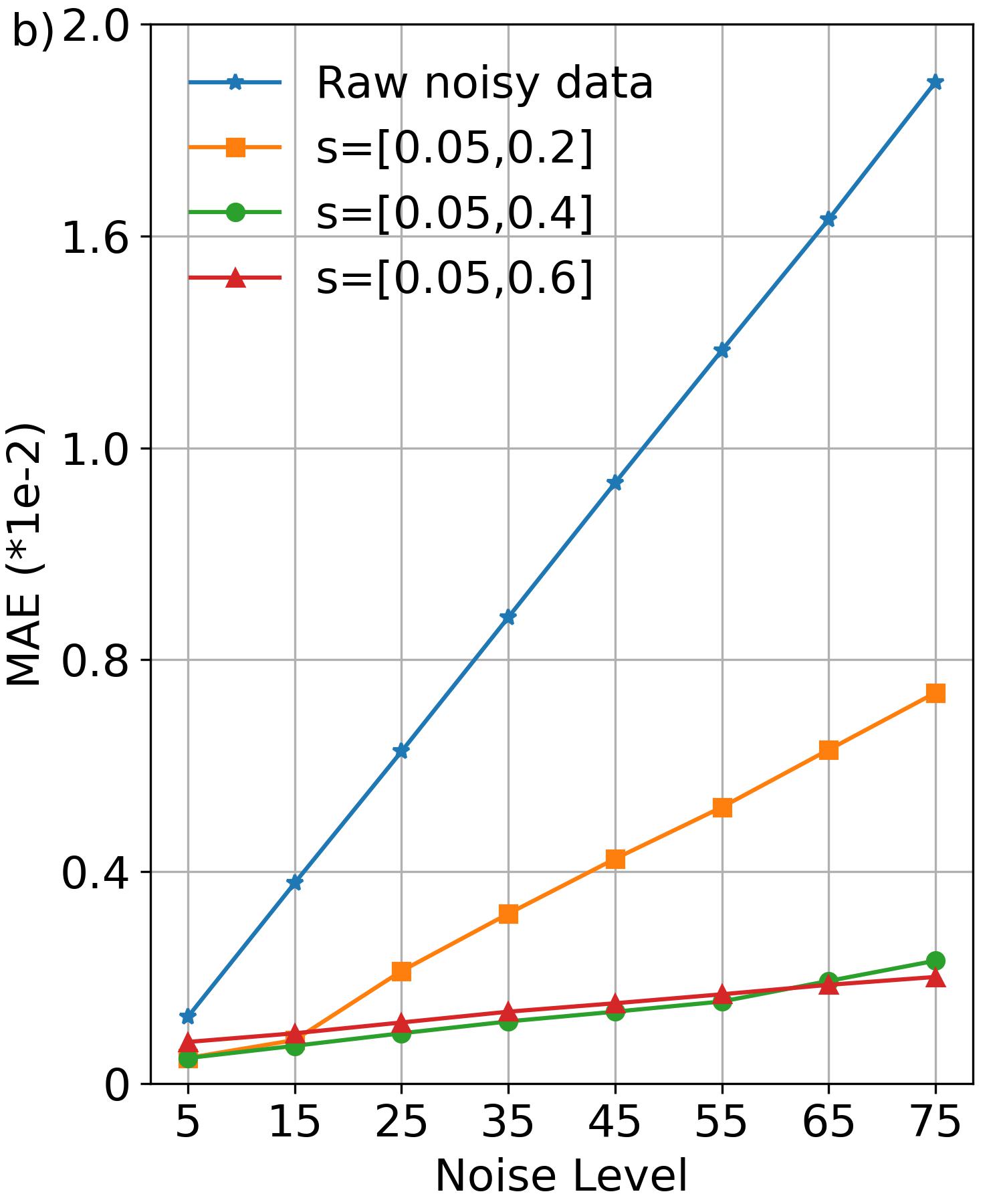}
\caption{The denoising performance comparison of the network trained utilizing the different levels of noise added: $s=[0.05, 0.2]$, $s=[0.05, 0.4]$, and $s=[0.05, 0.6]$. This is benchmarked against the test noisy data comprising 8 distinct noise levels. The blue line represents the raw noisy test data. (a) Denoised results using SNR metric. (b) Denoised results using MAE metric.}
\label{fig16}
\end{figure}

We use the example of random noise attenuation to illustrate how to set the noise level. In the numerical example section of the synthetic data, the noise levels added during the warm-up and IDR phases were $s=[0.05, 0.4]$. Here, we incorporate two additional tests that utilize the same training dataset and training configuration as the previous experiment. However, the noise levels added during the warm-up and IDR phases for these tests are $s=[0.05, 0.2]$ and $s=[0.05, 0.6]$, respectively. The denoising networks trained in these two tests, along with the previously trained denoising network, are employed to predict the denoised results for noisy data contaminated by random noise at eight different noise levels. The SNR and MAE metrics of the predicted results from these three denoising networks, in comparison to the original data's SNR and MAE metrics, are depicted in Fig. \ref{fig16}.

We can observe that when the added noise level is low, for example $s=[0.05, 0.2]$, the trained network exhibits superior denoising performance for weak noise. However, as the noise intensifies, its denoising capability significantly reduces. Conversely, when the added noise level aligns with the original noisy dataset, such as $s=[0.05, 0.4]$, the trained denoising network showcases enhanced generalizability and robustness within this noise level range, meaning it possesses commendable denoising capability for noisy data within this range. If the range of added noise levels is broad and encompasses high noise levels, as seen with $s=[0.05, 0.6]$, the trained denoising network's proficiency in denoising data with low noise levels might decline, yet its efficacy in attenuating noise in data with high noise levels tends to improve.

These findings indicate that the level of noise introduced influences the denoising performance of our method. To achieve optimal denoising results for the target noisy data, it is advisable to preliminarily estimate the degree to which the noisy data is contaminated, that is, to gauge the noise level. This is not particularly challenging. For example, for the field data contaminated with random noise, we can introduce random noise to the noisy data and then observe the changes in contamination before and after. By comparing and observing these changes, we can approximate the contamination level of the noise. For the field data affected by the backscattered noise and the blending noise, since we extract it directly from the noisy data, the level of the directly extracted noise is defined as 1. Thus, we should ensure the noise level introduced primarily matches the intensity of the extracted noise.

\subsection{Limitations}
While our method demonstrates superior performance in attenuating various types of noise, it is essential to acknowledge its limitations. One significant constraint is the reliance on prior knowledge of the noise model; that is, we need to accurately access noise information. This dependence stems from our adoption of the Noisier2Noise approach, which necessitates the introduction of additional noise to already noisy data, thereby generating noisier datasets. In our implementations, random noise is anticipated through prior knowledge and is directly simulated using Equation \ref{eq3}. In contrast, the backscattered noise and the blending noise are extracted directly from the noisy data. When the noise model is elusive or when noise extraction becomes challenging, the creation of a noisier-noisy dataset, and thus the realization of our method, could be intricate.

Fortunately, current research suggests that conventional denoising techniques can be employed as an initial step to extract the desired noise. When direct extraction from data becomes unfeasible, one might contemplate leveraging traditional methods like f-k or curvelet method to capture noise insights \cite{liu2018random, mandelli2019interpolation, yuqing2019random}. Alternatively, noise modeling techniques, which has been reported in some works \cite{dong2023potential}, could be employed to produce data resembling real-world noise. Such simulated noise can then be harnessed to construct a noisier-noisy dataset, facilitating the application of our denoising methodology.
\section{Conclusion}
We developed a seismic denoising method designed for the attenuation of various noise types in a self-supervised learning (SSL) fashion. Within our paradigm, the neural network (NN) undergoes both a warm-up and an iterative data refinement (IDR) stage, aiming to effectively enhance its denoising capability even in the absence of label data. During the warm-up phase, we generate a noisier-noisy dataset by introducing additional noise into the original noisy data. Here, the noisier data serves as the input, while the original noisy data stands as the corresponding label, allowing the NN to undergo a preliminary warm-up to ensure stability during the IDR phase. In the IDR stage, a new noisier-noisy training set is constructed at the onset of each epoch. The label for this dataset is derived from the previous epoch's network predictions on the original noisy data, with the input created by adding noise to these predicted labels. Throughout the IDR phase, we iteratively perform multiple epoch trainings to reduce the data bias between the noisier-noisy dataset and the noisy-clean dataset required for supervised training (SL), thereby achieving denoising performance competitive with SL. Through tests on three distinct noise attenuation scenarios, and by contrasting our approach with SL and the curvelet method, the efficacy of our method has been validated. It not only effectively removes noise but also minimizes signal leakage. This outstanding denoising capability offers a reliable strategy for achieving seismic attenuation in real-world scenarios using an SSL approach.
\section*{Acknowledgment}
The authors would like to thank Seismic Wave Analysis Group (SWAG) for the fruitful discussions. They also thank CGG for sharing the field seismic data.

\section*{Data availability statement}
Data associated with this research are available and can be obtained by contacting the corresponding author. \\

\bibliographystyle{unsrt}  
\bibliography{references}

\end{document}